\documentclass[pra,twocolumn,preprintnumbers,amsmath,amssymb,floatfix,superscriptaddress,showpacs]{revtex4}
\usepackage{graphicx}
\usepackage{graphics}
\usepackage{psfrag}
\usepackage{hyperref}
\usepackage{multirow}
\usepackage[sort&compress]{natbib}

\newcommand{\bphi}{\bar{\phi}}
\newcommand{\bpsi}{\bar{\psi}}
\newcommand{\vare}{\varepsilon}
\newcommand{\sgn}{\mbox{sgn}}
\newcommand{\rmi}{{\rm i}}

\begin{document}

\hypersetup{pdftitle={Critical temperature and superfluid gap of the Unitary Fermi Gas from Functional Renormalization}}
\title{Critical temperature and superfluid gap of the Unitary Fermi Gas\\ from Functional Renormalization}
\date{\today}
\author{Igor Boettcher}
\affiliation{Institute for Theoretical Physics, University of Heidelberg, D-69120 Heidelberg, Germany}
\author{Jan M. Pawlowski}
\affiliation{Institute for Theoretical Physics, University of Heidelberg, D-69120 Heidelberg, Germany}
\affiliation{ExtreMe Matter Institute EMMI, GSI Helmholtzzentrum f\"{u}r Schwerionenforschung mbH, D-64291 Darmstadt, Germany}
\author{Christof Wetterich}
\affiliation{Institute for Theoretical Physics, University of Heidelberg, D-69120 Heidelberg, Germany}

\begin{abstract}
We investigate the superfluid transition of the Unitary Fermi Gas by means of the Functional Renormalization Group, aiming at quantitative precision. We extract $T_{\rm c}/\mu=0.38(2)$ and $\Delta/\mu=1.04(15)$ for the critical temperature and the superfluid gap at zero temperature, respectively, within a systematic improvement of the truncation for the effective average action. The key new ingredient in comparison to previous approaches consists in the use of regulators which cut off both frequencies and momenta. We incorporate renormalization effects on both the bosonic and the fermionic propagator, include higher order bosonic scattering processes, and investigate the regulator and specification parameter dependence for an error estimate. The ratio $\Delta/T_{\rm c}=2.7(3)$ becomes less sensitive to the relative cutoff scale of bosons and fermions when improving the truncation. The techniques developed in this work are easily carried over to the cases of finite scattering length, lower dimensionality, and spin-imbalance.
\end{abstract}


\maketitle


\section{Introduction}
Precision measurements of the equation of state and the phase diagram of ultracold fermions \cite{Horikoshi22012010,Salomon,Navon07052010,Ku03022012} allow for a unique benchmarking of many-body theory. By comparing to experimental data, it is possible to identify advantages and shortcomings of particular theoretical methods and approximations. This provides important insights into the mechanisms which govern interacting many-body systems. One also gains information on the appropriate manner to compute observables for them, and hence sheds light on a wide variety of other physics systems. Indeed, the clear picture obtainable from ultracold quantum gases can, for instance, be applied in the context of solid state physics, heavy ion collisions, or  astrophysics.

The superfluid phase transition of two-component fermions in the BCS-BEC crossover  changes its character as the interaction strength is varied by means of the application of an external magnetic field. Whereas the features of the weakly attractive superfluid are well-captured by BCS theory, the ground state of the weakly repulsive gas can be described as a condensate of bosonic molecules. In the intermediate region of large scattering length the superfluid is strongly correlated, and the pairing mechanism does not fit into one of the two perturbative cases. A deeper understanding of strongly correlated superfluids is expected to have a great impact on research on both high-temperature superconductors and neutron stars. For an introduction to the physics of the BCS-BEC crossover and a complete account of experimental and theoretical references see Refs. \cite{RevModPhys.80.885,Zwerger,Randeria:2013kda}.

The order parameter of the superfluid transition is given by the gap $\Delta$, which we define here as the constant part of the anomalous fermion self-energy. For zero temperature, this observable manifests itself as a gap in the energy spectrum of fermionic excitations, hence the name. The presence of a nonvanishing anomalous self-energy indicates the spontaneous breaking of U(1)-invariance of global phase rotations in the system. The order parameter $\Delta(T)$ diminishes due to thermal fluctuations as temperature is increased, and eventually vanishes continuously at the critical temperature $T_{\rm c}$ in a second order phase transition. In the presence of a pseudogap $\Delta_{\rm pg}$, the excitation spectrum of fermions keeps memory of the gap even in the normal phase. Consequences of a pseudogap are a suppression of both the density of states and the entropy just above the critical temperature.

The equilibrium physics of interacting many-body systems is accessible from a path integral formulation of the corresponding Euclidean quantum field theory. The evaluation of the path integral, however, is necessarily bound to approximations and truncations. In particular, in the nonperturbative regime of strong correlations no simple ordering principle governs the analysis. Typically, the error estimation for theoretical predictions is problematic in these cases, as there is no obvious way of how to improve a given approximation scheme. This is most prominent for functional methods like the Functional Renormalization Group (FRG) or Dyson--Schwinger equations (DSE), which rely on a truncation of the underlying space of functionals.

The purpose of this work is to obtain a quantitatively accurate picture of the BCS-BEC crossover by means of the FRG, thereby keeping the truncation as simple and physically clear as possible. This can be achieved by the use of regulators which cut off both frequencies and momenta. A main objective of this paper is a comparison of the effects of truncations, regulators, and specification prescriptions. For this purpose we concentrate on a few observables for the Unitary Fermi Gas. Examples are the critical temperature and the gap for which we display our results in Table \ref{TableResults}.  We emphasize, however, that the procedure can be applied to the whole range of interaction parameters, as well as to the lower-dimensional and spin-imbalanced setting. (The latter requires a more sophisticated treatment of the effective potential to resolve the first order phase transition.)

\begin{table}
\begin{tabular}[t]{|c||c|c|c|}
 \hline
  Observable & \hspace{3mm} $T_{\rm c}/\mu$  \hspace{3mm} &  \hspace{3mm} $\Delta/\mu$  \hspace{3mm} &  \hspace{3mm} $\Delta/T_{\rm c}$  \hspace{3mm} \\
\hline \hline
 Mean field & 0.664 & 1.16 & 1.75 \\
 \hline
 \ Truncation 1 \ & \ 0.38(2) \ & \ 1.17(6) \ & \ 2.9(2) \ \\
 Truncation 2 & 0.376(14) & 1.18(6) &  3.0(2)  \\
 Truncation 3 & 0.385(20) & 1.04(5) & 2.6(1) \\
 Truncation 4 & 0.38(2) & 0.89(5) & 2.4(1)\\
 \hline
 \ Best estimate \ & 0.38(2) & 1.04(15) & 2.7(3) \\
\hline
\end{tabular}
\label{TableResults}
\caption{Critical temperature and superfluid gap of the Unitary Fermi Gas obtained by successivley extending the truncation of the effective average action. The truncations with a smaller number are contained in the ones with a larger number. The particular choices of running couplings for each truncation are explained at the beginning of Sec. \ref{estimate}. The error in brackets gives the systematic error within the given truncation. The errors of $T_{\rm c}/\mu$ and $\Delta/\mu$ in truncations 3 and 4 result from an uncertainty in the chemical potential $\mu$.}
\end{table}

The Unitary Fermi Gas has been studied theoretically with a variety of methods including quantum Monte Carlo \cite{PhysRevLett.91.050401,PhysRevLett.93.200404, PhysRevLett.95.060401,PhysRevLett.95.230405,PhysRevLett.96.160402, PhysRevLett.100.150403, PhysRevA.78.023625,PhysRevA.81.033619,PhysRevA.82.053621,PhysRevLett.106.205302, PhysRevA.84.061602}, T-matrix approaches \cite{HaussmannZPhys91,PhysRevB.49.12975,PhysRevA.75.023610,PhysRevA.80.063612,PhysRevA.82.021605,Enss2011770,HuDrummond}, $\epsilon$-expansion \cite{PhysRevLett.97.050403,PhysRevA.75.063617,PhysRevA.75.063618}, $1/\mathcal{N}$-expansion \cite{PhysRevA.86.013616}, DSE \cite{PhysRevA.73.033615,Diehl:2005ae,PhysRevA.77.023626}, Wilsonian renormalization group \cite{PhysRevA.75.033608,PhysRevLett.100.140407,PhysRevA.84.013610,Gubbels:2013mda}, and the FRG \cite{Birse:2004ha,PhysRevA.76.021602,PhysRevA.76.053627,PhysRevB.78.174528,ANDP:ANDP201010458,Tanizaki:2013doa,Tanizaki:2013fba,Tanizaki:2013yba}. This allows for a solid benchmarking of our results by comparing to values for the critical temperature and superfluid gap from other approaches. In our conclusions section, we give an overview of the reference values for $T_{\rm c}/\mu$ and $\Delta/\mu$ in Tables VI and VII, respectively.

We investigate the physics of the Unitary Fermi Gas by means of a partially bosonized model, where the momentum dependence of the particle-particle loop is resolved in terms of a composite bosonic degree of freedom, described by a complex scalar field $\phi$. This has several advantages in comparison to a purely fermionic treatment. First, it allows in a straightforward manner to describe regimes with spontaneously broken symmetry and to resolve the corresponding Goldstone boson fluctuations. The latter are particularly important at the second order phase transition, where the correct universal physics of the O(2)-universality class is recovered. 

Moreover, bosonic interactions are encoded in the effective potential $U(\phi)$. Expanding the latter in powers of the field to order $\phi^{2N}$ includes interactions between $N$ bosons, and, accordingly, $2N$ fermions. For instance, a $\phi^8$-truncation, which is often applied in the following, describes interactions of eight atoms, which are difficult to treat in a purely fermionic description. This is less important for the perturbative regimes of the BCS-BEC crossover, but necessary to capture the quantitative features of the strongly correlated Unitary Fermi Gas.

The bosonization of the particle-particle loop neglects the contribution of the particle-hole loop on the flow of the four-fermion interaction.  Previous FRG studies on particle-hole fluctuations in the BCS-BEC crossover \cite{PhysRevB.78.174528} have shown how a flowing bosonization scheme \cite{PhysRevD.65.065001,Floerchinger:2010da,Pawlowski20072831} can undo this shortcoming, and, for instance, reproduce the Gorkov correction on the the BCS side of the crossover. However, it was also found that the effect of particle-hole fluctuations on the Unitary Fermi Gas is rather small due to the less pronounced Fermi surface. Therefore, we do not apply a flowing bosonization in this work.

This paper is organized as follows. We discuss how frequency and momentum shells help to truncate the functional flow equation for the effective average action in Sec. \ref{SecShells}. An explicit setting using frequency and momentum cutoffs for the BCS-BEC crossover is put forward. The different truncations employed in this work are introduced in Sec. \ref{SecCoupl}. Therein we also explain the physical significance of the individual running couplings. Our results are summarized and discussed in Sec. \ref{SecResults}. In Sec. \ref{SecConcl} we draw our conclusions and give an outlook on possible extensions of our investigation. 

Technical aspects of our analysis are presented in detail in the appendices. The first three appendices contain a comprehensive introduction to the functional flow equation (App. \ref{appA}), and how the running of couplings can be obtained from a few constitutive equations (App. \ref{appB}) by means of appropriate projection descriptions (App. \ref{appC}). At the beginning of App. \ref{appA} we also fix our notation.  In Apps. \ref{AppIni} and \ref{AppEstimate} we extend the discussion of initial conditions and the chemical potential which is given in the main text. The numerical implementation of regularized loop integrals and the finite temperature flow is outlined in Apps. \ref{appD} and \ref{appE}, respectively. We recall the derivation of the mean field result for the Unitary Fermi Gas in App. \ref{appF}, and show its relation to the truncation with only fermion diagrams in the FRG setting.

\section{Frequency and momentum shell integration}\label{SecShells}

In this section we discuss how the path integral of an interacting many-body system can be integrated approximately by successively including fluctuations in narrow frequency and momentum shells. In this way, physics at all scales is incorporated while retaining a manageable truncation inside each individual shell. The parametrization of these effects in terms of running couplings is introduced in the next section.

\subsection{General strategy}

For ultracold fermions close to a broad Feshbach resonance, details of the atomic interaction potential become irrelevant for the macroscopic physics, and the system is solidly described in terms of the universal local action
\begin{align}
 \label{shell1} S= \int_X \Biggl[ \sum_{\sigma={1,2}} \psi_\sigma^* \Bigl(\partial_\tau-\nabla^2-\bar{\mu}\Bigr) \psi_\sigma+ \bar{\lambda}_\psi \psi_1^*\psi_2^* \psi_2\psi_1  \Biggr].
\end{align}
Herein, $\psi_\sigma^*$ and $\psi_\sigma$ are Grassmann fields describing an atom in a hyperfine state $\sigma$. We define $X=(\tau,\vec{x})$ with Euclidean time $\tau$ and $\int_X=\int_0^\beta \mbox{d}\tau\int\mbox{d}^3x$. After an appropriate renormalization in vacuum, the microscopic coupling constant $\bar{\lambda}_\psi$ is related to the scattering length  according to $\lambda_\psi=8 \pi a$. Similarly, the relation between $\bar{\mu}$ and the physical chemical potential $\mu$ may involve renormalization effects. We employ units such that $\hbar=k_{\rm B}=2M=1$, where $M$ is the mass of the atoms.

Due to many-body effects (density, temperature, interactions), the contact interaction of the microscopic theory becomes a frequency and momentum dependent function on large length scales. In particular, for sufficiently low temperatures, the system develops a many-body instability signalled by a diverging vertex function for a certain combination of frequencies and momenta. In these cases, a macroscopic anomalous self-energy $\Delta$ builds up. It is related to a nonvanishing expectation value $\langle\psi_1\psi_2\rangle$, which signals the breaking of the U(1)-invariance of the theory. The determination of the quantitative features of this transition in the strongly coupled regime is complicated by the need to resolve the full frequency and momentum dependence of the interaction vertex. Moreover, the vertex function has a nontrivial feedback on the self energy and higher correlation functions, and vice versa.

The apparent complexity of the situation just described can often (at least partially) be untangled by treating the system in a scale-dependent fashion. This procedure emerges naturally in a path integral formulation of the grand canonical partition function,
\begin{align}
 \label{shell2} Z(\mu,T) = \int \mbox{D}\psi \ e^{-S[\psi]} = \int \Bigl(\prod_{Q} \mbox{d} \psi(Q)\Bigr) e^{-S[\psi]}.
\end{align}
In the second equality we inserted the definition of the functional measure, where $Q=(q_0,\vec{q})$ summarizes Matsubara frequencies and spatial momenta. Following Wilson's idea we arrange the integration such that the values of $Q$ are grouped in ``frequency and momentum shells'' which satisfy
\begin{align}
 \label{shell3}  q_0^2+(\vec{q}^2-\mu)^2 \simeq k^4
\end{align}
for an external momentum scale $k$. Note that frequencies have dimension of momentum squared, as we have set $2M=1$. Denoting by $\int\mbox{d}\psi_k$ the integration over modes belonging to the shell one arrives at
\begin{align}
 \label{shell4} Z(\mu,T) \simeq \Bigl(\prod_{k=0}^\infty \int \mbox{d}\psi_k\Bigr)  e^{-S[\psi]}.
\end{align}

Now the integration is performed consecutively, starting from $k=\infty$, and eventually arriving at $k=0$. At a given scale $k$ the momentum shells with $k'>k$ are already included. Conceptually, one may define
\begin{align}
 Z_k =\prod_{k'>k} \int \mbox{d}\psi_{k'} e^{-S} = e^{-S_k}
\end{align}
such that 
\begin{align}
Z = \prod_{k'<k} \int \mbox{d}\psi_{k'} e^{-S_k}.
\end{align}
The action $S_k$ serves as a classical action for computing the physics at momentum scales below the flowing scale $k$. Furthermore, the latter functional integral contains an effective (UV) cutoff since only shells with $k'<k$ are included. A continuum version of this idea can be implemented by forming smooth averages of fields \cite{Wetterich:1989xg}. An infinitesimal change of $S_k$ from $k$ to $k-\mbox{d}k$ can be described by an exact flow equation \cite{Wetterich:1989xg,Wetterich:1993ne} which corresponds to Polchinski's flow equation \cite{Polchinski:1983gv}. Approximate solutions of this equation with the aim of precision are difficult, however. This is due to the complicated form of $S_k$ and the flow equation.

\subsection{Functional renormalization}

The strategy outlined above can be realized by applying the FRG for the effective average action. The method is based on the regularized generating functional
\begin{align}
 \label{shell5} Z_k[j] = \int \mbox{D} \psi e^{-S[\psi]-\Delta S_k[\psi]+j\psi}.
\end{align}
The $k$-dependence of this expression originates from adding
\begin{align}
 \label{shell6} \Delta S_k[\psi] = \int_Q \sum_\sigma \psi_\sigma^*(-Q) R_k(Q)\psi_\sigma(Q)
\end{align}
to the classical action. This term provides for a cutoff which effectively excludes momentum shells with $k'<k$ from the integration. It is quadratic in the fields and required to vanish for $k\to 0$. Accordingly, we recover the full path integral in the limit $Z_{0}[j]=Z[j]$. Instead of performing the momentum integration in shells for one given model one considers here a family of models characterized by the IR-cutoff scale $k$. For every $k$ one can construct an effective action (free energy). This effective average action $\Gamma_k[
\bar{\psi}]$ is defined such that $\Gamma_{\infty}=S$ and $\Gamma_{0}=\Gamma$, where $\Gamma$ is the effective action, i.e. the Legendre transform of the generating functional $\log Z[j]$. The evolution of $\Gamma_k$ as $k$ is lowered is given by the exact functional flow equation \cite{Wetterich1993}
\begin{align}
 \label{shell7} \partial_k \Gamma_k = \frac{1}{2}\mbox{STr} \Biggl(\Bigl(\Gamma_k^{(2)}+R_k\Bigr)^{-1}\partial_k R_k\Biggr).
\end{align}
Herein, $\Gamma^{(2)}_k$ is the second functional derivative of $\Gamma_k$. This makes the equation nonlinear.

We have to find an appropriate ansatz for the effective average action $\Gamma_k$ to reduce the functional flow equation (\ref{shell7}) to a manageable set of flow equations for the $k$-dependence of a finite set of running couplings $\{g_k\}$. A convenient way of treating the fermionic interaction vertex consists in replacing, at a large cutoff scale, the particle-particle channel with a bosonic degree of freedom, $\phi$, by means of a Hubbard--Stratonovich transformation. The frequency and momentum dependence of the interaction channel is then encoded in the frequency and momentum dependence of the boson propagator (or T-matrix).

Separating the fluctuation effects into distinct frequency and momentum shells allows for an efficient parametrization of the path integral, as $\Gamma_k$ itself might have a simple form for every individual $k$. The sum over all $k$, however, yields a highly nontrivial result.

As an example for the simplification which arises from frequency and momentum shell integration, we discuss the behavior of the inverse fermion propagator during the renormalization group flow. Its microscopic form $P_{\psi,\rm mic}(Q)=\rmi q_0 +q^2-\bar{\mu}$ receives contributions from fluctuation effects. This is encoded in the self-energy $\Sigma_\psi(Q)=P_\psi(Q)-P_{\psi,\rm mic}(Q)$, where $P_\psi(Q)$ is the full (dressed) inverse propagator of the theory. The $Q$-dependence of the latter is complicated, as it reflects the behavior of the system on different momentum scales set by density, temperature, and scattering length. On the other hand, for any given $k$, we can expand the self-energy according to
\begin{align}
  \Sigma_{\psi,k} = \rmi \delta Z_{\psi,k} q_0 + \delta A_{\psi,k} q^2 + \delta m^2_{\psi,k} +\dots
\end{align}
The self-energy for a given value of $q_0^2+q^4$ is well approximated by $\Sigma_{\psi,k}$ with $k^4\simeq q_0^2+q^4$. Therefore, the frequency and momentum dependence of $\Sigma_\psi(Q)=\Sigma_{\psi,k=0}(Q)$, which is induced by the running couplings $\delta Z_{\psi,k}$ and $\delta A_{\psi,k}$, can describe a rather complex momentum dependence, far beyond a simple derivative expansion in powers of $\rmi q_0$ and $q^2$.

For introductions to the FRG in various contexts see Refs. \cite{Berges:2000ew,Wetterich:2001kra,Pawlowski20072831,Gies:2006wv,Schaefer:2006sr,Delamotte:2007pf,Kopietz2010,Metzner:2011cw,Braun:2011pp}. A particular emphasis on the BCS-BEC crossover is placed in Refs. \cite{Scherer:2010sv,Boettcher:2012cm}.

\subsection{Regularization scheme}

For choosing the cutoff function $R_k(Q)$, mainly two strategies can be applied. On the one hand, since the flow equation (\ref{shell7}) is valid for every appropriate regulator, one may choose a particularly simple function $R_k(Q)$ which is sufficient to yield finite loop-integrals. A convenient choice consists in the three-dimensional optimized cutoff, which only cuts off spatial momenta $q^2=|\vec{q}|^2$. For bosons and fermions, respectively, it is given by
\begin{align}
 \nonumber &q^2\text{-opt:}\\
 \label{shell8} &R_{\phi,k}(Q) = \Bigl(k^2-\frac{q^2}{2}\Bigr)\theta\Bigl(k^2-\frac{q^2}{2}\Bigr),\\
 \label{shell9} &R_{\psi,k}(Q) = \Bigl[\sgn(q^2-\mu)k^2-(q^2-\mu)\Bigr]\theta\Bigl(k^2-|q^2-\mu|\Bigr),
\end{align}
where $\theta$ is the step function. (The notation $q^2$-opt shall indicate that this is a purely momentum cutoff.) The bosonic regulator $R_\phi$ takes into account that the boson mass is twice the fermion mass, whereas the fermionic function $R_\psi$ regularizes around the Fermi surface. For a $k$-dependent running Fermi surface, one has to replace the chemical potential $\mu$ with a running coupling.

Whereas the $q^2$-opt regulators provide an efficient regularization of spatial momenta, they do not limit the range of summation for the Matsubara frequencies. Thus, at every scale $k$, both very large and very small frequencies contribute to the flow of $\Gamma_k$. This, however, spoils the separation of scales discussed above. As a result, the frequency and momentum structure of the $k$-dependent propagators is complicated. Precision then requires a sophisticated (numerical) treatment. For successful implementations with the nonperturbative RG in the context of the Kardar-Parisi-Zhang equation see Refs. \cite{PhysRevLett.104.150601,PhysRevE.84.061128,PhysRevE.86.051124}.

A second possibility consists in the use of regulators which implement the idea of frequency and momentum shells, thereby depending on $Q=(q_0,\vec{q})$. In order to implement such regulators for a nonrelativistic system, we face the problem that the Galilean invariants for bosons and fermions are given by $\rmi q_0+q^2/2$ and $\rmi q_0+q^2-\mu$, respectively. Due to the imaginary frequency dependence, the regulators frequently employed for Lorentz invariant relativistic systems cannot be applied here. We choose the regulators
\begin{align}
 \nonumber &Q\text{-exp:}\\
 \label{shell10}&R_{\phi,k}(Q) = \Bigl(\rmi q_0 +\frac{q^2}{2}\Bigr) r\Bigl(\frac{q_0^2+q^4/4}{c_\phi k^4}\Bigr),\\
 \label{shell11} &R_{\psi,k}(Q) = \Bigl(\rmi q_0 +q^2-\mu\Bigr) r\Bigl(\frac{q_0^2+(q^2-\mu)^2}{k^4}\Bigr),
\end{align}
with an exponential shape function
\begin{align}
\label{shell12} r(X) = (e^{X}-1)^{-1}.
\end{align}
(The notation is again chosen to indicate that the $R_k$ are now frequency and momentum cutoffs.) This particular choice respects all requirements on appropriate FRG regulators, cuts off frequencies efficiently, and has shown to be numerically convenient.

A different frequency and momentum cutoff has been proposed in Ref. \cite{Floerchinger:2011sc}. The latter choice allows to analytically perform Matsubara summations and preserves the few-body hierarchy \cite{Floerchinger:2013tp} of the underlying field theory. In contrast to the $Q$-exp regulator, however, it only decays algebraically.

The relative cutoff scale $c_\phi$ in Eq. (\ref{shell10}) allows to regularize the bosons and fermions on slightly different scales $\sim k$. Given the different shapes of the dispersion relations and the somewhat arbitrary parametrization of the regulator functions, $c_\phi=1$ is not necessarily a natural or distinguished choice. In particular, earlier works found rather strong dependences of observables on relative cutoff scales in two-species systems \cite{PhysRevC.78.034001,PhysRevA.81.043628,PhysRevA.83.023621}. Since exact results do not depend on $c_\phi$, the residual dependence found with a given truncation gives some indication of the error due to the truncation \cite{Schnoerr:2013bk}. For the truncations employed in this work, we will see below that $\Delta/\mu$ and $T_{\rm c}/\mu$ show only five percent variations with respect to $0.2 \leq c_\phi\leq 1$.

\section{Running of couplings}\label{SecCoupl}

Every truncation of the renormalization group flow can be characterized in terms of a set of running coupling $\{g_k\}$, accompanied by a corresponding set of beta functions $\{\beta_g\}$ and initial values $\{g_\Lambda\}$. The latter ensure $\Gamma_\Lambda=S$ in the beginning of the flow. Here, $\Lambda$ is a large UV momentum cutoff scale. It has to be chosen much larger than the physical scales set by chemical potential, temperature, and scattering length. On the other hand, $\Lambda$ has to be much smaller than the momentum scale where details of the atomic interactions are resolved, typically given by the inverse van-der-Waals length.

We present a systematic truncation scheme for the effective average action, starting with only a few running couplings which are necessary to describe the superfluid transition, then including more couplings which are expected to give only subleading corrections. As the proposed set of improvements is based on a physical picture of the mechanisms in the crossover, the convergence of results verifies the corresponding intuition. Deviations, on the other hand, hint on missing ingredients, and can be employed for error estimates. We emphasize that the truncations described here are expected to work best for a sufficiently local RG flow, i.e. shells in \emph{both} frequencies and momenta. 

\subsection{Truncation}\label{trunc}

The microscopic action, after the Hubbard--Stratonovich transformation on the microscopic scale, is given by
\begin{align}
 \nonumber S[\psi,\phi] = &\int_X \Biggl[ \sum_{\sigma={1,2}} \psi_\sigma^*\Bigl(\partial_\tau-\nabla^2+m_{\psi}^2\Bigr) \psi_\sigma \\
 \label{run1} &\mbox{ } + m^2_\phi \phi^*\phi - h\Bigl(\phi^*\psi_1\psi_2-\phi\psi_1^*\psi_2^*\Bigr)\Biggr].
\end{align}
Performing the Gaussian integration over $\phi$ one sees the equivalence with Eq. (\ref{shell1}), identifying $\bar{\lambda}_\psi=-h^2/m^2_\phi$. The microscopic action serves as the initial condition for the effective average action for $k=\Lambda$. The sign of the Feshbach coupling $h$ is not important, as only $h^2$ enters diagrams. The ``fermion mass term'' $m_\psi^2$ is related to the chemical potential $\mu$ by possible additive and multiplicative renormalization, as we will explain later. The bosonic degrees of freedom are not dynamical on the microscopic level (i.e. no derivative terms for $\phi$), which is a result of the point-like interaction of the fermions. After a few momentum shell integrations, however, boson dynamics and a corresponding $Q$-dependence of the boson propagator emerge from the RG flow. We discuss the initial conditions for the flow equation at the end of this section.

Our ansatz for the effective average action $\Gamma_k$ consists of a kinetic part, which comprises the fermion and boson propagators, and an interaction part:
\begin{align}
 \label{run2} \Gamma_k = \Gamma_{\rm kin} + \Gamma_{\rm int}.
\end{align}
In terms of the renormalized fields $\psi=A_\psi^{1/2}\bar{\psi}$ and $\phi=A_\phi^{1/2}\bar{\phi}$, the kinetic part is given by
\begin{align}
 \nonumber \Gamma_{\rm kin}[\psi,\phi] = &\int_X \Biggl( \sum_{\sigma=1,2} \psi^*_\sigma \Bigl(S_\psi \partial_\tau-\nabla^2+m^2_\psi\Bigr)\psi_\sigma\\
 \label{run3} &+\phi^*\Bigl(S_\phi\partial_\tau-V_\phi \partial_\tau^2-\frac{1}{2}\nabla^2\Bigr)\phi\Biggr).
\end{align}
Here and in the following, the $k$-dependence of the couplings is understood implicitly. The coefficients of the gradient terms are normalized to constants by means of the wave function renormalizations $A_\psi$ and $A_\phi$. The latter two quantities do not enter the RG flow, but only their beta functions given by the anomalous dimensions
\begin{align}
 \label{run4} \eta_\psi=-k \partial_k \log A_\psi,\ \eta_\phi=-k \partial_k \log A_\phi.
\end{align}
In the following, unrenormalized quantities are denoted with an overbar, renormalized ones without an overbar. We refer to App. \ref{appA} for a more detailed definition.

Interactions are parametrized according to
\begin{align}
 \label{run5} \Gamma_{\rm int}[\psi,\phi] = \int_X \Biggl( U(\phi^*\phi) - h\Bigl(\phi^*\psi_1\psi_2-\phi\psi_1^*\psi_2^*\Bigr)\Biggr).
\end{align}
The effective average potential $U(\rho)$ only depends on the U(1)-invariant $\rho=\phi^*\phi$. It describes higher order bosonic scattering processes. A nonzero minimum $\rho_0$ of $U_{k=0}(\rho)$ indicates the spontaneous breaking of U(1)-invariance, and, thus, superfluidity. We write
\begin{align}
 \label{run6} U(\rho) = m^2_\phi(\rho-\rho_0)+\frac{\lambda_\phi}{2}(\rho-\rho_0)^2+\sum_{n=3}^N \frac{u_n}{n!}(\rho-\rho_0)^n.
\end{align}
In the following, we will call a truncation of the effective potential of order $N$ simply a $\phi^{2N}$-truncation. We always work at least with a $\phi^4$-truncation.

It is important to note that, besides an ansatz for $\Gamma_k$, the truncation of the flow equation also consists in projection prescriptions for the running couplings. The corresponding equations are given in App. \ref{appC}.

\begin{figure}[t]
\centering
\includegraphics[width=8cm]{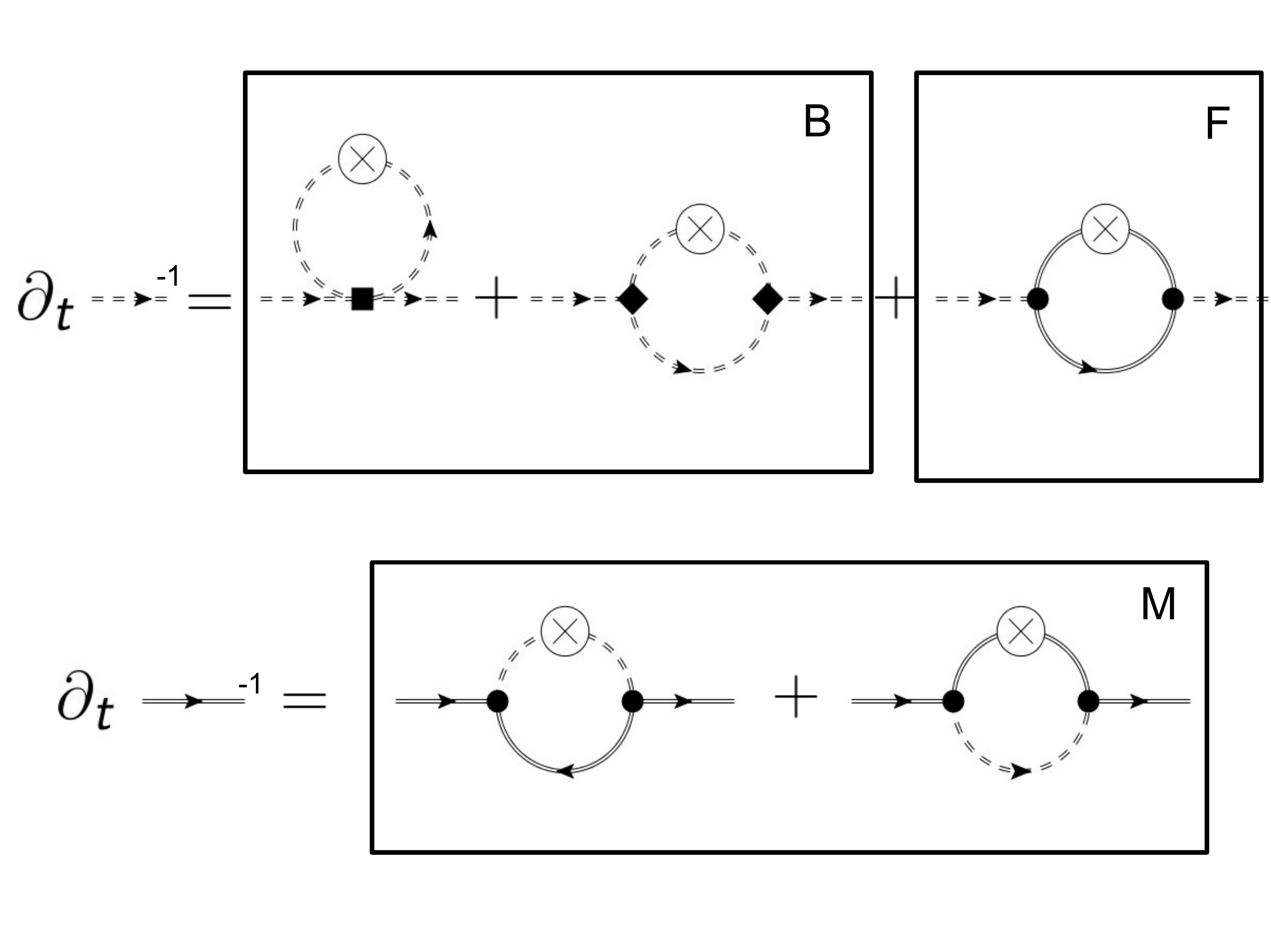}
\caption{The truncations employed in this work can be classified according to the diagrams which appear on the right hand side of the flow equation. Here, as an example, we show the flow equations for the boson propagator (dashed line) and the fermion propagator (solid line), respectively. All lines and vertices are fully dressed. The cross indicates a regulator insertion $\dot{R}_k$. The particle-particle loop of fermionic (F) atoms corresponds to the F-truncation. Within the latter, the fermion propagator does not get renormalized. The same holds for the FB-truncations, where, in addition, purely bosonic diagrams (B) are also taken into account. Eventually, our highest truncations also include mixed diagrams (M) with both a fermion and a boson line.}
\label{FeynmanDiagrams}
\end{figure}

We can classify our truncations by means of the diagrams which are included on the right hand side of the flow equation. Those containing only fermionic (F) lines (or propagators) reproduce the mean field result. Including those with two bosonic (B) lines is important to resolve the impact of boson fluctuations on the critical temperature. The renormalization effects on the fermion propagator are given by mixed (M) diagrams with both boson and fermion lines. We visualize this hierarchy of diagrams in Fig. \ref{FeynmanDiagrams}. By elaborating the truncation according to the inclusion F $\to$ FB $\to$ FBM, we successively incorporate higher order terms while keeping the physical content of the lower truncations in this hierarchy. Within a class of diagrams, we still have the freedom to keep several couplings at their classical level.

In this work we restrict to the following five truncation schemes:
\begin{itemize}
 \item{F}: Fermion diagrams,\\
 Running couplings: $U(\rho), A_\phi$
 \item{FB$_0$}: Fermion and boson diagrams,\\
 Running couplings: $U(\rho), A_\phi, S_\phi$
 \item{FB}: Fermion and boson diagrams,\\
 Running couplings: $U(\rho), A_\phi, S_\phi, V_\phi$
 \item{FBM$_0$}: Fermion, boson, and mixed diagrams,\\
 Running couplings: $U(\rho), A_\phi, S_\phi, V_\phi, m^2_\psi$
  \item{FBM}: Fermion, boson, and mixed diagrams,\\
 Running couplings: $U(\rho), A_\phi, S_\phi, V_\phi, m^2_\psi, A_\psi, h^2.$
\end{itemize}
The subscript $0$ in the second and fourth truncation indicates that we leave out some running couplings which are included at a higher level of the truncation hierarchy with the same diagrams. For the last truncation we still keep $S_\psi=1$. The effective potential $U(\rho)$ can be elaborated independently of the other running couplings. We will mostly restrict to a $\phi^4$- or $\phi^8$-truncation.

\subsection{Physical content}

We now discuss the physical content of the running couplings introduced above. The momentum dependence of the (renormalized) inverse bosons propagator 
\begin{align}
P_\phi(Q)=\rmi S_\phi q_0 +V_\phi q_0^2+q^2/2
\end{align}
is generated during the early stages of the RG flow. It parametrizes the $Q$-dependence of the particle-particle channel of the four-fermion vertex. The coefficients $S_\phi, V_\phi$ can be regarded as the expansion coefficients of a derivative expansion of the boson self-energy for each scale $k$ individually. Again, the frequency and momentum regulators ensure that only modes with $(q_0^2+q^4)^{1/4}\approx k$ contribute to the flow.

\begin{figure}[t]
\centering
\includegraphics[width=8cm]{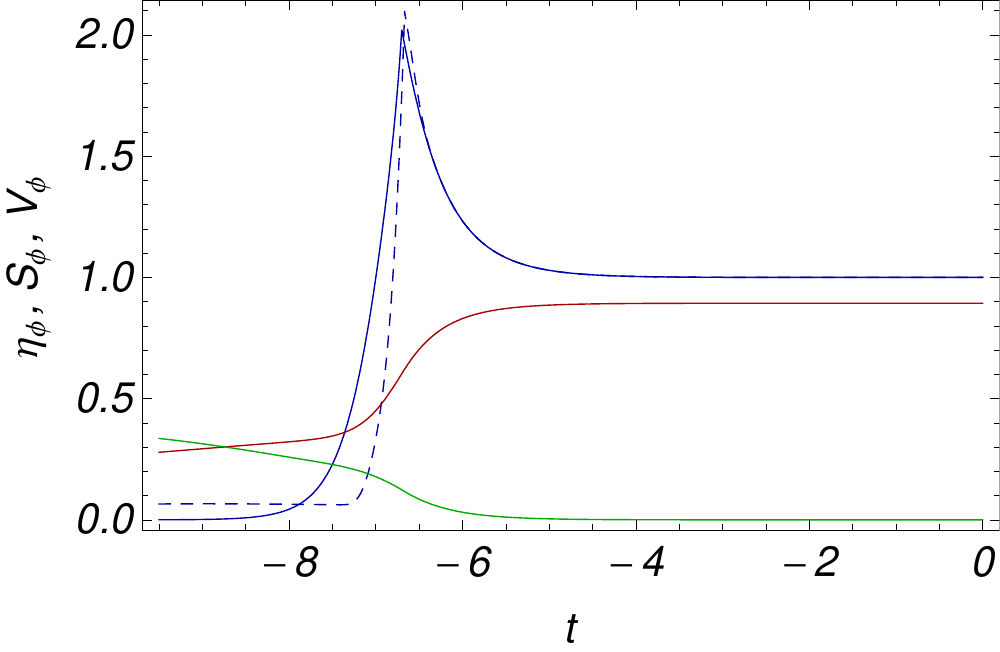}
\caption{Typical running of the couplings which parametrize the boson propagtor $P_\phi(Q)$. Units are such that $\mu=1$. From top to bottom we show the $t=\log(k/\Lambda)$-dependence of $\eta_\phi$ (blue, solid), $S_\phi$ (red), and $V_\phi$ (green) at $T=0$. The initial values correspond to $t=0$ ($k=\Lambda$), and the infrared regime is found for $t\to -\infty$ ($k\to 0$). Physical observables like $\Delta/\mu$ saturate at a sufficiently small $t$ such that we can stop the flow at a finite $t$. Many-body effects strongly influence the flow at $k^2\simeq \mu$, which corresponds to $t=-6.9$ in this plot. We also show the anomalous dimension $\eta_\phi$ for $T=T_{\rm c}$ (blue, dashed), which does not vanish in the infrared but settles at the critical exponent $\eta = 0.05(1)$ for a $\phi^4$-truncation. This value is expected in the O(2)-universality class within this order of the truncation.}
\label{plot_boson_couplings}
\end{figure}

For large $k$, and a sufficiently large initial Feshbach coupling $h^2$ (i.e. the situation of a broad Feshbach resonance), the flow drives the couplings to a universal vacuum fixed point with $\eta_{\phi\star}=1$. Then $A_\phi$ scales according to
\begin{align}
 A_{\phi,k} \sim k^{-\eta_{\phi\star}} \sim k^{-1}.
\end{align}
Accordingly, $A_{k} q^2 \simeq A_q q^2 \sim q$ in the early stages of the flow. In fact, the inverse boson propagator can be integrated analytically in this regime, yielding $P_\phi(Q) \sim \sqrt{\rmi q_0/2+q^2/4-\mu}$, which indeed scales linear in $q$. We see that, although we pushed the truncation into a $q^2$-dependence of the propagator, the running couplings react in such a manner as to undo this forcing. The running of $\eta_\phi$, $S_\phi$, and $V_\phi$ with $t=\log(k/\Lambda)$ is visualized in Fig. \ref{plot_boson_couplings}.

The quadratic frequency dependence $V_\phi q_0^2$ in the boson propagator constitutes the first nonvanishing frequency dependent term of the real part of the boson self-energy. Deep in the infrared Goldstone regime, it is expected to be dominant over the linear frequency term. This is well-known from purely bosonic systems, most pronounced in reduced dimensionality $d\leq 2$ \cite{PhysRevB.77.064504,PhysRevLett.102.190401}. We expect the $V_\phi$-term to be important when turning to the two-dimensional BCS-BEC crossover in future work. Here we find a rather mild dependence of the overall flow on the presence of $V_\phi$, indicating that the frequency and momentum regulators work sufficiently well, such that already the truncation without $V_\phi$ captures the leading  frequency dependence. In contrast, for purely momentum regulators, the influence of $V_\phi$ is stronger. We refer to Sec. \ref{SecResults} and our conclusions for the results and a detailed discussion.

\begin{figure}[t]
\centering
\includegraphics[width=8cm]{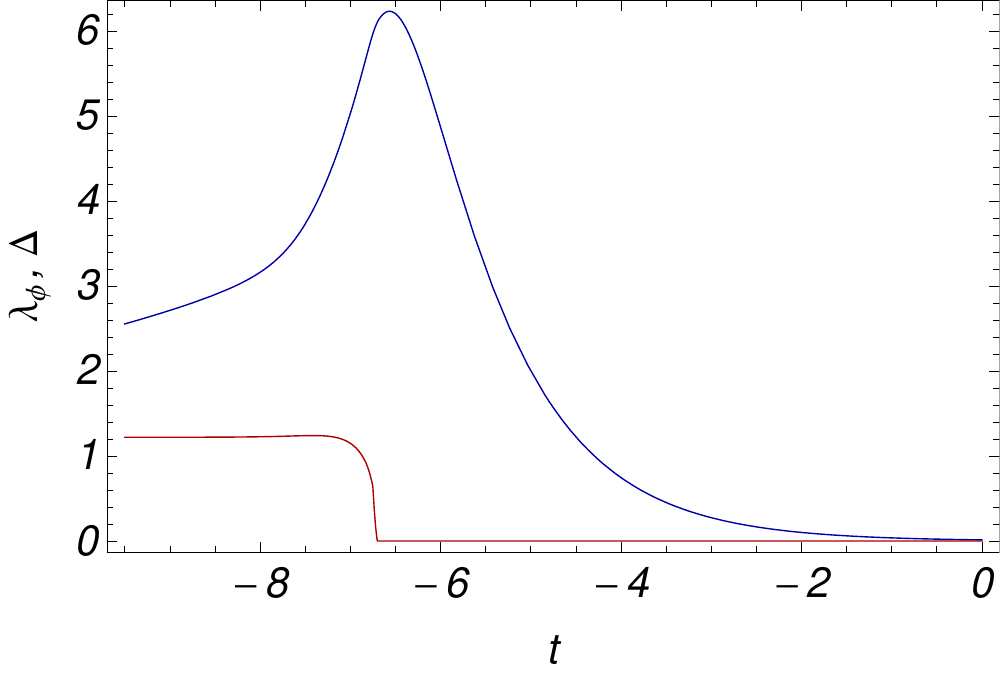}
\caption{Zero temperature running of the boson-boson coupling $\lambda_\phi$ (blue, upper curve), which is the coefficient of the term $(\rho-\rho_0)^2$ in a power series expansion of the effective potential $U(\rho)$. Units are such that $\mu=1$. We can clearly identify three regimes in the flow: For $t\simeq 0$ the coupling follows the scaling solution with constant $\tilde{\lambda}_\phi=k\lambda_\phi$. This behavior would continue in vacuum, where $\mu=T=0$. For a nonzero $\mu >0$, however, the flow is influenced by many-body effects at $k^2\simeq\mu$ ($t\simeq -6.9$). For smaller scales, the flow enters the Goldstone regime, where all contributions to the running of couplings come from infrared Goldstone fluctuations, whereas the chemical potential $\mu/k^2$ is gapped out due to $k\to 0$. The behavior exemplified here for $\lambda_\phi$ is found for all running couplings. We also show the running of $\Delta_k=(h^2\rho_0)^{1/2}$ in the red lower curve. It has a nonvanishing value below the symmetry breaking scale $t_{\rm sb}=-6.7$.}
\label{plot_running_lambda}
\end{figure}

In order to capture the physics of the second order superfluid phase transition, the effective potential $U(\rho)$ needs to be at least of order $\phi^4$. We start at large $k$ with a non-vanishing boson ``mass term'' (or detuning) $m^2_\phi>0$. The field expectation value is zero in this symmetric regime of the flow, $\rho_{0,k}=0$. Scattering between bosons is described by the boson-boson coupling $\lambda_\phi$, and $n$-boson scattering processes are encoded in $u_n$ for $n\geq 3$. Due to a nonzero chemical potential, the boson mass term decreases during the flow and may reach zero for a nonzero symmetry breaking scale $k_{\rm sb}$. This is equivalent to the Thouless criterion of a diverging four-fermion vertex, however, at a given scale $k$. Typically, the symmetry breaking scale is slightly above the chemical potential, $k_{\rm sb}^2  \gtrsim\mu$. The origin of this divergence can be rooted in large contributions of fermion fluctuations in case of an approximate zero of the inverse fermion propagator $\rmi q_0+q^2-\mu$, which occurs as soon as the typical momenta become of the order of the chemical potential. The flow of the boson-boson coupling $\lambda_\phi$ at $T=0$ is displayed in Fig. \ref{plot_running_lambda}.

Fort $k<k_{\rm sb}$ a nonzero expectation value $\rho_{0,k}$ of the boson field indicates local order on length scales $\sim k^{-1}$. Within our truncation, this results in an anomalous fermion self-energy 
\begin{align}
\Sigma_{\rm an,k}(Q=0)=\Delta_k=(h^2\rho_0)^{1/2},
\end{align}
which enters the fermion propagator as a gap, hence removing the zero in the denominator. The fate of the $k$-dependent gap $\Delta_k$ depends on the temperature of the system. For $T=0$ one finds a superfluid ground state with $\rho_0\neq 0$ at $k=0$. For sufficiently high temperatures,  thermal fluctuations may destroy the local order, such that $\rho_{0,k=0}=0$ at the end of the flow. We call this temperature range the precondensation region. The local expectation value $\rho_{0,k}$ of bosons can then be seen as a strong bosonic correlation on scales of order $k$, which do not yet suffice to produce a true long-range order. It is closely related to the notion of a pseudogap. For temperatures above the precondensation temperature, no local order emerges during the flow and we have $\rho_{0,k}=0$ for all $k$.

\begin{figure}[t]
\centering
\includegraphics[width=8cm]{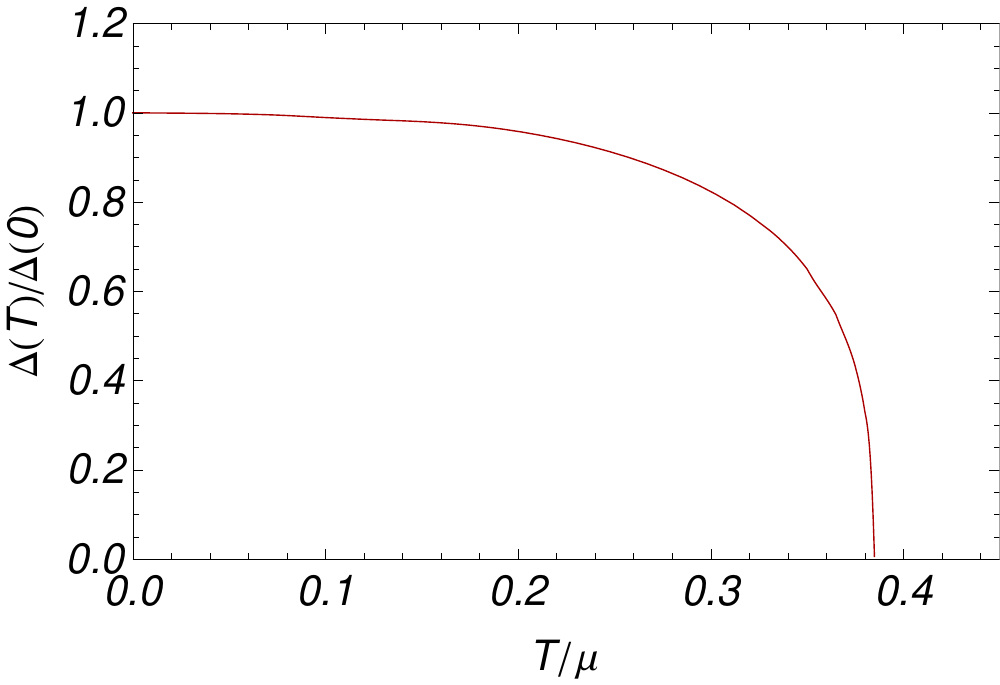}
\caption{Temperature dependence of the superfluid gap normalized by its zero temperature value. As we approach the critical temperature, the order parameter diminishes continuously, and eventually vanishes at $T_{\rm c}$. The continuous behavior at $T_{\rm c}$ is found for all of our truncations and results from the inclusion of bosonic fluctuations, which become dominant close to criticality.}
\label{plot_gapT}
\end{figure}

The superfluid gap is defined as
\begin{align}
 \label{run7} \Delta = \lim_{k\to 0} \Delta_k.
\end{align}
Superfluidity is equivalent to a nonzero gap in our truncation. The critical temperature for the phase transition to superfluidity is defined as the highest temperature, such that a precondensate $\rho_{0,k}>0$ appearing during the flow survives at $k=0$. We have $\Delta=0$ exactly at $T=T_{\rm c}$ as the order parameter becomes arbitrarily small at $k=0$. On the BCS side of the crossover, no precondensation occurs, and the rise of bosonic correlations is in one-to-one correspondence with bosonic condensation. On the BEC side, in contrast, the precondensation region is huge. The running of $\Delta_k$ for the zero temperature Unitary Fermi Gas is shown in Fig. \ref{plot_running_lambda}. The continuous behavior of $\Delta(T)$ for all $T$ is shown in Fig. \ref{plot_gapT}. It indicates a second order phase transition.

Whereas the phase transition and thus the critical temperature is mainly driven by the bosons, we expect the renormalization of the fermion propagator to be important for the density of the system due to the Tan contact effect \cite{PhysRevA.87.023606}. When employing purely momentum regulators, the diagrams renormalizing the fermion propagator vanish in the symmetric regime of the flow. This is a result of the possibility to analytically perform the Matsubara summations, which have all poles in one half-plane for vanishing anomalous self-energies. When employing frequency and momentum regulators,  the fermion propagator gets renormalized both in the symmetric many-body regime and in vacuum. This effect can be controlled in vacuum by means of an appropriate vacuum renormalization of the initial ``fermion mass term'', $m^2_{\psi\Lambda}(\mu=0)=C\Lambda^2$, where $C$ is a regulator-dependent constant $C\simeq -0.2$. 

The physical chemical potential $\mu$ vanishes in vacuum, where $\mu=0$. The association of $\bar{m}^2_{\psi}-\bar{m}^2_{\psi,\rm vac}$ with the chemical potential $\mu$ is meaningful only for a constant difference
\begin{align}
 \label{run8}\bar{m}^2_{\psi,k} - \bar{m}^2_{\psi,k,\rm vac} = - \mu
\end{align}
in the early stages of the many-body flow, i.e. for large $k$. The vacuum flow for the fermion mass term is given by the canonical running $m^2_{\psi,\rm vac}=Ck^2$ for the Unitary Fermi Gas. If Eq. (\ref{run8}) is not satisfied, the relation between the chemical potential and the fermion mass term $m^2_\psi$ is not obvious.

The Feshbach coupling $h^2$ receives small corrections in the ordered regime of the flow where $\rho_0>0$. We include this effect in our highest truncation. In this work, we only employ truncations with $S_\psi=1$.

\subsection{Universality}

To initialize the set of ordinary differential equations for the running couplings $\{g_k\}$, we have to equip the system with appropriate initial conditions $\{g_\Lambda\}$. This is particularly simple for the Unitary Fermi gas, where the initial conditions for the running couplings are found as the zeros of the beta functions for the dimensionless renormalized running couplings. 

The standard procedure for solving the flow equation for the effective average action consists in choosing the initial $\Gamma_\Lambda$ to agree with the microscopic action $S$. For instance, this would result in the initial values $A_\phi=S_\phi=V_{\phi}=0$. The parameters of the microscopic action, here the boson and fermion mass terms $m^2_{\phi\Lambda}$ and $m^2_{\psi\Lambda}$, have to be chosen such that we arrive at the right vacuum scattering physics, namely
\begin{align}
 \label{run9} m^2_{\phi,k=0,\rm vac}&=-\frac{h^2}{8\pi a}\theta(-a),\\
  \label{run10} m^2_{\psi,k=0,\rm vac} &= -\frac{1}{2}\vare_{\rm b} =\frac{1}{a^2}\theta(a),
\end{align}
where $a$ is the scattering length and $\vare_{\rm b}$ the binding energy of a bosonic molecule. Accordingly, bosons (fermions) are gapped on the BCS (BEC) side of the crossover in vacuum. In the Unitary limit, both masses vanish. 

For a sufficiently broad Feshbach resonance, the running couplings are attracted to a scaling solution \cite{PhysRevA.76.021602,PhysRevA.76.053627,ANDP:ANDP201010458}, for which the renormalized dimensionless couplings take constant values. The anomalous dimensions obey $4-d-\eta_\phi-2\eta_\psi=0$, as for this choice $\partial_k \tilde{h}^2=0$ in our truncation. Depending on the deviation of the relevant boson and fermion mass terms from their Unitary Fermi Gas initial values, the system will stay sufficiently long (i.e. many $k$-steps) at this fixed point. At the fixed point, all other couplings acquire their corresponding fixed point values. We call this regime the scaling regime. For any nonzero $a^{-1}$, $\mu$, or $T$, the system will eventually leave the scaling solution. However, memory of the precise initial condition is lost, as all couplings acquired their fixed point values. Accordingly, we might as well start directly at the scaling solution.

Due to this property, the initial conditions for the Unitary Fermi Gas can now be found for any given truncation-, regularization- and specification-prescription by simply solving a fixed point equation.  Indeed, for $a^{-1}=0$, Eqs. (\ref{run9}) and (\ref{run10}) can be solved by ensuring the dimensionless boson and fermion mass terms to be constant and given by their fixed point values, $\tilde{m}^2_\phi=m^2_\phi/k^2=\tilde{m}^2_{\phi\star}$ and $\tilde{m}^2_\psi=m^2_\psi/k^2=\tilde{m}^2_{\psi\star}$. (Dimensionless running couplings, which are divided by their canonical power of $k$, will be denoted by a tilde. The subscript $\star$ indicates the fixed point.) We then trivially have $m^2_\phi,m^2_\psi \sim k^{2} \to 0$ for $k\to 0$. The other running couplings are attracted to their fixed point values. To simplify the ultraviolet flow, we let them start directly at the scaling solution: $\Gamma_\Lambda=\Gamma_\star$.

Thus, given a truncation in terms of a set of dimensionless running couplings $\{\tilde{g}_k\}$ with beta functions $\{\tilde{\beta}_g\}$, the initial conditions for the Unitary Fermi Gas are found from
\begin{align}
  \label{run11} 0 = \tilde{\beta}_g(\tilde{g}_\star)
\end{align}
for $g\neq m^2_\psi$, and $m^2_{\psi\Lambda}=\tilde{m}^2_{\psi\star}\Lambda^2-\bar{\mu}$. The chemical potential defines a relevant perturbation which drives the system away from the fixed point. The initial scale $\Lambda$ has to be chosen large enough, such that many-body and interaction effects do not influence the ultraviolet flow. The initial conditions for the remaining running couplings do not need a modification due to the chemical potential, as these terms are generated automatically during the ultraviolet flow. To simplify the latter, however, we choose them to be on their corresponding fixed point values as well. Thus we arrive at
\begin{align}
 \tilde{g}_\Lambda = \tilde{g}_{\star}+\tilde{g}_{\mu\star} \bar{\mu}/\Lambda^2,
\end{align}
where the subscript $\mu$ indices the $\mu$-derivative of the respective running coupling, with $\tilde{m}^2_{\psi\mu}=-1$. The procedure of determining the terms proportional to $\mu$ is described in detail in App. \ref{AppIni}.

\section{Results}\label{SecResults}
\subsection{Recovering mean field theory}\label{mf}
The simplest truncation capturing the superfluid phase transition for all values of the scattering length consists in mean field theory. The latter is built on a saddle-point approximation to the effective action. We review the mean field predictions for the zero temperature gap $\Delta/\mu$ and the critical temperature $T_{\rm c}/\mu$ in App. \ref{appF}. Here, we discuss how the mean field approximation is recovered in an FRG framework by taking into account fermionic diagrams (F), but neglecting bosonic fluctuations.

In the mean field limit, there is no feedback of bosonic fluctuations onto the flow of running couplings. Therefore, neither the inverse boson propagator $P_\phi(Q)$, nor the boson regulator $R_\phi(Q)$ appear in the loop integrals. Hence, the mean field limit is a good testbed for benchmarking the implementation of the fermionic diagrams. The set of running couplings consists of the boson anomalous dimension $\eta_\phi=-k\partial_k \log A_\phi$ and the effective potential $U_k(\rho)$. The inverse fermion propagator remains in its initial shape given by $P_\psi(Q)=\rmi q_0 + q^2-\mu$.

Considering only fermion diagrams, the $n$-th derivative of the flow equation for the effective average potential is given by
\begin{align}
 \label{mf2} \partial_k\bar{U}^{(n)}_k(\bar{\rho}) = -n!(-\bar{h}^2)^n \int_Q \frac{L_\psi^Q \partial_kR_\psi^{-Q}+L_\psi^{-Q}\partial_kR_\psi^Q}{(L_\psi^QL_\psi^{-Q}+h^2\rho)^n},
\end{align}
with $L_\psi^Q=P_\psi(Q)+R_\psi(Q)$. As discussed in Eq. (\ref{appC2}), the flow of $\bar{u}_n=\bar{U}^{(n)}(\bar{\rho}_0)$ receives an additional contribution proportional to $\partial_k\bar{\rho}_0$. The flow of the $n$-th expansion coefficient of the effective potential is thus given by
\begin{align}
 \label{mf3} \partial_k\bar{u}_n = \partial_k\bar{U}_k^{(n)}(\bar{\rho}_0) + \bar{u}_{n+1} \partial_k\bar{\rho}_0
\end{align} 
for $ n\geq 2$. We emphasize that only due to the second term there is a feedback of the higher couplings $u_3, u_4,\dots$ onto the remaining couplings. The flow equations for $\bar{\rho}_0$ and $A_\phi$ are given in Apps. \ref{appB} and \ref{appC}.

The mean field zero temperature gap $\Delta/\mu=1.162$ provides a benchmark for testing the regulator and truncation dependence of the fermionic contributions to the flow. We find that the quantitative difference between a $\phi^4$- and a $\phi^8$-truncation is at the 10 percent level. By further extension to a $\phi^{2N}$-expansion, the results converge quickly to the expected value. We summarize our findings in Table II.

\begin{table}\label{TableF}
\begin{tabular}[t]{|c||c|c||c|c|}
\hline
 $\mbox{ }$ & \multicolumn{2}{|c||}{ $\Delta/\mu$} & \multicolumn{2}{|c|}{$T_{\rm c}/\mu$}\\
 \hline \hline
 Truncation & \ $q^2$-opt \ & \ $Q$-exp \  & \ $q^2$-opt \ &\ $Q$-exp \  \\
 \hline
 F, $\phi^4$ & 1.045 & 1.056 & 0.665 & 0.664 \\
 F, $\phi^8$ & 1.133 & 1.141 & 0.665 & 0.664 \\
 F, $\phi^{12}$ & 1.154 & 1.157 & 0.665 & 0.664 \\
 F, $\phi^{16}$ & 1.160 & 1.160 & 0.665 & 0.664 \\
 F, $\phi^{20}$ & 1.162 & 1.161 & 0.665 & 0.664 \\
 F, $\phi^{24}$ & 1.162 & 1.161 & 0.665 & 0.664 \\
\hline \hline
 Mean field & \multicolumn{2}{|c||}{1.162} & \multicolumn{2}{|c|}{0.6646}\\
 \hline
\end{tabular}
\caption{Critical temperature and superfluid gap obtained from the inclusion of fermionic diagrams (F). When including higher order terms in the effective potential, we recover the mean field result for both types of regulators considered in this work. In particular, a $\phi^8$-truncation already yields a good approximation to the exact result. The value of $T_{\rm c}/\mu$ is unaffected by this change in truncation, as is discussed in the main text.}
\end{table}

Due to the absence of a precondensation regime in the mean field treatment, the critical temperature $T_{\rm c}/\mu$ at the mean field level is not affected by terms proportional to $\partial_k\bar{\rho}_0$. Indeed, whenever $\rho_{0,k}>0$ for some $k$, we also have $\rho_{0,k=0}>0$. Accordingly, the correct value is already found in a $\phi^4$-truncation. Higher orders in a $\phi^{2N}$-expansion of the effective potential influence the critical temperature once bosonic diagrams are included in the RG flow.

\subsection{Bosonic fluctuations}

The bosonic dynamics emerge in the crossover due to the F-diagram containing two fermion lines. Once built up, the boson propagator has an important impact on the flow of running couplings due to diagrams containing two bosonic lines (B). These bosonic fluctuations are particularly important for an accurate description of the superfluid phase transition. We find here that the effect of B-diagrams is most prominent on the value of the critical temperature, whereas mixed diagrams change the latter only moderately.

\begin{table}\label{TableB0}
\begin{tabular}[t]{|c||c|c||c|c|}
\hline
 $\mbox{ }$ & \multicolumn{2}{|c||}{ $\Delta/\mu$} & \multicolumn{2}{|c|}{$T_{\rm c}/\mu$}\\
 \hline \hline
 Truncation & \ $q^2$-opt \  & \ $Q$-exp \ & \ $q^2$-opt \ & \ $Q$-exp \  \\
 \hline 
 FB$_0$, $\phi^4$ & 1.09 & 1.244(5) & 0.441 & 0.399(2) \\
 FB$_0$, $\phi^8$ & 1.13 & 1.227 &0.424 & 0.380 \\
 FB$_0$, $\phi^{10}$ & 1.16 & - & 0.427 & 0.394 \\
\hline \hline
 FB, $\phi^4$ & 1.05 & 1.228(10) & 0.405 & 0.389(2) \\
 FB, $\phi^8$ & - & 1.240 & - & 0.380 \\
 FB, $\phi^{10}$ & - & - & - & 0.386 \\
\hline
\end{tabular}
\caption{Influence of the regularization scheme and higher orders in the effective potential when including bosonic diagrams (B), where B$_0$ and B correspond to a boson propagator without and with the term $V_\phi q_0^2$, respectively. We observe that the scheme with a frequency and momentum cutoff is less sensitive to the inclusion of this term. The values for this table have been obtained for $c_\phi=1$. The errors in brackets estimate the numerical error.}
\end{table}

The truncation FB$_0$ has been studied in previous works by means of the optimized momentum $q^2$-opt regulator \cite{PhysRevA.76.053627,PhysRevB.78.174528,ANDP:ANDP201010458}. Here we aim at comparing these results to the application of a $Q$-exp regulator. Moreover, we include the $V_\phi q_0^2$-term, which has been left out so far. We further increase the truncation of the effective potential in order to estimate the effect of higher order bosonic scattering processes on physical observables. As discussions with the $q^2$-opt regulator usually employ a relative cutoff scale $c_\phi=1$, we choose this value here. Below, we will discuss the relative cutoff scale dependence for the $Q$-exp regulator in more detail. The results of our investigation are summarized in Table III.

By including bosonic fluctuations we observe the critical temperature to drop dramatically as compared to its mean field value. This behavior is expected as bosons generically tend to wash out the ordering and thus to decrease the critical temperature. In this context, it is interesting to study the influence of the emergent ``relativistic'' term $V_\phi q_0^2$ in the boson propagator. Whereas the effect of including this running coupling is strong for the purely momentum $q^2$-opt regulator, its effect is only moderate for a regulator which cuts off both frequencies and momenta. This is an indication for the efficiency of the latter cutoff, which incorporates the frequency behavior of the boson propagator already within a simple truncation. In contrast, the $q^2$-opt cutoff needs a higher resolution of the nontrivial $q_0$-dependence in order to obtain reliable results.

The superfluid gap comes out substantially larger for a $Q$-exp regulator. This is also true when including higher terms in the effective potential, see also Fig. \ref{plot_delta}.

The effects of higher orders in a series expansion of the effective potential are less conclusive as in the mean field case. We find a trend to decrease the critical temperature by applying an order $\phi^8$-truncation, but this effect is almost cancelled at order $\phi^{10}$. The critical temperature in the FB-truncation is particularly stable with variations of a few percent. For the $q^2$-opt regulator we did not implement higher truncations of $U(\rho)$ in the FB-truncation.

When going to higher orders in the effective potential one eventually expects the results to converge to a fixed value. Within our investigation, however, we found that the series expansion of $U(\rho)$ in powers of $\rho-\rho_0$ breaks down during the flow, indicating the nonanalytic shape of the effective potential. The latter is well-known to be reproduced with the FRG \cite{Litim:2006nn}. Thus we cannot report on values beyond $\phi^8$ for the superfluid gap, and $\phi^{10}$ for the critical temperature. As will be discussed in more detail in our conclusion, this shortcoming can be resolved by an expansion about a field value $\rho \neq \rho_0$, or by incorporating the full function $U(\rho)$ on a grid of $\rho$-values.

\subsection{Renormalization of the fermion propagator}
We now proceed by discussing the highest truncations employed in this work. By including mixed diagrams (M) containing both a boson and a fermion line, we can resolve renormalization effects on the fermion propagator and the Feshbach coupling $h^2$.

When applying a truncation with $V_\phi=0$ and a purely momentum cutoff (such as the $q^2$-opt regulator), the fermion propagator does not get renormalized in vacuum or in the symmetric regime of the flow where $\rho_{0,k}=0$. This property is due to the analytic structure of the regularized propagators, which have both poles lying in the same half-plane for the M-diagrams. Accordingly, the contour of the frequency integration can be closed in the other half-plane, thereby yielding a vanishing beta function.

This simple behavior is spoiled by the application of the $Q$-exp regulator or the inclusion of the $V_\phi q_0^2$-term. This is not problematic for the FB-truncations, as an appropriate renormalization in vacuum removes the corresponding unphysical flow. However, as we allow for a running of the fermion mass term $m^2_\psi$, the interpretation of the chemical potential $\bar{\mu}$, which enters the initial conditions through 
\begin{align}
m^2_{\psi\Lambda} = C\Lambda^2 -\bar{\mu},
\end{align}
is complicated. $C=\tilde{m}^2_{\psi\star}$ is a bare renormalization constant fixed in vacuum. The value of $C$ only depends on the truncation and the regularization scheme.

In general, for the truncations FBM$_0$ and FBM, we do not have $\bar{\mu}=\mu$, whereas this is true for all other truncations discussed so far. To see this, we vary $\bar{\mu}$ and check whether $\Delta(\bar{\mu})/\bar{\mu}$ or $T_{\rm c}(\bar{\mu})/\bar{\mu}$ are independent of $\bar{\mu}$. We then find a logarithmic $(\bar{\mu}/\Lambda^2)$-dependence of both observables when including M-diagrams. However, we checked that the ratio $\Delta/T_{\rm c}$ is indeed independent of $\bar{\mu}$. This shows that the uncertainty in the ratios $\Delta/\mu$ and $T_{\rm c}/\mu$ dominantly results from the inequality $\bar{\mu}\neq\mu$. 

As is discussed in App. \ref{AppIni}, for the FBM$_0$- and FBM-truncations, the initial conditions only allow to interpret $\bar{\mu}=\mu$ for a special choice of $c_\phi$ within our setting, which is $c_0=0.2454\simeq 1/4$. For $c_\phi=c_0$, the observables $\Delta/\mu$ and $T_{\rm c}/\mu$ are independent of $\bar{\mu}$. Therefore, the values of $T_{\rm c}/\mu$ and $\Delta/\mu$ can only be read off for this particular choice. Results for $c_\phi=c_0$ are summarized in Table IV.

We emphasize that the renormalization group flow in the truncations with M-diagrams is well-defined for every choice of $c_\phi$. However, it requires to determine the function $\mu(\bar{\mu},c_\phi)$, which is $\mu=\bar{\mu}$ for $c_\phi=c_0$. For other values of $c_\phi$, an appropriate infrared renormalization condition has to relate the initial value $\bar{\mu}$ to the physical chemical potential. The determination of this condition is postponed to future work. Here we restrict ourselves to the simpler task of discussing the physical point $c_\phi=c_0$. An error estimate in the FBM-truncations, however, can still be obtained by means of the subtraction prescription discussed in App. \ref{AppEstimate}.

\begin{table}\label{TableGap}
\begin{tabular}[t]{|c|c|c|c|c|c|}
\hline
 $\mbox{ }$ & \multicolumn{5}{|c|}{ $\Delta/\mu$}\\
 \hline \hline
 \  Truncation \ & \ F \ & \ FB$_0$ \ & \ FB \ & \ FBM$_0$ \ & \ FBM \ \\
 \hline 
 $\phi^4$ & \ 1.04 \  & 0.97 & 0.99 & 0.94 & 0.82 \\
 $\phi^8$ & 1.13 & 1.11 & 1.13 & 1.04 & 0.89\\
\hline\hline
 $\mbox{ }$ & \multicolumn{5}{|c|}{ $T_{\rm c}/\mu$}\\
 \hline \hline
 \  Truncation \ & \ F \ & \ FB$_0$ \ & \ FB \ & \ FBM$_0$ \ & \ FBM \ \\
 \hline 
 $\phi^4$ & \ 0.664 \  & \ 0.381 \ & \ 0.381 \ & \ 0.385\ & \ 0.383\  \\
 \hline
\end{tabular}
\caption{Critical temperature and superfluid gap for all truncations applied in this work. By improving the truncation of the effective potential to order $\phi^8$, the gap is increased by approximately 10 percent. This is independent of the given truncation scheme. The values for this table have been obtained for a $Q$-exp regulator and with $c_\phi=c_0$, which also allows to compare to truncations which include mixed diagrams (M) with both fermionic and bosonic lines.}
\end{table}

\begin{figure}[t]
\centering
\includegraphics[width=8cm]{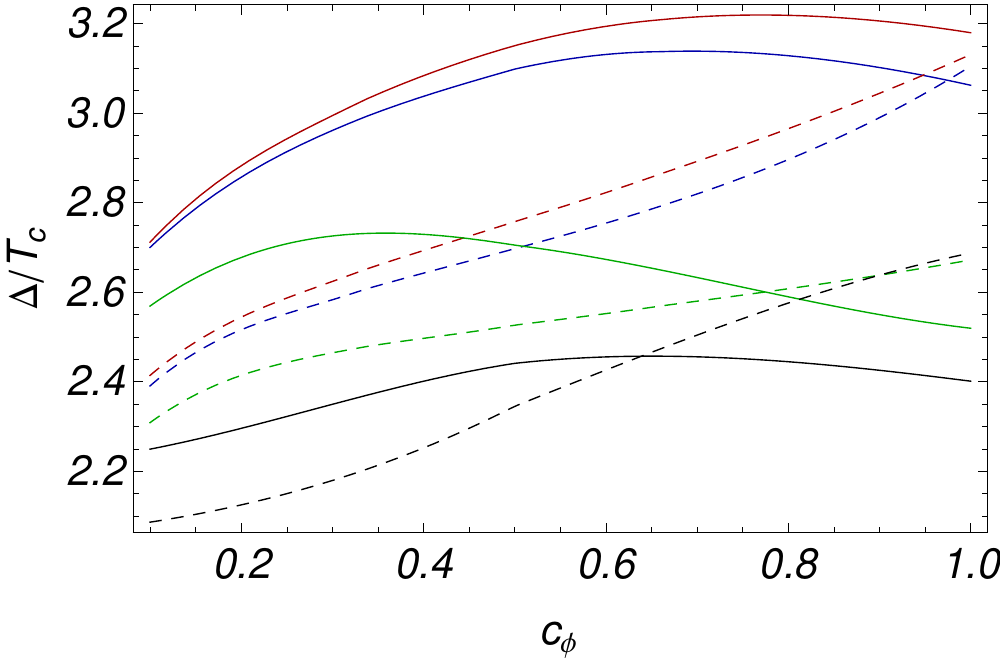}
\caption{Dependence of the ratio $\Delta/T_{\rm c}$ on the choice of the relative cutoff scale $c_\phi$. We employ the $Q$-exp regulator. Here and in Figs. \ref{plot_delta} and \ref{plot_Tc}, we choose the following labelling of the curves: Solid lines correspond to a $\phi^8$-truncation for computing the gap, whereas dashed lines give the results for the gap in a $\phi^4$-truncation. The critical temperature is always computed in a $\phi^4$-truncation. The colors correspond to the truncations (from top to bottom) FB (red), FB$_0$ (blue), FBM$_0$ (green), and FBM (black). Upon including higher orders in the effective potential, the gap increases by 10 percent in all truncation schemes.}
\label{plot_delta_Tc}
\end{figure}

The ratio $\Delta/T_{\rm c}$ can be computed consistently for every truncation. We display our results in Fig. \ref{plot_delta_Tc} and Table V. Whereas the $c_\phi$-dependence is rather strong in a $\phi^4$-truncation, we find the $\phi^8$-truncation to flatten the curve for all truncations. Moreover, there is a systematic increase of $\Delta/T_{\rm c}$ when going to order $\phi^8$. We find that all four curves in a $\phi^8$-truncation show a maximum in the interval $I_c=[0.2,1]$. It is a generic finding of our analysis that observables tend to have minima or maxima within this interval. Accordingly, we can use the variation within $I_c$ for an error estimate. This procedure is applied below.

\begin{table}\label{Tablecphi2}
\begin{tabular}[t]{|c|c|c|c|c|}
\hline
 $\mbox{ }$ & \multicolumn{4}{|c|}{$\Delta/T_{\rm c}$}\\
 \hline \hline
 \ Truncation \ & \ FB$_0$ \ & \ FB \ &\  FBM$_0$ \ & \ FBM \ \\
 \hline 
 \ $c_\phi=1$ \ & \ 3.1 \ & \ 3.2 \ & \ 2.5 \ & \ 2.4 \ \\
 \ $c_\phi=c_0$ \ & \ 2.9 \  & \ 2.9 \ & \ 2.7 \ & \ 2.3 \ \\
\hline
\end{tabular}
\caption{Relative cutoff scale $c_\phi$-dependence of the ratio $\Delta/T_{\rm c}$ for a $Q$-exp regularization scheme. The values of the gap are obtained for a $\phi^8$-truncation, whereas for the critical temperature we have chosen a $\phi^4$-truncation of the effective potential. We observe a substantial lowering when including mixed diagrams (M).}
\end{table}

The fermion anomalous dimension $\eta_{\psi k}$ receives strong corrections at the symmetry breaking scale $k^2 \simeq \mu$, where it becomes of order $0.2$. For $k\to 0$, it eventually vanishes. The renormalization effects on the Feshbach coupling $h^2$ are found to be small.

\subsection{Error estimates}\label{estimate}
\begin{figure}[t]
\centering
\includegraphics[width=8cm]{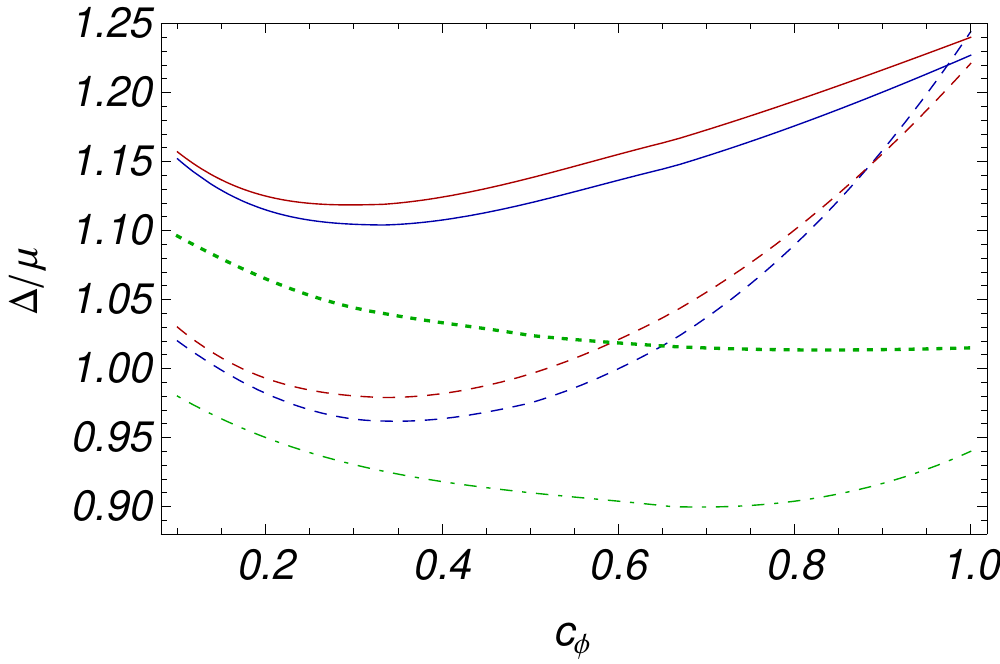}
\caption{Relative cutoff scale dependence of the superfluid gap $\Delta/\mu$. Colors are as in Fig. \ref{plot_delta_Tc}, and solid (dashed) lines correspond to a $\phi^8$- ($\phi^4$-) truncation. The FB$_0$- and FB-truncations become much more stable when including the higher terms in the effective potential. For an error estimate we show the FBM$_0$-truncation where we subtracted the anomalous running of the chemical potential according to Eq. (\ref{EQetamu}). This corresponds to the dotted (dotdashed) curve for a $\phi^8$- ($\phi^4$-) truncation. We emphasize that the latter two curves are applied here only for estimating the error.}
\label{plot_delta}
\end{figure}

\begin{figure}[t]
\centering
\includegraphics[width=8cm]{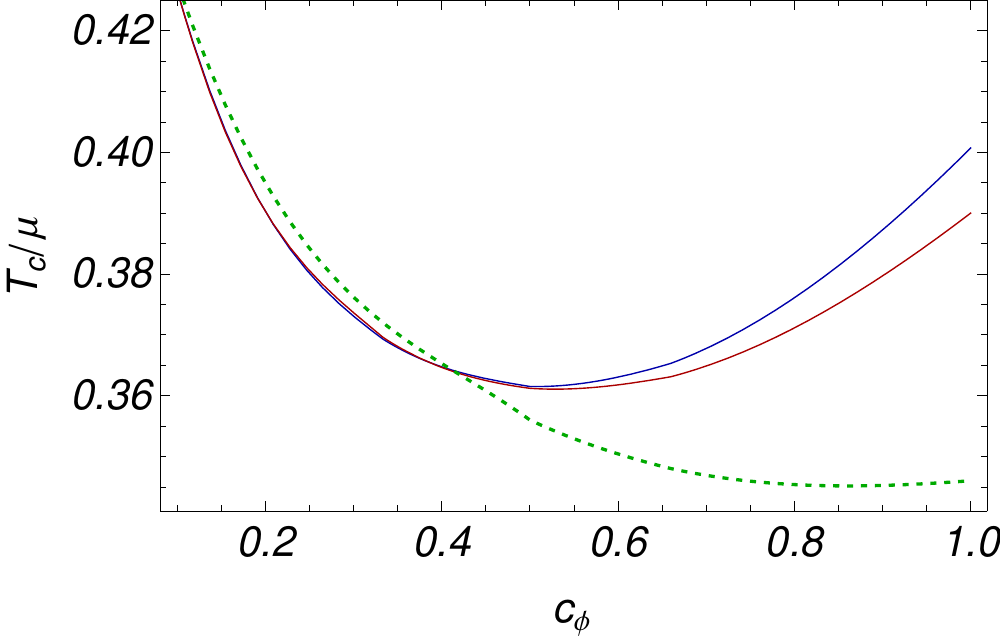}
\caption{Relative cutoff scale dependence of the critical temperature $T_{\rm c}/\mu$. Colors are as in Fig. \ref{plot_delta}. The critical temperatures shown here have been obtained in a $\phi^4$-truncation. The dotted curve gives the error estimate for the FBM$_0$-truncation according to the subtraction of the anomalous running of the chemical potential in Eq. (\ref{EQetamu}). }
\label{plot_Tc}
\end{figure}

Now we can estimate the errors of $\Delta/\mu$ and $T_{\rm c}/\mu$ within each individual truncation. We estimate the errors from the variation with $c_\phi$ shown in Figs. \ref{plot_delta_Tc}, \ref{plot_delta}, and \ref{plot_Tc}. From the figures presented in this section it is apparent that observables show pronounced features like minima or maxima inside the interval $c_\phi\in I_c=[0.2,1]$. It is therefore reasonable to concentrate on this interval to estimate the error. 

Our final results of the error analysis are summarized in Table \ref{TableResults}, where we apply the following notation:
\begin{itemize}
 \item{Mean field:} F-truncation to order $\phi^{24}$
 \item{Truncation 1:} FB$_0$-truncation to order $\phi^8$ ($\phi^4$) for $\Delta/\mu$ ($T_{\rm c}/\mu$)
 \item{Truncation 2:} FB-truncation to order $\phi^8$ ($\phi^4$) for $\Delta/\mu$ ($T_{\rm c}/\mu$)
 \item{Truncation 3:} FBM$_0$-truncation to order $\phi^8$ ($\phi^4$) for $\Delta/\mu$ ($T_{\rm c}/\mu$)
 \item{Truncation 4:} FBM-truncation to order $\phi^8$ ($\phi^4$) for $\Delta/\mu$ ($T_{\rm c}/\mu$)
\end{itemize}
The running couplings associated to the truncations are listed at the end of Sec. \ref{trunc}.

For the FB$_0$- and FB-truncations, the $c_\phi$-dependence of the superfluid gap is small within the $\phi^8$-truncation. Moreover, the improvement $\phi^4\to\phi^8$ in the effective potential seems to equilibrate the values, since the value at $c_\phi=1$ remains almost unchanged, whereas the values for smaller $c_\phi$ are increased. We choose the central value in the interval $I_c$ and find $\Delta/\mu=1.17(6)$ and $\Delta/\mu=1.18(6)$ for the FB$_0$- and FB-truncations, respectively. The error is given by the distance from the minimum and maximum inside the interval $I_c$. In the same fashion we find $T_{\rm c}/\mu=0.38(2)$ and $T_{\rm c}/\mu=0.376(14)$ for FB$_0$ and FB, respectively. Applying this procedure to the ratio $\Delta/T_{\rm c}$ in Fig. \ref{plot_delta_Tc}, we find $\Delta/T_{\rm c}=2.9(2)$ and $3.0(2)$, respectively. The relative error of all observables is thus consistently given by 5 percent.

Estimating the error within the FBM$_0$- and FBM-truncations is complicated by the fact that $\bar{\mu}\neq\mu$ for $c_\phi\neq c_0$. We therefore choose the physical point $c_0$ to obtain our central values for $\Delta/\mu$ and $T_{\rm c}/\mu$. These values are given in Table V. A good estimate of the relative error can be obtained from $\Delta/T_{\rm c}$, which is independent of $\bar{\mu}$. To estimate the error we apply the procedure described in App. \ref{AppEstimate}.

We find $\Delta/T_{\rm c}=2.6(1)$ for the FBM$_0$-truncation, which corresponds to a 4 percent error. Averaging $\Delta/\mu$ which is obtained by means of the $\eta_\mu$-subtraction procedure of Eq. (\ref{EQetamu}) over the interval $I_c$ yields $1.04(3)$. This coincides with the central value. Given the fact that $\Delta/T_{\rm c}$ is indeed very flat as a function of $c_\phi$, we conclude that $\Delta/\mu=1.04(5)$ is a reasonable error estimate. 

The dotted curve in Fig. \ref{plot_Tc} gives the error estimate of $T_{\rm c}/\mu$ when subtracting the anomalous running of the chemical potential in the FBM$_0$-truncation. The rather strong dependence on $c_\phi$ (when compared to the $c_\phi$-dependence of $\Delta/T_{\rm c}$ and $\Delta/\mu$) can be explained by the fact that the subtraction procedure in Eq. (\ref{EQetamu}) strongly influences the flow at $k^2\simeq \mu$. As this is precisely the scale where precondensation occurs and decides over the value of $T_{\rm c}$, the critical temperature is strongly affected by Eq. (\ref{EQetamu}). Hence, the error estimate also contains unphysical contributions and should not be extrapolated too far into the region where $\eta_\mu$ is large. A reasonable error estimate is thus again found to be $T_{\rm c}/\mu=0.385(20)$. This is also in harmony with the relative variation of $\Delta/T_{\rm c}$ in dependence of $c_\phi$.

Finally, for the FBM-truncation we find within the interval $I_c$ that $\Delta/T_{\rm c}=2.4(1)$. The insensitivity of this result with respect to $c_\phi$ is similar to the FBM$_0$-truncation. Since both truncations are similar to each other, we assume a 5 percent error within the FBM-truncation just like for the FBM$_0$-case. The central values are taken at $c_\phi=c_0$. We then arrive at $\Delta/\mu=0.89(5)$ and $T_{\rm c}/\mu=0.38(2)$.

Our best estimates for $\Delta/T_{\rm c}$ and $\Delta/\mu$ are obtained as the central value within the 4 truncations, with the error being given by the distance to the maximum (minimum). This yields $\Delta/\mu=1.04(15)$ and $\Delta/T_{\rm c}=2.7(3)$. For the critical temperature we obtain $0.381(6)$ with this procedure of averaging. However, this underestimates the error of the individual truncations, so we choose $T_{\rm c}/\mu=0.38(2)$ which is valid in all four truncations.

\section{Conclusions and outlook}\label{SecConcl}
In this paper we have studied the physics of the Unitary Fermi Gas by employing the FRG with frequency and momentum regulators. The advantage of this choice of cutoff functions is a flow of running couplings which is influenced only by local frequency and momentum shells. Accordingly, a scale-dependent derivative expansion is expected to be a good approximation to the full RG flow. We have implemented general frequency and momentum cutoffs for non-relativistic models, which represent a flexible tool  to study the many-body physics of the Unitary Fermi Gas. These cutoffs lead to flows that are local in frequency and momentum space. This is mandatory within the approximation schemes if aiming at quantitative precision. Moreover, the  conceptually simple set-up allows to systematically study the effects of higher order truncations and specification parameters like the relative cutoff scale $c_\phi$. Hence, error estimates for observables computed with the FRG can be provided, which is an important step towards comparison with other methods and experimental data, where error estimates are standard.

In Tables VI and VII we display reference values on $T_{\rm c}/\mu$ and $\Delta/\mu$ from other theoretical approaches, and from experiment. The stability of our result  $T_{\rm c}/\mu=0.38(2)$ within all truncations considered in this work indicates the efficiency of the $Q$-exp regulator to incorporate the required physical effects. In particular, we found a strong decrease of the critical temperature in comparison to previous FRG calculations with the $q^2$-regulator, which places our calculation in the range $T_{\rm c}/\mu=0.3-0.4$ of the reference values. For the superfluid gap we obtain $\Delta/\mu=1.04(15)$, which is smaller than all of the reference values. In particular, by improving the truncation due to the inclusion of the running fermion propagator, we do not find a convergence of results, in contrast to the critical temperature. Thus, the limit $T=0$ requires additional running couplings, which are less important at criticality. We expect the inclusion of the full effective potential $U(\rho)$ to significantly improve our results at zero temperature.

\begin{table}\label{TableRefTc}
\begin{tabular}[t]{|c|c|c|c|c|}
 \hline
  $\mbox{ }$ &   $T_{\rm c}/\vare_{\rm F}$  &   $\mu_{\rm c}/\vare_{\rm F}$   &  $T_{\rm c}/\mu$  \\
\hline \hline
 Burovski \emph{et al.} (DDMC \cite{PhysRevLett.96.160402}) & 0.152(7) & 0.493(14) & 0.308 \\
 Haussmann \emph{et al.} (LW \cite{PhysRevA.75.023610}) & 0.160 & 0.394 & 0.406 \\
 Bulgac \emph{et al.} (QMC \cite{PhysRevA.78.023625}) & 0.15(1) & 0.43(1) & 0.35 \\
 Nascimbene \emph{et al.} (Exp \cite{Salomon}) & 0.157(15) & 0.49(2) & 0.32(3) \\
 Horikoshi \emph{et al.} (Exp \cite{Horikoshi22012010}) & 0.17(1) & 0.43(1) & 0.40 \\
 Goulko, Wingate (DDMC \cite{PhysRevA.82.053621}) & 0.171(5) & 0.429(9) & 0.399 \\
 Floerchinger \emph{et al.} (FRG \cite{PhysRevA.81.063619}) & 0.248 & 0.55 & 0.45 \\
 Ku \emph{et al.} (Exp \cite{Ku03022012}) & 0.167(13) & 0.42* & 0.40 \\
 \hline
 This work & - & - & 0.38(2) \\
 \hline
\end{tabular}
\caption{Reference values for the critical temperature $T_{\rm c}/\mu$ from different theoretical and experimental works. The abbreviations correspond to Diagrammatic determinant Monte Carlo (DDMC), Quantum Monte Carlo (QMC), Self-consistent T-matrix approach or Luttinger-Ward formalism (LW), and experiment (Exp). (*We estimated the critical chemical potential of the MIT data \cite{Ku03022012} by the maximal chemical potential $\mu_{\rm max}/\vare_{\rm F}=0.42(1)$ at $T/\vare_{\rm F}=0.171(10)$.)}
\end{table}

\begin{table}\label{TableRefGap}
\begin{tabular}[t]{|c|c|c|c|c|}
 \hline
  Method &   $\Delta/\vare_{\rm F}$  &   $\mu/\vare_{\rm F}$   &  $\Delta/\mu$  \\
\hline \hline
 Carlson \emph{et. al} (QMC \cite{PhysRevLett.91.050401}) & 0.55(5) & 0.44(1) & 1.3\\
 Carlson, Reddy (QMC \cite{PhysRevLett.95.060401}) & 0.50(5) & 0.42(1) & 1.2 \\
 Haussmann \emph{et al.} (LW \cite{PhysRevA.75.023610}) & 0.46 & 0.36 & 1.3 \\
 Carlson, Reddy (\cite{PhysRevLett.100.150403}) & 0.45(5) & - & - \\
 Bulgac \emph{et al.} (QMC \cite{PhysRevA.78.023625}) & - & 0.37(5) & - \\
 Schirotzek \emph{et al.} (Exp \cite{PhysRevLett.101.140403}) & 0.44(3) & - & - \\
 Bartosch \emph{et al.} (FRG \cite{PhysRevB.80.104514}) & 0.61 & 0.32 & 1.9 \\
 Floerchinger \emph{et al.} (FRG \cite{PhysRevA.81.063619}) & 0.46 & 0.51 & 0.90\\
 Carlson et. al (QMC, \cite{PhysRevA.84.061602}) & - & \ 0.372(5) \ & - \\
 Ku \emph{et al.} (Exp \cite{Ku03022012}) & - & 0.376(4) & - \\
 \hline
 This work & - & - & 1.04(15) \\
 \hline
\end{tabular}
\caption{Reference values for the superfluid gap $\Delta/\mu$ at $T=0$. The abbreviations are as in Table VI. The values in Ref. \cite{PhysRevLett.100.150403} have been extracted from measured density distributions from partially spin-polarized trapped atoms. The overall trend of the Monte Carlo and experimental data indicates a preferred value $\Delta/\mu \simeq 1.2-1.3$. }
\end{table}

Besides the use of a new class of cutoff functions, this work extended previous treatments of the Unitary Fermi Gas by means of the FRG in several aspects. We included the emergent ``relativistic'' term $V_\phi q_0^2$ in the boson propagator, which is expected to be important in the deep infrared (Goldstone) regime. As this regime is irrelevant for most practical purposes in three dimensions, the effect on physical observables is rather small. However, we find that the inclusion of the coupling $V_\phi$ makes the flow less sensitive to variations in the regularization and specification scheme. The effect of $V_\phi$ is more pronounced for momentum only cutoffs like the $q^2$-opt regulator. This can be interpreted as follows: The frequency-dependence of the boson propagator with a momentum only regulator has to be known for a large frequency interval, because no narrow frequency shell is singled out by the regulator $R_k(q^2)$. In contrast, for a frequency and momentum regulator $R_k(q_0,q^2)$, the inclusion of $V_\phi$ leads only to slight modifications, because the leading contributions to the real and imaginary parts of the boson self-energy are already incorporated in $A_\phi$ and $S_\phi$, respectively. In particular, for some choices of the relative cutoff scale $c_\phi$, the truncations FB$_0$ and FB (i.e., without and with $V_\phi$) are practically equivalent. The role of $c_\phi$ is discussed below in more detail.

The bosonized model of the BCS-BEC crossover employed in this work allows to include higher order bosonic processes by means of a more elaborate treatment of the effective potential $U(\rho)$. We have included in this work for the first time higher powers in a series expansion in orders of $(\rho-\rho_0)^n$. As a generic feature we found that the superfluid gap is increased by 10 percent when going from a $\phi^4$- to a $\phi^8$-truncation. Within the F-truncation we showed that a successive inclusion of higher orders yields the correct mean field prediction. When also including bosonic fluctuations, there is a competition between the corresponding diagrams, and an expansion in $\phi^{2N}$ shows fast apparent convergence. It is not obvious whether higher orders in a $\phi^{2N}$-truncation actually give an improved results. We have found in some cases, like the critical temperature for the FB-truncation, that the prediction from a $\phi^4$-truncation is already quite stable towards the inclusion of higher terms in $U(\rho)$.

The series expansion of $U(\rho)$ in powers of $(\rho-\rho_0)^n$ has some drawbacks. First of all, first order phase transitions are potentially difficult to access within such an  expansion. (A successful recent study within such an expansion has been performed in Ref. \cite{Eichhorn:2013zza}.) For the spin-balanced case discussed in this work, we do not expect the transition to be of first order. However, when including a deviation $\delta\mu=\mu_1-\mu_2$ in the chemical potentials of species 1 and 2, the superfluid phase transition can change from second to first order. To resolve the corresponding physics, it is necessary to compute the effective average potential $U_k(\rho)$ for each $k$ on a grid of $\rho$-values. Studies of the quark-meson-model and related relativistic field theories with the FRG \cite{Schaefer:2004en,Herbst:2010rf,PhysRevD.88.014007} have already shown to be feasible of such an improvement. Another possibility to resolve the $\rho$-dependence of $U_k(\rho)$ consists in an expansion around a fixed value $\rho \neq \rho_0$. Again, this has been found to be an efficient method in the quark-meson-model with the FRG in Ref. \cite{Fabian}, which also highlights the importance of the expansion point for the evaluation of the wave function renormalizations $A$ and $S$. 

Every truncation of the functional flow equation for $\Gamma_k$ depends on some specification parameters which cannot be determined easily from physical grounds. For instance, these can be the widths of a finite difference projection of $A$ and $S$ in a derivative expansion \cite{Schnoerr:2013bk}. In our case, we investigated the influence of the relative cutoff scale $c_\phi$ between bosons and fermions as an error estimate. A good approximation to the full flow of the effective average action should be independent of the regulators $R_\phi(Q)$ and $R_\psi(Q)§$, and, consequently, independent of $c_\phi$. Within a truncation of $\Gamma_k$, however, a spurious dependence of observables at $k=0$ on $c_\phi$ shows up. Clearly, the limiting values $c_\phi\to 0$ and $c_\phi\to \infty$ are unphysical, whereas a value of order unity seems reasonable. Accordingly, only for $c_\phi=\mathcal{O}(1)$ a sensitivity study can be applied.

The parameter $c_\phi$ does not influence the F-truncations, as no boson propagator appears. Applying optimization theory \cite{Litim:2000ci,Litim:2001up,Pawlowski20072831} in the present local approximation scheme leads to the demand of maximally local FRG flows: One has to minimize the frequency and momentum transfer in the flow. As this is a demanding task we leave it to future work. Here we apply the principle of minimum sensitivity which agrees with the full optimization in specific cases \cite{Litim:2001fd}. It states that 
the physical value may correspond to the minimum with respect to the variation of an external parameter. For the FB$_0$- and FB-truncations we find that $\Delta/\mu$ and $T_{\rm c}/\mu $ show the tendency to develop a minimal value at $c_\phi \simeq 0.5$. However, as this interpretation is not unambiguous, we only use this feature for a benchmark point to perform an error estimate.

The price to pay when employing frequency and momentum cutoffs is a nontrivial flow of running couplings in the symmetric regime of the flow. In fact, diagrammatic simplifications, which are characteristic for non-relativistic field theories in vacuum and disordered regimes \cite{Floerchinger:2013tp}, are no longer present. This is not problematic in the sense that all physical quantities are defined for $k=0$, where the cutoff function $R_k(Q)$ is removed from the flow. For instance, the correct UV-renormalization of the initial four-fermion coupling $\bar{\lambda}_{\psi}$ (or, equivalently, the boson detuning $m^2_{\phi\Lambda}$), can be implemented by the requirements in Eqs. (\ref{run9}) and (\ref{run10}). Accordingly, the scattering length $a$ is well-defined for every truncation and regularization scheme. The correct determination of the chemical potential is more involved, as we have discussed in detail.

In this work we did not address the question of determining the density, as this observable has no feedback on the remaining running couplings which parametrize the effective action $\Gamma$. To understand the latter, only the dependence on the grand canonical variables $\mu$ and $T$ has be resolved. We leave the determination of the density, and thus the equation of state $P(\mu,T)$, for future work. The results of Refs. \cite{PhysRevA.85.063607,PhysRevB.84.174513,PhysRevA.86.043624} on the Bose--Hubbard model have shown the capability of the nonperturbative RG to provide all the necessary thermodynamic observables for cold atom experiments.

The techniques presented in this work are easily carried over to systems which go beyond the Unitary Fermi Gas, and which can be realized in cold atom experiments. To include a finite scattering length is straightforward. Here, we concentrated on $a^{-1}=0$ as the initial conditions of the flow are particularly simple in this case. This allows for a systematic study of truncation, regularization, and specification effects without the need to redo the fine-tuning of $m^2_{\phi\Lambda}$ and $m^2_{\psi\Lambda}$ for each individual case. However, once an agreeable set of running couplings is singled out, the whole BCS-BEC crossover can be studied in a unified fashion. We found that the critical temperature is rather robust already in the FB-truncation, whereas the strong renormalization of the effective chemical potential indicates that an accurate determination of the density requires to include mixed (M) diagrams.

The extension of the analysis to the case of the two-dimensional (2D) BCS-BEC crossover is also straightforward. In this case, one always has a bound state in vacuum, such that the initial conditions are technically similar to the BEC-side of the corresponding three-dimensional case. The pronounced infrared flow which is well-known from non-relativistic bosons in two dimensions will also be important to resolve the physics of the 2D-crossover. The advantage of the FRG in this respect lies in the absence of logarithmic divergences and the proper treatment of Goldstone fluctuations. Within our truncations with emergent $V_\phi q_0^2$-term and higher orders in the effective potential, we expect a quantitatively good account of the 2D BCS-BEC crossover.

Interesting questions arise in 3D and 2D when a chemical potential imbalance shifts the relative Fermi surfaces of atoms in states 1 and 2. When incorporating a grid of $\rho$-values for the effective average potential $U_k(\rho)$, the full parameter space of the spin-imbalanced systems can be studied with the methods presented in this work. In particular, our treatment can be applied to the BEC-side of the 3D crossover, and the whole range of binding energies in the 2D-case, where the phase diagram of the imbalanced crossover is not yet fully understood.

In our view the present work establishes the FRG as a quantitatively precise method for ultracold atom gases. This includes an estimate of the error of the results. The precision obtained so far is comparable to other methods. It can be increased by extended truncations. Perhaps most importantly it has been demonstrated that the FRG successfully can describe all regions and limits of the phase diagram, including, for example, the critical phenomena of a second order phase transition. We expect that within the truncation of the present paper the accuracy for other regions of the phase diagram is at least as good as the one obtained in the present work for the Unitary Fermi Gas, which is, in some sense, the most difficult case.

\begin{center}
 \textbf{Acknowledgements}
\end{center}

\noindent We thank N. Christiansen, S. Diehl, N. Dupuis, S. Floerchinger, T. K. Herbst, S. Moroz, F. Rennecke, M. Scherer, D. Schnoerr, and N. Strodthoff for discussions. I. B. acknowledges funding from the Graduate Academy Heidelberg. This work is supported by the Helmholtz Alliance HA216/EMMI and the grant ERC-AdG-290623.

\appendix

\section{Flow equation}\label{appA}
The flow equations for the running couplings which parametrize the effective average action can be obtained by means of suitable projection prescriptions from a few constitutive or master equations. In this section, we outline the general structure of the functional flow equation and construct the hierarchy of flow equations for correlation functions in a homogeneous setting. In Appendix \ref{appB}, we show how the constitutive equations can be obtained from the latter, whereas the projections are specified in Appendix \ref{appC}.

\emph{Notation.} To express the running of couplings $\{g_k\}$ with $k$, we write $t=\log(k/\Lambda)$, where $\Lambda$ is the large UV cutoff scale defined at the beginning of Sec. \ref{SecCoupl}. Flow equations are then given by the beta functions $\beta_g= k \partial_k g_k = \partial_t g =:\dot{g}$. The regime of large $k$ ($t\simeq0$) will be referred to as UV, whereas we call $k\to0$ ($t\to -\infty$) the infrared (IR). Dimensionless running couplings $\{\tilde{g}_k\}$ are rescaled with powers of $k$ according to their canonical dimensions, and indicated by a tilde. The effective average action $\bar{\Gamma}[\bar{\phi},\bar{\psi}]$ is parametrized in terms of the unrenormalized fields $\bar{\phi}=\phi /A_\phi^{1/2}$ and $\bar{\psi}=\psi /A_\psi^{1/2}$. However, we can rescale all running couplings by appropriate powers of $A_\phi$ and $A_\psi$ such that $\bar{\Gamma}[\bar{\phi},\bar{\psi}]=\Gamma[\phi,\psi]$. Accordingly, we denote unrenormalized (not rescaled) couplings by an overbar, $\bar{g}$, and the renormalized (rescaled) couplings without an overbar, $g$. The projection will be performed for the unrenormalized quantities, but the flow equations can solely be expressed in terms of renormalized couplings. 

The flow equation for the full effective average action \cite{Wetterich1993} has a supertrace one-loop structure according to
\begin{align}
 \nonumber \dot{\bar{\Gamma}}_k[\bar{\Psi}] &= \frac{1}{2} \mbox{STr} \Bigl[ \Bigl(\bar{\Gamma}^{(2)}_k + \bar{R}_k\Bigr)^{-1} \dot{\bar{R}}_k\Bigr]\\
 \nonumber &= \frac{1}{2} \mbox{STr} \Biggl[ \begin{pmatrix} \bar{\Gamma}^{(2)}_{\rm BB}[\bar{\Psi}]  + \bar{R}_{\rm BB} & \bar{\Gamma}^{(2)}_{\rm BF}[\bar{\Psi}]  \\ \bar{\Gamma}^{(2)}_{\rm FB}[\bar{\Psi}]  & \bar{\Gamma}^{(2)}_{\rm FF}[\bar{\Psi}]  + \bar{R}_{\rm FF} \end{pmatrix}^{-1} \\
 \label{appA1} &\mbox{ }\times \begin{pmatrix} \dot{\bar{R}}_{\rm BB} & 0 \\ 0 & \dot{\bar{R}}_{\rm FF}\end{pmatrix} \Biggr],
\end{align}
where we collected the ($X$-dependent) field variables in the list $\bar{\Psi}=(\bphi,\bpsi_\sigma) = (\bphi,\bphi^*,\bpsi_1,\bpsi_2,\bpsi_1^*,\bpsi_2^*)$. The first two elements of this set constitute complex numbers, whereas the latter four are Grassmann variables. Due to this fact, computations are most easily performed in a superalgebraic formulation: A supermatrix $M$ can be decomposed into four block matrices,
\begin{align}
 \label{appA2} M = \begin{pmatrix} M_{\rm BB} & M_{\rm BF} \\  M_{\rm FB} & M_{\rm FF} \end{pmatrix},
\end{align}
where $M_{\rm BB}$ and $M_{\rm FF}$ are of even Grassmann parity, and $M_{\rm BF}$ and $M_{\rm FB}$ are Grassmann odd, respectively. In our case, $M_{\rm BB}$ is a $2\times 2$-matrix, and $M_{\rm FF}$ has a $4\times 4$-structure. The supertrace is then defined as
\begin{align}
 \label{appA3} \mbox{STr} M = \mbox{Tr} M_{\rm BB} - \mbox{Tr} M_{\rm FF},
\end{align}
with the remaining traces being over the block structure of $M_{\rm BB}$ and $M_{\rm FF}$, respectively. Note that the fermionic part appears with the characteristic minus sign.

The right hand side of Eq. (\ref{appA1}) depends on the scale dependent propagator, related to the effective action by means of the second functional derivative
\begin{align}
 \label{appA4} \bar{\Gamma}^{(2)}_k[\bar{\Psi}] = \Bigl(\frac{\stackrel{\rightarrow}{\delta}}{\delta \bar{\Psi}} \bar{\Gamma}_k \frac{\stackrel{\leftarrow}{\delta}}{\delta \bar{\Psi}}\Bigr)[\bar{\Psi}] = \begin{pmatrix} \bar{\Gamma}^{(2)}_{\rm BB}[\bar{\Psi}]  & \bar{\Gamma}^{(2)}_{\rm BF}[\bar{\Psi}]  \\ \bar{\Gamma}^{(2)}_{\rm FB}[\bar{\Psi}]  & \bar{\Gamma}^{(2)}_{\rm FF}[\bar{\Psi}]  \end{pmatrix}.
\end{align}
The regulator matrix 
\begin{align}
 \label{appA5} \bar{R}_k = \begin{pmatrix} \bar{R}_{\rm BB} & 0 \\ 0 & \bar{R}_{\rm FF} \end{pmatrix}
\end{align}
is independent of $\bar{\Psi}$ and diagonal in superspace.

Since the flow equation is an equation for a functional, the arguments $\bpsi_\sigma(X)$ and $\bphi(X)$ are arbitrary and possibly inhomogeneous in $X$, thereby describing physical situations where the fields acquire nonvanishing mean field values due to appropriately chosen sources. For this reason, the second functional derivative $\bar{\Gamma}_k^{(2)}$ has nonvanishing off-diagonal entries, and the inversion of $(\bar{\Gamma}^{(2)}_k+\bar{R}_k)$ in nontrivial. In particular, taking functional derivatives of Eq. (\ref{appA1}) with respect to $\bar{\Psi}$, we generate a hierarchy of flow equations for the one-point irreducible vertices, $\dot{\bar{\Gamma}}_k^{(n)}$, which depend on $\bar{\Gamma}_k^{(3)}$, $\bar{\Gamma}_k^{(4)}$, etc. 

The equilibrium configuration $\bar{\Psi}_0(X)=(\bphi_0,\bpsi_{\sigma0})$, which minimizes the effective action in the absence of external sources, will be homogeneous in $X$. (We neglect here the possibility of inhomogeneous phases like the FFLO-state, since they are not expected to be stable in the population-balanced three-dimensional system.) Moreover, only the bosonic field can be occupied macroscopically, whereas the fermionic mean field vanishes: $\bpsi_{\sigma0}=0$. Since the effective action inherits the $U(1)$-symmetry of the microscopic action, the former, when evaluated for homogeneous bosonic configurations, can only depend on the $U(1)$-invariant $\bar{\rho}=\bphi^*\bphi$. The ground state $\bphi_0$ breaks $U(1)$-symmetry spontaneously by choosing, for fixed $\bar{\rho}_0$, one direction in the $(\bphi,\bphi^*)$-plane. For convenience, we assume the bosonic ground state to be real valued, which corresponds to the choice $(\bphi,\bphi^*)=(\sqrt{\bar{\rho}},\sqrt{\bar{\rho}})$. Often it will be useful to decompose the boson field into its real and imaginary parts according to
\begin{align}
 \label{appA8} \bphi^{(*)} = \frac{1}{\sqrt{2}} \Bigl( \bphi_1 \pm \rmi \bphi_2 \Bigr).
\end{align}
In the presence of a nonvanishing real expectation value $(\bphi_{10},\bphi_{20})=(\sqrt{2\bar{\rho}},0)$, the real and imaginary components, respectively, correspond to the radial and Goldstone modes.

When evaluated for homogeneous mean fields, the $n$-th functional derivative $\bar{\Gamma}^{(n)}_k(X_1,\dots,X_n)$ of the effective average action is proportional to $n$ delta functions $\delta(X_1-X_2)\dots \delta(X_1-X_n)$, where $X_i$ is the argument of the $i$-th field derivative. Within the momentum representation of the fields,
\begin{align}
 \label{appA9} \bar{\Psi}(P) = \int_X \bar{\Psi}(X) e^{-\rmi PX},
\end{align}
the vertices transform according to
\begin{align}
 \nonumber &\bar{\Gamma}^{(n)}_k[\bar{\Psi}](P_1,\dots,P_n) \\
 \label{appA10} &= \int_{X_1,\dots,X_n} \bar{\Gamma}^{(n)}_k[\bar{\Psi}](X_1,\dots,X_n) e^{\rmi P_1X_1}\dots e^{\rmi P_nX_n}.
\end{align}
This result in an overall delta function $\delta(P_1+\dots+P_n)$ for a homogeneous setting. The hierarchy of functional flow equations is thus reduced to the hierarchy of flow equations for correlation functions $\bar{\gamma}^{(n)}_k$ according to
\begin{align}
 \label{appA11} \dot{\bar{\Gamma}}^{(n)}_k(\bar{\rho})(P_1,\dots,P_n) =\delta(P_1+\dots+P_n) \dot{\bar{\gamma}}^{(n)}_k(\bar{\rho},P_2,\dots,P_{n}).
\end{align}
This set of equations is equivalent to Eq. (\ref{appA1}) for homogeneous situations and provides a good starting point for physically sound truncations of the correlation functions $\bar{\gamma}^{(n)}_k$. 

The supermatices $\bar{\Gamma}_k^{(2)}$ and $\bar{R}_k$ become diagonal for constant bosonic background field, and can be written as
\begin{align}
 \label{appA12} \Bigl(\bar{\Gamma}^{(2)}_k(\bar{\rho})+\bar{R}_k\Bigr)(Q',Q) &= \delta(Q'+Q) \begin{pmatrix} \bar{G}^{-1}_\phi(Q) & 0 \\ 0 & \bar{G}^{-1}_{\psi}(Q) \end{pmatrix},\\
 \label{appA13} \bar{R}_k(Q',Q) &= \delta(Q'+Q) \begin{pmatrix} \bar{R}_\phi(Q) & 0 \\ 0 & \bar{R}_\psi(Q) \end{pmatrix}
\end{align}
with regulators
\begin{align}
 \label{appA14} \bar{R}_\phi(Q) &= A_\phi R_\phi(Q) = \begin{pmatrix} 0 & R_\phi^{-Q} \\ R_\phi^Q & 0 \end{pmatrix},\\
 \label{appA15} \bar{R}_\psi(Q) &= A_\psi R_\psi(Q) = \begin{pmatrix} 0 & - R_\psi^{-Q}\textbf{1}  \\ R_\psi^Q\textbf{1} & 0 \end{pmatrix}.
\end{align}
To keep notation short, we will often denote the $Q$-dependence of functions by a superscript $Q$.

\section{Constitutive equations}\label{appB}
In order to solve the set of flow equations (\ref{appA11}) in practice, we need to truncate them such that we arrive at a finite set of equations, which can then be solved numerically. For this purpose, the field and momentum dependence of the correlation functions $\bar{\gamma}^{(n)}_k(\bar{\rho},\{P_i\})$ has to be approximated suitably. The scheme used throughout this work is built on an expansion around the $k$-dependent minimum $\rho_{0k}$ and zero external momenta and frequencies.

The regularized propagators can be parametrized according to
\begin{align}
 \label{appB6} \bar{G}_\phi^Q &= \frac{1}{A_\phi} G_\phi^Q = \frac{1}{A_\phi} \frac{1}{\mbox{det}_B^Q} \begin{pmatrix} -\rho U'' & L_\phi^{-Q} \\ L_\phi^Q & - \rho U''\end{pmatrix},\\
 \label{appB7} \bar{G}_\psi^Q &= \frac{1}{A_\psi} G_\psi^Q = \frac{1}{A_\psi} \frac{1}{\mbox{det}_F^Q} \begin{pmatrix} (h^2\rho)^{1/2}\vare & L_\psi^{-Q} \textbf{1} \\ -L_\psi^Q \textbf{1} & - (h^2\rho)^{1/2} \vare \end{pmatrix},
\end{align}
where $\vare=((0,1),(-1,0))$ is the fully antisymmetric tensor, $\textbf{1}$ is the $2\times 2$ unit matrix, and we introduced the notation
\begin{align}
 \label{appB8} L_\psi^Q &= P_\psi^Q + R_\psi^ Q,\\
 \label{appB9} \mbox{det}_F^Q &= L_\psi^QL_\psi^{-Q}+h^2\rho,\\
 \label{appB10} L_\phi^Q &= P_\phi^Q +R_\phi^Q + U' + \rho U'',\\
 \label{appB11} \mbox{det}_B^Q &= L_\phi^Q L_\phi^{-Q}-(\rho U'')^2.
\end{align}
Herein, a prime denotes a derivative with respect to $\rho = A_\phi \bar{\rho}$, and $U(\rho)=\bar{U}(\bar{\rho})$. The flow equation for the effective average potential $\bar{U}_k(\bar{\rho})=\bar{\gamma}^{(0)}_k(\bar{\rho})$ now takes the simple form
\begin{align}
 \nonumber \dot{\bar{U}}_k(\bar{\rho}) &= \frac{1}{2}\mbox{tr} \int_Q \bar{G}_\phi^Q\dot{\bar{R}}_\phi^Q - \frac{1}{2}\mbox{tr} \int_Q \bar{G}_\psi^Q \dot{\bar{R}}_\psi^Q\\
 \nonumber &= \frac{1}{2}\int_Q \frac{1}{A_\phi}\frac{L_\phi^Q\dot{\bar{R}}_\phi^{-Q}+L_\phi^{-Q}\dot{\bar{R}}_\phi^Q}{\mbox{det}_B^Q} \\
 \label{appB14} &\mbox{ }- \int_Q \frac{1}{A_\psi} \frac{L_\psi^Q\dot{\bar{R}}_\psi^{-Q}+L_\psi^{-Q}\dot{\bar{R}}_\psi^Q}{\mbox{det}_F^Q}.
\end{align}

The flow of the two-point function $\bar{\gamma}^{(2)}(Q)$ depends on $\bar{\gamma}_k^{(3)}$ and $\bar{\gamma}_k^{(4)}$. To close the set of equations, we assume momentum independent higher vertices according to
\begin{align}
 \label{appB20b} \bar{\gamma}^{(n)}_k(Q_2,\dots,Q_n) = \bar{\gamma}^{(n)}_k \ \text{ for } n\geq 3.
\end{align}

The flow equations for the inverse boson and fermion propagators are then most easily parametrized by introducing 
\begin{align}
 \label{appB15} X_\psi^Q &= \frac{(h^2\rho)^{1/2}}{A_\psi}\Bigl( L_\psi^Q \dot{\bar{R}}_\psi^{-Q}+L_\psi^{-Q}\dot{\bar{R}}_\psi^Q\Bigr),\\
 \label{appB16} Y_\psi^Q &= \frac{1}{A_\psi}\Bigl((L_\psi^Q)^2 \dot{\bar{R}}_\psi^{-Q}-h^2\rho \dot{\bar{R}}_\psi^Q\Bigr),\\
 \label{appB17} X_\phi^Q &= \frac{- \rho U''}{A_\phi}\Bigl(L_\phi^Q \dot{\bar{R}}_\phi^{-Q}+L_\phi^{-Q}\dot{\bar{R}}_\phi^Q\Bigr),\\
 \label{appB18} Y_\phi^Q &= \frac{1}{A_\phi}\Bigl((L_\phi^Q)^2\dot{\bar{R}}_\phi^{-Q}+(\rho U'')^2 \dot{\bar{R}}_\phi^Q\Bigr),
\end{align}
such that
\begin{align}
 \label{appB19} \bar{G}_\phi^Q \dot{\bar{R}}_\phi^Q \bar{G}_\phi^Q &= \frac{1}{A_\phi} \frac{1}{\mbox{det}_B^2(Q)}\begin{pmatrix} X_\phi^Q & Y_\phi^{-Q} \\ Y_\phi^Q & X_\phi^Q \end{pmatrix},\\
 \label{appB20} \bar{G}_\psi^Q \dot{\bar{R}}_\psi^Q \bar{G}_\psi^Q &= \frac{1}{A_\psi} \frac{1}{\mbox{det}_F^2(Q)} \begin{pmatrix} X_\psi^Q \vare & Y_\psi^{-Q}\textbf{1} \\ - Y_\psi^Q \textbf{1} & - X_\psi^Q \vare\end{pmatrix}.
\end{align}

The flow of the fermion self-energy reads
\begin{align}
 \nonumber \dot{\bar{G}}_{\psi_1^*\psi_1}^{-1}(P) &=  - A_\psi h^2 \int_Q \frac{Y_\phi^Q L_\psi^{Q+P}}{(\mbox{det}^Q_B)^2\mbox{det}_F^{Q+P}} \\
 \label{appB21}&\mbox{ }- A_\psi h^2 \int_Q  \frac{Y_\psi^Q L_\phi^{Q-P}}{(\mbox{det}_F^Q)^2\mbox{det}_B^{Q-P}},
\end{align}
and the anomalous fermion self-energy flow is given by
\begin{align}
 \nonumber \dot{\bar{G}}_{\psi_1\psi_2}^{-1}(P) &=A_\psi (h^2 \rho)^{3/2}U''\Biggl( \int_Q \frac{1}{A_\phi} \frac{L_\phi^Q \dot{\bar{R}}_\phi^{-Q}+L_\phi^{-Q} \dot{\bar{R}}_\phi^Q}{(\mbox{det}_B^Q)^2\mbox{det}_F^{Q+P}}\\
 \label{appB22} &\mbox{ }+\int_Q \frac{1}{A_\psi} \frac{L_\psi^Q \dot{\bar{R}}_\psi^{-Q}+L_\psi^{-Q}\dot{\bar{R}}_\psi^Q}{(\mbox{det}_F^Q)^2\mbox{det}_B^{Q+P}}\Biggr).
\end{align}
The flow of the Feshbach coupling $\bar{h}$ is given by
\begin{align}
 \nonumber \dot{\bar{h}} &= -A_\psi A_\phi^{1/2} h^3 U''\rho\Biggl( \int_Q \frac{1}{A_\phi} \frac{L_\phi^Q \dot{\bar{R}}_\phi^{-Q}+L_\phi^{-Q} \dot{\bar{R}}_\phi^Q}{(\mbox{det}_B^Q)^2\mbox{det}_F^{Q}}\\
 \label{appBfesh} &\mbox{ }+\int_Q \frac{1}{A_\psi} \frac{L_\psi^Q \dot{\bar{R}}_\psi^{-Q}+L_\psi^{-Q}\dot{\bar{R}}_\psi^Q}{(\mbox{det}_F^Q)^2\mbox{det}_B^{Q}}\Biggr).
\end{align}

For the flow of the inverse boson propagator we turn to the real $(\bphi_1,\bphi_2)$-basis introduced in Eq. (\ref{appA8}). The advantage of this choice is explained in Appendix \ref{appC}. We have
\begin{align}
\nonumber &\dot{\bar{G}}_{\phi_2\phi_2}^{-1}(P) =  - A_\phi U''\int_Q  \frac{-X_\phi^Q+Y_\phi^Q+Y_\phi^{-Q}}{(\mbox{det}_B^Q)^2} \\
 \nonumber &\mbox{ }+A_\phi 2\rho (U'')^2 \int_Q\frac{2\rho U'' X_\phi^Q+L_\phi^{-(Q+P)}Y_\phi^Q+L_\phi^{Q+P}Y_\phi^{-Q}}{(\mbox{det}_B^Q)^2\mbox{det}_B^{Q+P}}\\
 \label{appB23} &\mbox{ }+A_\phi h^2 \int_Q\frac{2(h^2\rho)^{1/2}X_\psi^Q+L_\psi^{-(Q+P)}Y_\psi^Q +L_\psi^{Q+P}Y_\psi^{-Q}}{(\mbox{det}_F^Q)^2\mbox{det}_F^{Q+P}}
\end{align}
for the 22-component. The tadpole diagram in the first line is momentum independent due to our choice of momentum independent vertices. For the 12-component we find
\begin{widetext}
\begin{align}
 \nonumber \dot{\bar{G}}_{\phi_1\phi_2}^{-1}(P) &=  2 \rmi A_\phi \rho  U'' \int_Q \frac{1}{(\mbox{det}_B^Q)^2\mbox{det}_B^{Q+P}}\Biggl[-L_\phi^{Q+P}\Bigl((2U''+\rho  U^{(3)})X_\phi^Q+(U''+\rho  U^{(3)})Y_\phi^{-Q}\Bigr)\\
 \nonumber &\mbox{ }+L_\phi^{-(Q+P)}\Bigl((2U''+\rho  U^{(3)})X_\phi^Q+(U''+\rho  U^{(3)})Y_\phi^Q\Bigr)-\rho U''(2U''+\rho U^{(3)})\Bigl(Y_\phi^Q-Y_\phi^{-Q}\Bigr) \Biggr]\\
 \label{appB24} &\mbox{ }-\rmi A_\phi h^2 \int_Q \frac{Y_\psi^{-Q}L_\psi^{Q+P}-Y_\psi^QL_\psi^{-(Q+P)}}{(\mbox{det}_F^Q)^2\mbox{det}_F^{Q+P}}.
\end{align}
\end{widetext}

\section{Projection of running couplings}\label{appC}
Given the constitutive or master equations from Appendix \ref{appB}, the flow of running couplings is obtained through suitable projection prescriptions. Typically, there are several candidates for these projections, which superficially seem equivalent within a truncation, but result in distinct flow equations. The reason for this ambiguity is that the flow equation incorporates all terms in the full effective average action, in particular all higher order terms. Therefore, when specifying a particular projection procedure, we always neglect certain higher order couplings in a particular way.

The dependence of the running of couplings on the projection can be used to estimate the accuracy of a given truncation. Within a truncation which includes the most important effects, the precise projection should only result in minor modifications of observables. A strong dependence, however, signals a shortcoming of a particular truncation. 

We employ the flow equation for the effective potential in Eq. (\ref{appB14}) to project onto the coefficients in an expansion
\begin{align}
 \label{appC1} U_k(\rho) = m^2_\phi (\rho-\rho_0) + \frac{\lambda_\phi}{2}(\rho-\rho_0)^2 + \sum_{n\geq 2} \frac{u_n}{n!}(\rho-\rho_0)^n.
\end{align}
 In the symmetric regime of the flow, where $\bar{\rho}_{0,k},\dot{\bar{\rho}}_{0,k}=0$, we have $\dot{\bar{m}}_\phi^2 = \dot{\bar{U}}'(0)$, whereas this flow equation is replaced by $ \dot{\bar{\rho}}_0 = - \dot{\bar{U}}'(\bar{\rho}_0)/\bar{\lambda}_\phi$ in the ordered regime. (We write $u_2=\lambda_\phi$.) We have $\bar{u}_n = \bar{U}^{(n)}_k(\bar{\rho}_{0,k}).$ Accordingly, for $n\geq 2$,
\begin{align}
 \label{appC2} \dot{\bar{u}}_n = \partial_t \Bigl( \bar{U}^{(n)}_k(\bar{\rho}_{0,k}) \Bigr) = \dot{\bar{U}}^{(n)}_k(\bar{\rho}_0) + \bar{u}_{n+1} \dot{\bar{\rho}}_0.
\end{align}
As explained in the main text, the second term is important to obtain quantitative precision of the results. The flow of the renormalized couplings
\begin{align}
 \label{appC3} m^2_\phi = \frac{\bar{m}^2_\phi}{A_\phi}, \ \rho_0 = A_\phi \bar{\rho}_0,\ u_n = \frac{\bar{u}_n}{A_\phi^n}
\end{align}
is given by
\begin{align}
 \label{appC4} \dot{m}^2_\phi &= \eta_\phi m^2_\phi +\frac{\dot{\bar{m}}^2_\phi}{A_\phi},\ \dot{\rho}_0 = -\eta_\phi \rho_0 + A_\phi \dot{\bar{\rho}}_0,\\
 \label{appC5} \dot{u}_n &= n \eta_\phi u_n + \frac{\dot{\bar{u}}_n}{A_\phi^n}.
\end{align}

The running couplings entering the fermion propagator are projected according to
\begin{align}
 \label{appC6} \dot{\bar{m}}^2_\psi &= \dot{\bar{G}}_{\psi_1^*\psi_1}^{-1}(P)\Bigr|_{P=0,\rho_0},\\
 \label{appC7} \dot{\bar{S}}_\psi &= \frac{1}{\rmi} \frac{\partial}{\partial p_0} \dot{\bar{G}}_{\psi_1^*\psi_1}^{-1}(P)\Bigr|_{P=0,\rho_0},\\
 \label{appC8}\dot{A}_\psi &= \frac{\partial}{\partial p^2} \dot{\bar{G}}_{\psi_1^*\psi_1}^{-1}(P)\Bigr|_{P=0,\rho_0} = \frac{1}{2} \frac{\partial^2}{\partial p^2} \dot{\bar{G}}_{\psi_1^*\psi_1}^{-1}(P)\Bigr|_{P=0,\rho_0}.
\end{align}
Similar to the expansion coefficients of the effective potential, one could first start from $\bar{m}^2_\psi=\bar{G}^{-1}_{\psi_1^*\psi_1}(P=0,\bar{\rho}_0)$ and then take the $t$-derivative of this expression. In Addition to Eq. (\ref{appC6}), this generates a term $\partial_{\bar{\rho}}\bar{G}^{-1}_{\psi_1^*\psi_1}(0,\bar{\rho}_0) \dot{\bar{\rho}}_0$ in the ordered regime. However, we will neglect such contributions in this work. The fermion anomalous dimension is defined by $\eta_\psi = -\dot{A}_\psi/A_\psi$, and the flow of the renormalized couplings $m^2_\psi=\bar{m}^2_\psi/A_\psi$, $S_\psi=\bar{S}_\psi/A_\psi$ reads
\begin{align}
 \label{appC9}  \dot{m}^2_\psi = \eta_\psi m^2_\psi + \frac{\dot{\bar{m}}^2_\psi}{A_\psi},\ \dot{S}_\psi = \eta_\psi S_\psi + \frac{\dot{\bar{S}}_\psi}{A_\psi}.
\end{align}

For our parametrization of the boson dynamics we employ
\begin{align}
 \label{appC10} \dot{\bar{S}}_\phi &= -\frac{\partial}{\partial p_0} \dot{\bar{G}}_{\phi_1\phi_2}^{-1}(P)\Bigr|_{P=0,\rho_0},\\
 \label{appC11} \dot{\bar{V}}_\phi &= \frac{\partial}{\partial p_0^2}  \dot{\bar{G}}_{\phi_2\phi_2}^{-1}(P)\Bigr|_{P=0} = \frac{1}{2}\frac{\partial^2}{\partial p_0^2}  \dot{\bar{G}}_{\phi_2\phi_2}^{-1}(P)\Bigr|_{P=0,\rho_0},\\
 \label{appC12} \dot{A}_\phi &= 2 \frac{\partial}{\partial p^2}   \dot{\bar{G}}_{\phi_2\phi_2}^{-1}(P)\Bigr|_{P=0,\rho_0} = \frac{\partial^2}{\partial p^2}   \dot{\bar{G}}_{\phi_2\phi_2}^{-1}(P)\Bigr|_{P=0,\rho_0}.
\end{align}
The boson anomalous dimension is again given by $\eta_\phi=-\dot{A}_\phi/A_\phi$, and the renormalized couplings $S_\phi=\bar{S}_\phi/A_\phi$ and $V_\phi=\bar{V}_\phi/A_\phi$ flow according to
\begin{align}
 \label{appC13} \dot{S}_\phi = \eta_\phi S_\phi + \frac{\dot{\bar{S}}_\phi}{A_\phi},\ \dot{V}_\phi = \eta_\phi V_\phi + \frac{\dot{\bar{V}}_\phi}{A_\phi}.
\end{align}

The reason for choosing the $(\bphi_1,\bphi_2)$-basis to project onto the boson coefficients consists in the following. To be consistent with our truncation of $\bar{\rho}$-independent couplings $Z\in\{\bar{S}_\phi, \bar{V}_\phi, A_\phi\}$, we have to project them such that terms which arise from $Z'(\bar{\rho})$ are absent. In fact, if we would include  the latter, we should also incorporate momentum dependent vertices which are proportional to $Z'(\bar{\rho})$. Now, if we start from the more general ansatz
\begin{align}
 \label{appC14} \bar{\Gamma}_{\rm kin}[\bphi] = \int_X \frac{1}{2} Z(\bar{\rho}) \Bigl(\bphi^* P_\phi(\partial_\tau,-\rmi \nabla) \bphi + \bphi P_\phi(-\partial_\tau,\rmi \nabla)\bphi^*\Bigr)
\end{align}
for the kinetic term of the bosons, where $P_\phi(\pm\partial_\tau,\mp\rmi \nabla)\to P_\phi(\pm Q) = \pm \rmi q_0, q_0^2, q^2$, we obtain from a second functional derivative
\begin{align}
 \nonumber &\bar{G}_\phi^{-1,\{\phi,\phi^*\}}(Q,\bar{\rho}) \\
 \label{appC15} &= \begin{pmatrix} \bar{\rho} Z'(\bar{\rho}) P_\phi^S(Q) & \Bigl(Z(\bar{\rho})+\bar{\rho} Z'(\bar{\rho})\Bigr) P_\phi(-Q)\\ \Bigl(Z(\bar{\rho})+\bar{\rho} Z'(\bar{\rho})\Bigr) P_\phi(Q) & \bar{\rho} Z'(\bar{\rho}) P_\phi^S(Q)\end{pmatrix}
\end{align}
in the $(\bphi,\bphi^*)$-basis, and
\begin{align}
 \nonumber & \bar{G}_\phi^{-1,\{\phi_1,\phi_2\}}(Q,\bar{\rho}) \\
 \label{appC16} &= \begin{pmatrix} \Bigl(Z(\bar{\rho})+2\bar{\rho} Z'(\bar{\rho})\Bigr)P_\phi^S(Q) & \rmi\Bigl(Z(\bar{\rho})+\bar{\rho} Z'(\bar{\rho})\Bigr)P_\phi^A(Q) \\ -\rmi\Bigl(Z(\bar{\rho})+\bar{\rho} Z'(\bar{\rho})\Bigr)P_\phi^A(Q) & Z(\bar{\rho}) P_\phi^S(Q) \end{pmatrix}
\end{align}
in the $(\bphi_1,\bphi_2)$-basis. Herein,
\begin{align}
 \label{appC17} P^{S,A}_\phi(Q) = \frac{1}{2}\Bigl( P_\phi(Q) \pm P_\phi(-Q)\Bigr)
\end{align}
are the symmetrized and anti-symmetrized kinetic terms, respectively. Since the terms proportional to $Z'(\bar{\rho})$ are included in the full flow equation (\ref{appA1}), they appear on the right hand side of the flow equation for $\bar{\Gamma}^{(2)}_k\sim \bar{G}^{-1}_\phi$. To avoid their influence on the beta functions, we project the coefficients $\bar{V}_\phi$, $A_\phi$ of the even functions $P^S_\phi(Q)=q_0^2, q^2$ from the 22-component of Eq. (\ref{appC16}). Equivalently, as can be seen from Eq. (\ref{appC15}), we may define them from the symmetrized part of $\dot{\bar{G}}_{\phi^*\phi}$ and subtract $\dot{\bar{G}}_{\phi\phi}$. The situation is less simple for $\bar{S}_\phi$, which appears as the coefficient of $P^A_\phi(Q)=\rmi q_0$. No projection is preferred in this case, and we choose the 12-component. Since the overall impact of $\bar{S}_\phi$ on the flow is rather small, this results in negligible error.

The regulators in this work are chosen such that $\bar{R}_\phi^Q$ depends on $A_\phi$, and $R_\psi^Q$ is both $A_\psi$- and $m^2_\psi$-dependent. This dependence is necessary to account for the right scaling of correlation functions and to regularize around a flowing Fermi surface. Moreover, it provides for an efficient resummation of flow equations. 

If we incorporate the flow of $A_\psi$ and $m^2_\psi$, the fermionic regulator insertion $\dot{\bar{R}}_\psi/A_\psi$ depends linearly on both $\eta_\psi$ and $\dot{m}^2_\psi$. Hence, we arrive at a set of equations
\begin{align}
 \label{appC20} \begin{pmatrix} \eta_\phi \\ \dot{m}^2_\psi \\ \eta_\psi \end{pmatrix} = \begin{pmatrix} A_0 \\ B_0 \\ C_0 \end{pmatrix} + \begin{pmatrix} A_1 & A_2 & A_3 \\ B_1 & B_2 & B_3 \\ C_1 & C_2 & C_3 \end{pmatrix} \begin{pmatrix} \eta_\phi \\ \dot{m}^2_\psi \\ \eta_\psi \end{pmatrix},
\end{align}
where the coefficients are one-loop integrals which depend on the remaining running couplings. The linear set is solved by
\begin{align}
 \label{appC21} \begin{pmatrix} \eta_\phi \\ \dot{m}^2_\psi \\ \eta_\psi \end{pmatrix} = \begin{pmatrix} 1-A_1 & -A_2 & -A_3 \\ -B_1 & 1-B_2 & -B_3 \\ -C_1 & -C_2 & 1-C_3 \end{pmatrix}^{-1}\begin{pmatrix} A_0 \\ B_0 \\ C_0 \end{pmatrix}.
\end{align}
The need for evaluating more one-loop integrals numerically makes an inclusion of even more couplings into $\bar{R}_\phi$ and $\bar{R}_\psi$ (e.g. $S_\phi, S_\psi, V_\psi$) less attractive, although promising a further improved resummation of diagrams.

\section{Initial conditions}\label{AppIni}
As explained in the main text, the initial values of the running couplings for the Unitary Fermi Gas are fixed from the zeros $\tilde{g}_\star$ of the beta functions for the dimensionless running couplings in vacuum ($\mu=T=0)$. We have
\begin{align}
 \dot{\tilde{g}}(\mu=0) = \beta_{\tilde{g}}(\tilde{\mu}=0) =0
\end{align}
with $\tilde{\mu}=\mu/k^2$. If we do not allow for a running fermion mass (F, FB$_0$, and FB-truncations), we have $\tilde{m}^2_{\psi\star}=0$.

Upon introducing a chemical potential,
\begin{align}
 m^2_{\psi\Lambda}(\mu) = m^2_{\psi\star}-\mu,
\end{align}
we slightly deviate from this fixed point, and expect deviations from the scaling solution as soon as $k^2\simeq \mu$. (We assume $T=0$ throughout the following.)  In terms of beta functions, the UV flow (large $k$) is governed by the set of equations
\begin{align}
 \nonumber \dot{\tilde{g}}(\mu) &= \beta_{\tilde{g}}(\tilde{\mu}) \simeq \frac{\partial \beta_{\tilde{g}}}{\partial \tilde{\mu}}(0) \cdot\tilde{\mu} +\dots \\
 &= \sum_i \frac{\partial \beta_{\tilde{g}}}{\partial \tilde{g}_i}(0) \cdot \tilde{g}_{i,\mu} \cdot \tilde{\mu}+\dots
\end{align}
We used that the leading term vanishes and introduced the notation
\begin{align}
 g_\mu = \partial_\mu g.
\end{align}
The dimensionless running couplings $\tilde{g}_\mu$ are then found in the UV from the equations
\begin{align}
 \nonumber \dot{\tilde{g}}_\mu &= \partial_t\Bigl(\frac{\partial \tilde{g}}{\partial \tilde{\mu}}\Bigr) = 2\tilde{g}_\mu+\sum_i\frac{\partial \beta_{\tilde{g}}}{\partial \tilde{g}_i}(0) \cdot \tilde{g}_{i,\mu}. \\
 \label{EQ1} & \Rightarrow \dot{\tilde{g}}_{\mu,i} = A_{ij} \cdot \tilde{g}_{\mu,j}.
\end{align}
The matrix
\begin{align}
A=2\cdot\textbf{1}+\{\partial \beta/\partial \tilde{g}\}
\end{align} 
is solely determined by the fixed point values $\tilde{g}_\star$. We will be interested in a situation where 
\begin{align}
 \tilde{g}_{\mu} \simeq \tilde{g}_{\mu,\star} =\text{const.}\ \text{ in the UV.}
\end{align}
Most importantly, we have to ensure
\begin{align}
 \alpha_\psi = \tilde{m}^2_{\psi\mu} = \frac{\partial m^2_\psi}{\partial \mu} \simeq-1 \ \text{ in the UV},
\end{align}
because otherwise the value of the chemical potential loses its meaning due to an anomalous running in the UV. 

The behavior of the set of equations (\ref{EQ1}) depends on whether we allow for a running fermion mass or not. If we allow for a running fermion mass, then $\alpha_\psi$ is a free parameter in (\ref{EQ1}), and, consequently, $\dot{\tilde{g}}_\mu=A\cdot\tilde{g}_\mu=0$ will in general not have a nontrivial solution, as $A$ has full rank: We have a homogeneous set of equations, with the only solution $\tilde{g}_{\mu,\star}=0$. This conflicts $\alpha_\psi=-1$. In contrast, if we do not allow for a running fermion mass, thereby enforcing $\alpha_\psi=-1$ by hand, we find a scaling solution with fixed point values $\tilde{g}_{\mu,\star}$. The UV flow is then very simple:
\begin{align}
 \tilde{g}_k(\mu) &\simeq \tilde{g}_\star + \tilde{g}_{\mu\star}\cdot\mu/k^2.
\end{align}
In particular, $\tilde{m}^2_{\psi} \simeq  -\tilde{\mu}$ for large $k$.

In order to have $\alpha_\psi=-1$ in the UV, we need $A$ not to be of full rank, i.e. $A$ must have an eigenvalue 0. Then, the variables in $A\cdot \tilde{g}_\mu=0$ are not independent, and there are nontrivial solutions. The matrix $A$ is fully determined by the vacuum scaling values $\tilde{g}_\star$. The values of $\tilde{g}_\star$, however, depend on the regularization procedure. With the relative cutoff scale $c_\phi$ we have a knob to tune the behavior of the UV running of the couplings. In particular, we find that for $c_\phi=c_0=0.2454\simeq 1/4$ we have a zero eigenvalue of $A$, and $m^2_\psi-m^2_{\psi,\star}\simeq -\mu$ in the UV.

\section{Estimated effective chemical potential in FBM-truncations}\label{AppEstimate}

In the presence of the mixed diagrams, the nontrivial running of the fermion propagator in the symmetric regime spoils the interpretation of $\bar{\mu}$ as the chemical potential.  Due to the presence of the relevant initial perturbation $\Delta m^2_{\psi\Lambda}=m^2_{\psi\Lambda}-C\Lambda^2=-\bar{\mu}$, all beta functions of the remaining running couplings scale linear in $\Delta \tilde{m}^2_{\psi k}=\Delta m^2_{\psi k}/k^2$ for large $k$. However, the latter acquires an (unphysical) anomalous running due to the presence of the regulator function. We have
\begin{align}
 \partial_t \tilde{m}^2_\psi(\mu_\Lambda,c\phi) = 0 - \Bigl(2+\eta_\mu(c_\phi)\Bigr) \Bigl(-\frac{\bar{\mu}}{k^2}\Bigr)+ \dots
\end{align}
for large $k$. The leading term vanishes due to the fixing of the initial conditions at the vacuum fixed point, where $\partial_t \tilde{m}^2_\psi(\mu=0) =0$. The anomalous dimension $\eta_\mu$ of the term linear in $-\bar{\mu}/k^2$ leads to a nontrivial running of the supposed ``chemical potential'' with $k$. Consequently, $\bar{\mu}\neq \mu$.  For $c_\phi=c_0$ we have $\eta_\mu(c_0)=0$, such that the problem does not arise in this case. The value of $\eta_\mu$ can be extracted from the flow equation of the fermion mass term at $t=0$. It is given by $\eta_\mu(1)=0.21892$ for $c_\phi=1$, decreasing for smaller values of $c_\phi$. The value of $\eta_\mu$ also depends on the truncation and receives small corrections in a $\phi^8$-truncation.

To estimate the influence of the anomalous running induced by $\eta_\mu$, we can subtract the corresponding contribution to the flow equation by hand. This is not a self-consistent procedure and serves here only for an error estimate. For this purpose, we replace the flow equation for the running fermion mass term according to
\begin{align}
 \label{EQetamu} \partial_t \tilde{m}^2_\psi(\bar{\mu},c_\phi) \to \partial_t \tilde{m}^2_\psi(\bar{\mu},c_\phi) - \eta_\mu(c_\phi) f\Bigl(\frac{\bar{\mu}}{k^2}\Bigr),
\end{align}
where $f(x)$ is chosen such that $f(x)=x$ for $x\ll 1$ and $f(x)=0$ for $x \gg 1$. This choice of $f(x)$  removes the anomalous running for large $k$, whereas we leave the flow equation unchanged for small $k$ as soon as higher powers of $\bar{\mu}/k^2$ become relevant. The leading contribution is then given by these higher terms, which are well-behaved. We employ $f(x)=x/(e^{(x-1)/0.1}+1)$ for the following analysis, but the precise form of $f(x)$ is not important. In fact, also $f(x)=x$ gives almost the same values for $\Delta/\mu$ and $T_{\rm c}/\mu$ within the purpose of this error estimate.

The subtraction in Eq. (\ref{EQetamu}) could also be elaborated to a systematic renormalization of the field with anomalous dimension $\eta_\psi=\mathcal{O}(\bar{\mu}/k^2)$. Since the value of $\eta_\mu$ can be inferred from the flow at $t=0$ (or the flow of $\partial m^2_\psi/\partial \mu$ in vacuum), it is fixed a priori by the truncation and regularization scheme. 

With the $\eta_\mu$-subtraction (\ref{EQetamu}), we find that the observables $T_{\rm c}/\mu$ and $\Delta/\mu$ are indeed independent of $\bar{\mu}$ for all $c_\phi$. Moreover, the qualitative running of couplings compared to their vacuum values, $g_k-g_{\rm vac}=\mathcal{O}(\tilde{\mu})$, is similar to the well-understood case of $c_\phi=c_0$.

\begin{figure}[t]
\centering
\includegraphics[width=8cm]{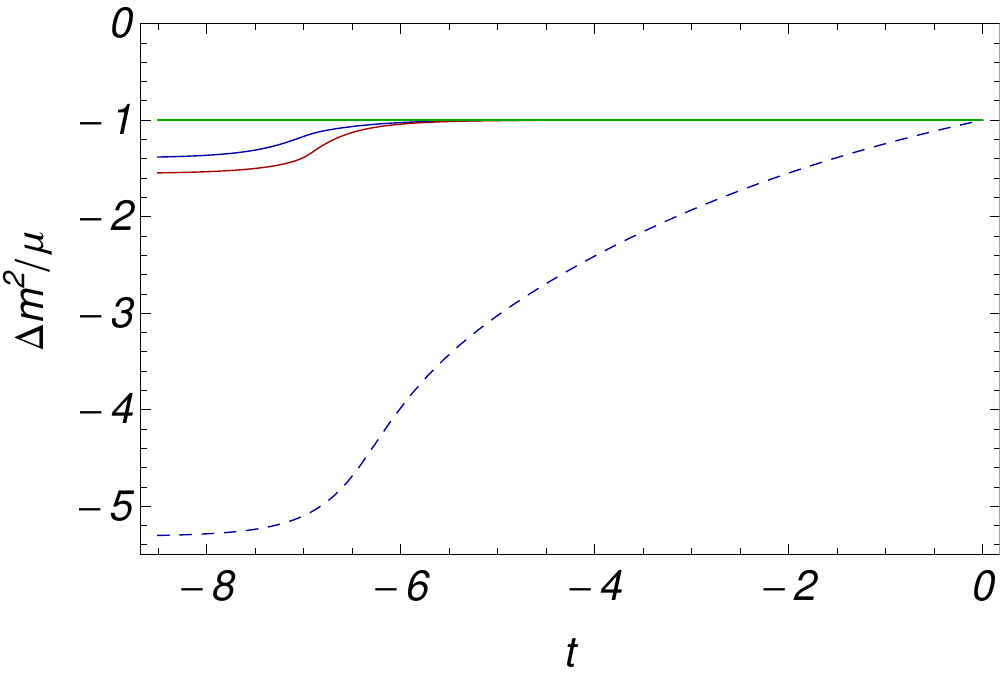}
\caption{Running of $\Delta m^2_\psi=m^2_\psi-m^2_{\psi,\rm vac}$ in units of the initial chemical potential $\mu$. The constant line (green) at the top corresponds to a F-, FB$_0$-, or FB-truncation, where $m^2_\psi$ does not flow, and hence $\Delta m^2_\psi=-\mu$ for all $k$. The remaining curves from top to bottom correspond to $c_\phi=1$ in the $\eta_\mu$-subtracted scheme of Eq. (\ref{EQetamu}) (blue, solid), $c_\phi=c_0$ (red), and $c_\phi=1$ without the $\eta_\mu$-subtraction (blue, dashed). Clearly, the anomalous running for large $k$ screens the physical effects in the latter case. For $c_\phi=c_0$ we see that renormalization effects on $\Delta m^2_\psi$ only show up at the many-body scales $k^2\simeq \mu$, and result in an effective chemical potential $\mu_{\rm eff}/\mu\simeq 1.5$.}
\label{plot_effective_mu}
\end{figure}

It is instructive to study the behavior of 
\begin{align}
 \Delta m^2_{\psi k} = m^2_{\psi k}-m^2_{\psi,\rm vac} = \mathcal{O}(\mu/k^2)
\end{align}
within the subtracted scheme just described. (Or, equivalently, for the physical point $c_\phi=c_0$ without the subtraction.) Herein, $m^2_{\psi,\rm vac}=\tilde{m}^2_{\psi\star}k^2$ only shows a canonical running fixed in vacuum. For large $k$ we have $\Delta m^2_{\psi,k} \simeq -\mu$. As $k$ is lowered towards the many-body scales given by $\mu$ and $T$, the absolute value of $|\Delta m^2_{\psi,k}|=\mu_{\rm eff}$ \emph{increases}, thereby yielding an effectively enhanced chemical potential appearing in the denominator of the fermionic propagator. We find the enhancement $\mu_{\rm eff}/\mu$ to be of order 50 percent. We show the behavior of $\Delta m^2_{\psi}$ in Fig. \ref{plot_effective_mu}.

Closing the $Q$-loop over $G_\psi(Q)=(\rmi q_0+q^2-\mu_{\rm eff})^{-1}$ yields a larger value $\sim \mu_{\rm eff}^{3/2}$ when compared to the loop over the microscopic propagator $(\rmi q_0+q^2-\mu)^{-1}$. As a result, the density of the system (which is related to this loop-integral) is enhanced. This increase of the density due to many-body effects is well-known for the Unitary Fermi Gas and can be attributed to the Tan effect. (See Ref. \cite{PhysRevA.87.023606} for a related FRG-study of the renormalization of the fermion propagator.) Hence we found evidence that the running of $m^2_\psi$ within our truncation correctly incorporates this effect on the density.

\section{Regularized Loop Integrals}\label{appD}
We consider the frequency and momentum regulator
\begin{align}
 \label{appD1} \bar{R}^Q = A (\rmi q_0 + \xi_q) r(Y),\ Y=\frac{q_0^2+\xi_q^2}{k^4},
\end{align}
where $\xi_q=q^2/2$ for bosons and $\xi_q=q^2+m^2_\psi$ for fermions. The $Q$-exp regulator employed in this works consists in the choice $ r(Y) = (e^Y-1)^{-1}$. All one-loop integrals entering the beta functions become UV finite due to the insertion of
\begin{align}
 \nonumber \dot{\bar{R}}^Q &= A (\rmi q_0 +\xi_q) \Bigl[-\eta r(Y) - 4 Y r'(Y)\Bigr] \\
 \label{appD2} &=: A (\rmi q_0 +\xi)\dot{r}(Y).
\end{align}
A typical diagram contributing to the flow has the form
\begin{align}
 \nonumber \int_Q \frac{\dot{\bar{R}}^Q}{(P^Q+R^Q)^n} &\propto \int_Q \frac{(\rmi q_0+\xi_q)\dot{r}(Y)}{(\rmi q_0+\xi_q)^n(1+r)^n} \\
 \label{appD3} &\sim \int_Q \frac{\dot{r}(Y)}{Y^{(n-1)/2}(1+r)^n}.
\end{align}
The integrand is cut off for a sufficiently large value of $Y$, such that only terms with $Y\leq N$ contribute, where $N$ depends on the particular choice of the regulator. Accordingly, we find that the 	frequency and momentum integrals can be restricted to the domains
\begin{align}
  \nonumber &Y = \hat{q}_0^2 + \hat{\xi}_q^2 \leq N \\
 \label{appD4}  &\Rightarrow  \begin{cases} |\hat{q}_0| \leq N^{1/2},\ |\hat{q}^2+\tilde{m}^2_\psi| \leq N^{1/2} & \text{(fermions)},\\ |\hat{q}_0| \leq N^{1/2},\ |\hat{q}| \leq 2^{1/2}N^{1/4} & \text{(bosons)}.\end{cases}
\end{align}
Herein, $\hat{q}_0=q_0/k^2$, $\tilde{\xi}_q=\xi_q/k^2$, $\hat{q}^2=q^2/k^2$, and $\tilde{m}^2_\psi=m^2_\psi/k^2$. For the $Q$-exp cutoff, we found that $N=25$ is a good choice. Enlarging $N$ allows for a check of the stability of the numerical integrations. For the fermions, the spatial condition translates to
\begin{align}
  \label{appD5} q \in [K_{\psi,\rm min},K_{\psi,\rm max}],
\end{align}
where
\begin{align}
 \label{appD6}  K_{\psi,\rm min} &= \begin{cases} \sqrt{-(N^{1/2}+\tilde{m}^2_\psi)} & \text{if } N^{1/2}+\tilde{m}^2_\psi<0 \\ 0 & \text{else} \end{cases},\\
 \label{appD7} K_{\psi,\rm max} &= \begin{cases} \sqrt{N^{1/2}-\tilde{m}^2_\psi} & \text{if }N^{1/2}-\tilde{m}^2_\psi>0 \\ K_{\psi,\rm min} & \text{else} \end{cases}.
\end{align}

\section{Finite temperature flow}\label{appE}
At nonzero temperatures $T>0$, the loop integration over frequencies turns into an infinite Matsubara sum over frequencies $\omega_n=2\pi (n+1/2)T$ and $\omega_n=2\pi n T$ for fermions and bosons, respectively. However, it is an interesting property of the renormalization group that, for large $k^2 \gg T$, the flow can be approximated by the zero temperature flow. In fact, the system at scale $k$ can only resolve the actual value of the temperature once $k^2$ is comparable to $T$. We detail here how this feature is reflected in the flow equations and how it can be implemented numerically.

Given a function $f(\hat{q}_0) \propto \dot{\bar{R}}(Q)$ which has finite support due to the frequency and momentum regulator $\bar{R}(Q)$, the Matsubara summation is restricted to a finite domain according to
\begin{align}
 \label{appE1} \tilde{T} \sum_{n-=\infty}^\infty f(\hat{\omega}_n) = \tilde{T} \sum_{n=-M}^M f(\hat{\omega}_n),
\end{align}
where the number $M$ depends on the choice of the regulator and the value of $\tilde{T}=T/k^2$. With the choice of $N$ from Eq. (\ref{appD4}), the number of bins which are summed in Eq. (\ref{appE1}) is given by
\begin{align}
 \label{appE2} M(\tilde{T}) \simeq \frac{N^{1/2}}{\Delta \hat{\omega}_n} = \frac{N^{1/2}}{2\pi \tilde{T}} = \frac{k^2}{T} \mathcal{O}(1).
\end{align}
In particular, for large $T$ or small $k^2$, we have
\begin{align}
 \label{appE3} \tilde{T} \sum_{n=-\infty}^\infty f(\hat{\omega}_n) \stackrel{\tilde{T}\gg 1}{\longrightarrow} \begin{cases} \tilde{T} f(0) & \text{(bosons)} \\ 0 & \text{(fermions)}\end{cases}.
\end{align}

Our strategy is as follows: We search for the lowest number of bins $M(\tilde{T})$ such that the area of the Matsubara sum is still well approximated by the zero temperature integral. This will be the case in the early stages of the flow. Once the rescaled temperature $\tilde{T}$ gets too high, we evaluate the sum. However, due to the finite support of $\dot{\bar{R}}(Q)$, this will  involve only a few terms. Eventually, for $k^2\to 0$, we only have to take into account the lowest modes, say $n=0,\pm 1$.

We define a transition temperature $\tilde{T}_{\rm tr}$ such that
\begin{align}
 \label{appE4} \tilde{T} \sum_n f(\hat{\omega}_0) = \begin{cases} \int_{-N^{1/2}}^{N^{1/2}} \frac{d\hat{q}_0}{2\pi} f(\hat{q}_0) & \tilde{T} \leq \tilde{T}_{\rm tr}\\ \tilde{T} \sum_{n=-M(\tilde{T})}^{M(\tilde{T})} f(\hat{\omega}_n) & \tilde{T} > \tilde{T}_{\rm tr}\end{cases}
\end{align}
during the evolution of the flow, where
\begin{align}
 \label{appE5} M(\tilde{T}) &= \Bigl[ \frac{N^{1/2}}{2\pi \tilde{T}}\Bigr] >0.
\end{align}
Herein, $[x]$ defines the ceiling function, which maps $x$ to the smallest integer not less than $x$. With this choice, we keep more terms than are actually necessary from the consideration in Eq. (\ref{appE3}). For the $Q$-exp regulator, $\tilde{T}_{\rm tr}=0.01$ is a good choice. Thus, we switch from an integration over continuous frequencies to a sum over finite frequencies once the number of Matsubara frequencies has decreased to $M(\tilde{T}_{\rm tr}) =80$. By decreasing $\tilde{T}_{\rm tr}$, we can check for the stability of our numerical computations.

\section{Mean field analysis}\label{appF}
We derive the mean field results for the BCS-BEC crossover. The formulas serve for a comparison of the FRG computations in the mean field limit, where only F-diagrams are taken into account.

From a saddle-point expansion of the effective action we obtain the effective potential in mean field approximation
\begin{align}
 \label{appF1} U(\Delta^2,\mu,T) = -\frac{\Delta^2}{\lambda_{\psi,\Lambda}} -\int_Q^\Lambda \log\Bigl(q_0^2+(q^2-\mu)^2+\Delta^2\Bigr).
\end{align}
The UV-divergent integral is regularized by means of a sharp momentum cutoff enforcing $q^2\leq \Lambda^2$. We impose the vacuum renormalization condition
\begin{align}
  \label{appF2} - \frac{1}{\lambda_\psi} \stackrel{!}{=} \frac{\partial U}{\partial \Delta^2}(0,0,0) = -\frac{1}{\lambda_{\psi,\Lambda}}-\frac{1}{2} \int^\Lambda \frac{\mbox{d}^3q}{(2\pi)^3} \frac{1}{q^2},
\end{align}
where $\lambda_\psi = 4 \pi \hbar^2 a/M_{\psi}=8 \pi a$ is related to the fermion scattering length. The renormalized gap equation at zero temperature  reads
\begin{align}
 \nonumber 0 &= \frac{\partial U}{\partial \Delta^2}(\Delta_0^2,\mu,0) \\
 \nonumber &=  -\frac{1}{\lambda_\psi} -\int_Q^\Lambda \frac{1}{q_0^2+(q^2-\mu)^2+\Delta_0^2}+\frac{1}{2} \int^\Lambda \frac{\mbox{d}^3q}{(2\pi)^3} \frac{1}{q^2}\\
 \label{appF3} &= - \frac{1}{\lambda_\psi} - \frac{1}{4\pi^2} \int_0^\infty \mbox{d}q \Bigl( \frac{q^2}{\sqrt{(q^2-\mu)^2+\Delta_0^2}}-1\Bigr).
\end{align}
The integral is UV finite and we can send $\Lambda\to \infty$.

For $\mu >0$, we can rewrite the gap equation (\ref{appF3}) in the dimensionless form
\begin{align}
 \label{appF7} \frac{\pi}{2\mu^{1/2}a} = f\Bigl(\frac{\Delta_0}{\mu}\Bigr)
\end{align}
with
\begin{align}
 \label{appF8} f(y) = - \int_0^\infty \mbox{d} x \Bigl(\frac{x^2}{\sqrt{(x^2-1)^2+y^2}}-1\Bigr).
\end{align}
The BCS-formula for an exponentially small gap is obtained from $f(y)\to \log(e^2y/8)$ for $y\to 0$. We then find
\begin{align}
 \label{appF9} \frac{\Delta_{\rm BCS}}{\mu} = \frac{8}{e^2} \exp\Bigl(\frac{\pi}{2\mu^{1/2}a}\Bigr).
\end{align}
However, this asymptotic formula cannot be applied to the unitary Fermi gas with $a^{-1}=0$. In this case, the BCS-formula results in $\Delta_0/\mu=1.083$, whereas the correct solution to the gap equation (\ref{appF7}) is given by
\begin{align}
 \label{appF10} \Delta_0/\mu = 1.162 \text{ for } a^{-1}=0.
\end{align}

The gap vanishes at the critical temperature and thus we have
\begin{align}
 \nonumber 0 &= \frac{\partial U}{\partial \Delta^2}(0,\mu,T_{\rm c}) = - \frac{1}{\lambda_\psi}\\
 \label{appF11} &\mbox{ } - \frac{1}{2\pi^2} \int_0^\infty \mbox{d}q \Biggl[ \frac{q^2}{|q^2-\mu|}\Bigl(\frac{1}{2}-\frac{1}{e^{|q^2-\mu|/T_{\rm c}}+1}\Bigr)-\frac{1}{2}\Bigr)\Biggr].
\end{align}
This can be cast into the form
\begin{align}
 \label{appF12} \frac{\pi}{2\mu^{1/2}a} = g\Bigl(\frac{T_{\rm c}}{\mu}\Bigr),
\end{align}
where
\begin{align}
 \label{appF13} g(y) = - \int_0^\infty \mbox{d} x \Biggl[\frac{x^2}{|x^2-1|}\Bigl(1-\frac{2}{e^{|x^2-1|/y}+1}\Bigr)-1\Biggr].
\end{align}
For small critical temperatures we can apply $g(y)\to \log(\pi e^2 y/8 e^\gamma)$ for $y\to 0$ to obtain the BCS-formula
\begin{align}
 \label{appF14} \frac{T_{\rm c,BCS}}{\mu} = \frac{8 e^\gamma}{\pi e^2} \exp\Bigl(\frac{\pi}{2\mu^{1/2}a}\Bigr).
\end{align} 
An extrapolation of this asymptotic formula to the unitary point yields $T_{\rm c}/\mu=0.6138$. In contrast, the solution to Eq. (\ref{appF12}) is given by
\begin{align}
 \label{appF15} T_{\rm c}/\mu = 0.6646 \text{ for } a^{-1}=0.
\end{align}
The mean field prediction for the ratio $\Delta/T_{\rm c}$ for the Unitary Fermi gas is thus given by $\Delta/T_{\rm c} = 1.75.$

In an FRG treatment, no precondensation appears on the mean field level. Therefore, the symmetric phase is accessible from the symmetric regime of the flow, where $\rho_{0,k}=0$. In particular, the flow equation for the effective potential is given by the F-diagram with microscopic fermion propagator $\rmi q_0+q^2-\mu$. We then find in the symmetric regime
\begin{align}
  \label{appF16} \dot{\bar{U}}_k'(0)&= \bar{h}^2 \int_Q \frac{L_\psi^Q\dot{R}_\psi^{-Q}+L_\psi^{-Q}\dot{R}_\psi^Q}{(L_\psi^Q L_\psi^{-Q})^2}= -\partial_t \bar{h}^2_\Lambda \int_Q \frac{1}{L_\psi^Q L_\psi^{-Q}},
\end{align} 
since $\bar{h}^2_k=\bar{h}^2_\Lambda$ and the only $k$-dependence of $L_\psi^Q = \rmi q_0 + q^2 -\mu + R_\psi^Q$ arises from the regulator. Eq. (\ref{appF16}) can readily be integrated to yield
\begin{align}
 \label{appF17} &\frac{1}{\bar{h}^2_\Lambda}\Bigl[\bar{U}'(0,\mu,T) - \bar{U}_\Lambda'(0) \Bigr]\\
 &= -\int_Q \Biggl(\frac{1}{q_0^2+(q^2-\mu)^2}-\frac{1}{|\rmi q_0 + q^2-\mu+R_{\psi,\Lambda}(Q)|^2}\Biggr).
\end{align} 
We used that $R_{\psi,k=0}(Q)=0$. The regularization of the UV divergent integral is performed by means of the regulator $R_{\psi,\Lambda}(Q)$, which vanishes for $q_0,q^2 \geq \Lambda^2$ and thus gives finite support to the integration. We have
\begin{align}
  \bar{U}_\Lambda'(0) = \bar{m}^2_{\phi,\Lambda} = - \frac{\bar{h}^2_\Lambda}{\lambda_{\psi,\Lambda}}.
\end{align}
Accordingly, we reproduce the gap equation (\ref{appF11}) for $T=T_{\rm c}$ and $\bar{U}'(0,\mu,T_{\rm c})=0$. Therefore, the critical temperature found from the flow equation in the mean field limit trivially coincides with the standard mean field result.

In order to reproduce the gap equation (\ref{appF3}) we have to include higher orders terms $u_n$ ($n\geq3$) in the effective potential. This is outlined in more detail in the main text.


\bibliographystyle{apsrev4-1}
\bibliography{references_4d_regulators} 

\begin{thebibliography}{88}%
\makeatletter
\providecommand \@ifxundefined [1]{%
 \@ifx{#1\undefined}
}%
\providecommand \@ifnum [1]{%
 \ifnum #1\expandafter \@firstoftwo
 \else \expandafter \@secondoftwo
 \fi
}%
\providecommand \@ifx [1]{%
 \ifx #1\expandafter \@firstoftwo
 \else \expandafter \@secondoftwo
 \fi
}%
\providecommand \natexlab [1]{#1}%
\providecommand \enquote  [1]{``#1''}%
\providecommand \bibnamefont  [1]{#1}%
\providecommand \bibfnamefont [1]{#1}%
\providecommand \citenamefont [1]{#1}%
\providecommand \href@noop [0]{\@secondoftwo}%
\providecommand \href [0]{\begingroup \@sanitize@url \@href}%
\providecommand \@href[1]{\@@startlink{#1}\@@href}%
\providecommand \@@href[1]{\endgroup#1\@@endlink}%
\providecommand \@sanitize@url [0]{\catcode `\\12\catcode `\$12\catcode
  `\&12\catcode `\#12\catcode `\^12\catcode `\_12\catcode `\%12\relax}%
\providecommand \@@startlink[1]{}%
\providecommand \@@endlink[0]{}%
\providecommand \url  [0]{\begingroup\@sanitize@url \@url }%
\providecommand \@url [1]{\endgroup\@href {#1}{\urlprefix }}%
\providecommand \urlprefix  [0]{URL }%
\providecommand \Eprint [0]{\href }%
\providecommand \doibase [0]{http://dx.doi.org/}%
\providecommand \selectlanguage [0]{\@gobble}%
\providecommand \bibinfo  [0]{\@secondoftwo}%
\providecommand \bibfield  [0]{\@secondoftwo}%
\providecommand \translation [1]{[#1]}%
\providecommand \BibitemOpen [0]{}%
\providecommand \bibitemStop [0]{}%
\providecommand \bibitemNoStop [0]{.\EOS\space}%
\providecommand \EOS [0]{\spacefactor3000\relax}%
\providecommand \BibitemShut  [1]{\csname bibitem#1\endcsname}%
\let\auto@bib@innerbib\@empty
\bibitem [{\citenamefont {Horikoshi}\ \emph {et~al.}(2010)\citenamefont
  {Horikoshi}, \citenamefont {Nakajima}, \citenamefont {Ueda},\ and\
  \citenamefont {Mukaiyama}}]{Horikoshi22012010}%
  \BibitemOpen
  \bibfield  {author} {\bibinfo {author} {\bibfnamefont {M.}~\bibnamefont
  {Horikoshi}}, \bibinfo {author} {\bibfnamefont {S.}~\bibnamefont {Nakajima}},
  \bibinfo {author} {\bibfnamefont {M.}~\bibnamefont {Ueda}}, \ and\ \bibinfo
  {author} {\bibfnamefont {T.}~\bibnamefont {Mukaiyama}},\ }\href {\doibase
  10.1126/science.1183012} {\bibfield  {journal} {\bibinfo  {journal}
  {Science}\ }\textbf {\bibinfo {volume} {327}},\ \bibinfo {pages} {442}
  (\bibinfo {year} {2010})}\BibitemShut {NoStop}%
\bibitem [{\citenamefont {Nascimbene}\ \emph {et~al.}(2010)\citenamefont
  {Nascimbene}, \citenamefont {Navon}, \citenamefont {Jiang}, \citenamefont
  {Chevy},\ and\ \citenamefont {Salomon}}]{Salomon}%
  \BibitemOpen
  \bibfield  {author} {\bibinfo {author} {\bibfnamefont {S.}~\bibnamefont
  {Nascimbene}}, \bibinfo {author} {\bibfnamefont {N.}~\bibnamefont {Navon}},
  \bibinfo {author} {\bibfnamefont {K.~J.}\ \bibnamefont {Jiang}}, \bibinfo
  {author} {\bibfnamefont {F.}~\bibnamefont {Chevy}}, \ and\ \bibinfo {author}
  {\bibfnamefont {C.}~\bibnamefont {Salomon}},\ }\href {\doibase
  10.1038/nature08814} {\bibfield  {journal} {\bibinfo  {journal} {Nature}\
  }\textbf {\bibinfo {volume} {463}},\ \bibinfo {pages} {1057} (\bibinfo {year}
  {2010})}\BibitemShut {NoStop}%
\bibitem [{\citenamefont {Navon}\ \emph {et~al.}(2010)\citenamefont {Navon},
  \citenamefont {Nascimbene}, \citenamefont {Chevy},\ and\ \citenamefont
  {Salomon}}]{Navon07052010}%
  \BibitemOpen
  \bibfield  {author} {\bibinfo {author} {\bibfnamefont {N.}~\bibnamefont
  {Navon}}, \bibinfo {author} {\bibfnamefont {S.}~\bibnamefont {Nascimbene}},
  \bibinfo {author} {\bibfnamefont {F.}~\bibnamefont {Chevy}}, \ and\ \bibinfo
  {author} {\bibfnamefont {C.}~\bibnamefont {Salomon}},\ }\href {\doibase
  10.1126/science.1187582} {\bibfield  {journal} {\bibinfo  {journal}
  {Science}\ }\textbf {\bibinfo {volume} {328}},\ \bibinfo {pages} {729}
  (\bibinfo {year} {2010})}\BibitemShut {NoStop}%
\bibitem [{\citenamefont {Ku}\ \emph {et~al.}(2012)\citenamefont {Ku},
  \citenamefont {Sommer}, \citenamefont {Cheuk},\ and\ \citenamefont
  {Zwierlein}}]{Ku03022012}%
  \BibitemOpen
  \bibfield  {author} {\bibinfo {author} {\bibfnamefont {M.~J.~H.}\
  \bibnamefont {Ku}}, \bibinfo {author} {\bibfnamefont {A.~T.}\ \bibnamefont
  {Sommer}}, \bibinfo {author} {\bibfnamefont {L.~W.}\ \bibnamefont {Cheuk}}, \
  and\ \bibinfo {author} {\bibfnamefont {M.~W.}\ \bibnamefont {Zwierlein}},\
  }\href {\doibase 10.1126/science.1214987} {\bibfield  {journal} {\bibinfo
  {journal} {Science}\ }\textbf {\bibinfo {volume} {335}},\ \bibinfo {pages}
  {563} (\bibinfo {year} {2012})}\BibitemShut {NoStop}%
\bibitem [{\citenamefont {Bloch}\ \emph {et~al.}(2008)\citenamefont {Bloch},
  \citenamefont {Dalibard},\ and\ \citenamefont {Zwerger}}]{RevModPhys.80.885}%
  \BibitemOpen
  \bibfield  {author} {\bibinfo {author} {\bibfnamefont {I.}~\bibnamefont
  {Bloch}}, \bibinfo {author} {\bibfnamefont {J.}~\bibnamefont {Dalibard}}, \
  and\ \bibinfo {author} {\bibfnamefont {W.}~\bibnamefont {Zwerger}},\ }\href
  {\doibase 10.1103/RevModPhys.80.885} {\bibfield  {journal} {\bibinfo
  {journal} {Rev. Mod. Phys.}\ }\textbf {\bibinfo {volume} {80}},\ \bibinfo
  {pages} {885} (\bibinfo {year} {2008})}\BibitemShut {NoStop}%
\bibitem [{\citenamefont {Zwerger}(2012)}]{Zwerger}%
  \BibitemOpen
  \bibinfo {editor} {\bibfnamefont {W.}~\bibnamefont {Zwerger}},\ ed.,\
  \href@noop {} {\emph {\bibinfo {title} {{The BCS-BEC Crossover and the
  Unitary Fermi Gas}}}}\ (\bibinfo  {publisher} {Springer, Berlin},\ \bibinfo
  {year} {2012})\BibitemShut {NoStop}%
\bibitem [{\citenamefont {Randeria}\ and\ \citenamefont
  {Taylor}(2013)}]{Randeria:2013kda}%
  \BibitemOpen
  \bibfield  {author} {\bibinfo {author} {\bibfnamefont {M.}~\bibnamefont
  {Randeria}}\ and\ \bibinfo {author} {\bibfnamefont {E.}~\bibnamefont
  {Taylor}},\ }\href@noop {} {\  (\bibinfo {year} {2013})},\ \Eprint
  {http://arxiv.org/abs/1306.5785} {arXiv:1306.5785 [cond-mat.quant-gas]}
  \BibitemShut {NoStop}%
\bibitem [{\citenamefont {Carlson}\ \emph {et~al.}(2003)\citenamefont
  {Carlson}, \citenamefont {Chang}, \citenamefont {Pandharipande},\ and\
  \citenamefont {Schmidt}}]{PhysRevLett.91.050401}%
  \BibitemOpen
  \bibfield  {author} {\bibinfo {author} {\bibfnamefont {J.}~\bibnamefont
  {Carlson}}, \bibinfo {author} {\bibfnamefont {S.-Y.}\ \bibnamefont {Chang}},
  \bibinfo {author} {\bibfnamefont {V.~R.}\ \bibnamefont {Pandharipande}}, \
  and\ \bibinfo {author} {\bibfnamefont {K.~E.}\ \bibnamefont {Schmidt}},\
  }\href {\doibase 10.1103/PhysRevLett.91.050401} {\bibfield  {journal}
  {\bibinfo  {journal} {Phys. Rev. Lett.}\ }\textbf {\bibinfo {volume} {91}},\
  \bibinfo {pages} {050401} (\bibinfo {year} {2003})}\BibitemShut {NoStop}%
\bibitem [{\citenamefont {Astrakharchik}\ \emph {et~al.}(2004)\citenamefont
  {Astrakharchik}, \citenamefont {Boronat}, \citenamefont {Casulleras},
  \citenamefont {Giorgini},\ and\ \citenamefont {S.}}]{PhysRevLett.93.200404}%
  \BibitemOpen
  \bibfield  {author} {\bibinfo {author} {\bibfnamefont {G.~E.}\ \bibnamefont
  {Astrakharchik}}, \bibinfo {author} {\bibfnamefont {J.}~\bibnamefont
  {Boronat}}, \bibinfo {author} {\bibfnamefont {J.}~\bibnamefont {Casulleras}},
  \bibinfo {author} {\bibnamefont {Giorgini}}, \ and\ \bibinfo {author}
  {\bibnamefont {S.}},\ }\href {\doibase 10.1103/PhysRevLett.93.200404}
  {\bibfield  {journal} {\bibinfo  {journal} {Phys. Rev. Lett.}\ }\textbf
  {\bibinfo {volume} {93}},\ \bibinfo {pages} {200404} (\bibinfo {year}
  {2004})}\BibitemShut {NoStop}%
\bibitem [{\citenamefont {Carlson}\ and\ \citenamefont
  {Reddy}(2005)}]{PhysRevLett.95.060401}%
  \BibitemOpen
  \bibfield  {author} {\bibinfo {author} {\bibfnamefont {J.}~\bibnamefont
  {Carlson}}\ and\ \bibinfo {author} {\bibfnamefont {S.}~\bibnamefont
  {Reddy}},\ }\href {\doibase 10.1103/PhysRevLett.95.060401} {\bibfield
  {journal} {\bibinfo  {journal} {Phys. Rev. Lett.}\ }\textbf {\bibinfo
  {volume} {95}},\ \bibinfo {pages} {060401} (\bibinfo {year}
  {2005})}\BibitemShut {NoStop}%
\bibitem [{\citenamefont {Astrakharchik}\ \emph {et~al.}(2005)\citenamefont
  {Astrakharchik}, \citenamefont {Boronat}, \citenamefont {Casulleras},\ and\
  \citenamefont {Giorgini}}]{PhysRevLett.95.230405}%
  \BibitemOpen
  \bibfield  {author} {\bibinfo {author} {\bibfnamefont {G.~E.}\ \bibnamefont
  {Astrakharchik}}, \bibinfo {author} {\bibfnamefont {J.}~\bibnamefont
  {Boronat}}, \bibinfo {author} {\bibfnamefont {J.}~\bibnamefont {Casulleras}},
  \ and\ \bibinfo {author} {\bibfnamefont {S.}~\bibnamefont {Giorgini}},\
  }\href {\doibase 10.1103/PhysRevLett.95.230405} {\bibfield  {journal}
  {\bibinfo  {journal} {Phys. Rev. Lett.}\ }\textbf {\bibinfo {volume} {95}},\
  \bibinfo {pages} {230405} (\bibinfo {year} {2005})}\BibitemShut {NoStop}%
\bibitem [{\citenamefont {Burovski}\ \emph {et~al.}(2006)\citenamefont
  {Burovski}, \citenamefont {Prokof'ev}, \citenamefont {Svistunov},\ and\
  \citenamefont {Troyer}}]{PhysRevLett.96.160402}%
  \BibitemOpen
  \bibfield  {author} {\bibinfo {author} {\bibfnamefont {E.}~\bibnamefont
  {Burovski}}, \bibinfo {author} {\bibfnamefont {N.}~\bibnamefont {Prokof'ev}},
  \bibinfo {author} {\bibfnamefont {B.}~\bibnamefont {Svistunov}}, \ and\
  \bibinfo {author} {\bibfnamefont {M.}~\bibnamefont {Troyer}},\ }\href
  {\doibase 10.1103/PhysRevLett.96.160402} {\bibfield  {journal} {\bibinfo
  {journal} {Phys. Rev. Lett.}\ }\textbf {\bibinfo {volume} {96}},\ \bibinfo
  {pages} {160402} (\bibinfo {year} {2006})}\BibitemShut {NoStop}%
\bibitem [{\citenamefont {Carlson}\ and\ \citenamefont
  {Reddy}(2008)}]{PhysRevLett.100.150403}%
  \BibitemOpen
  \bibfield  {author} {\bibinfo {author} {\bibfnamefont {J.}~\bibnamefont
  {Carlson}}\ and\ \bibinfo {author} {\bibfnamefont {S.}~\bibnamefont
  {Reddy}},\ }\href {\doibase 10.1103/PhysRevLett.100.150403} {\bibfield
  {journal} {\bibinfo  {journal} {Phys. Rev. Lett.}\ }\textbf {\bibinfo
  {volume} {100}},\ \bibinfo {pages} {150403} (\bibinfo {year}
  {2008})}\BibitemShut {NoStop}%
\bibitem [{\citenamefont {Bulgac}\ \emph {et~al.}(2008)\citenamefont {Bulgac},
  \citenamefont {Drut},\ and\ \citenamefont {Magierski}}]{PhysRevA.78.023625}%
  \BibitemOpen
  \bibfield  {author} {\bibinfo {author} {\bibfnamefont {A.}~\bibnamefont
  {Bulgac}}, \bibinfo {author} {\bibfnamefont {J.~E.}\ \bibnamefont {Drut}}, \
  and\ \bibinfo {author} {\bibfnamefont {P.}~\bibnamefont {Magierski}},\ }\href
  {\doibase 10.1103/PhysRevA.78.023625} {\bibfield  {journal} {\bibinfo
  {journal} {Phys. Rev. A}\ }\textbf {\bibinfo {volume} {78}},\ \bibinfo
  {pages} {023625} (\bibinfo {year} {2008})}\BibitemShut {NoStop}%
\bibitem [{\citenamefont {Morris}\ \emph {et~al.}(2010)\citenamefont {Morris},
  \citenamefont {Lopez~Rios},\ and\ \citenamefont
  {Needs}}]{PhysRevA.81.033619}%
  \BibitemOpen
  \bibfield  {author} {\bibinfo {author} {\bibfnamefont {A.~J.}\ \bibnamefont
  {Morris}}, \bibinfo {author} {\bibfnamefont {P.}~\bibnamefont {Lopez~Rios}},
  \ and\ \bibinfo {author} {\bibfnamefont {R.~J.}\ \bibnamefont {Needs}},\
  }\href {\doibase 10.1103/PhysRevA.81.033619} {\bibfield  {journal} {\bibinfo
  {journal} {Phys. Rev. A}\ }\textbf {\bibinfo {volume} {81}},\ \bibinfo
  {pages} {033619} (\bibinfo {year} {2010})}\BibitemShut {NoStop}%
\bibitem [{\citenamefont {Goulko}\ and\ \citenamefont
  {Wingate}(2010)}]{PhysRevA.82.053621}%
  \BibitemOpen
  \bibfield  {author} {\bibinfo {author} {\bibfnamefont {O.}~\bibnamefont
  {Goulko}}\ and\ \bibinfo {author} {\bibfnamefont {M.}~\bibnamefont
  {Wingate}},\ }\href {\doibase 10.1103/PhysRevA.82.053621} {\bibfield
  {journal} {\bibinfo  {journal} {Phys. Rev. A}\ }\textbf {\bibinfo {volume}
  {82}},\ \bibinfo {pages} {053621} (\bibinfo {year} {2010})}\BibitemShut
  {NoStop}%
\bibitem [{\citenamefont {Drut}\ \emph {et~al.}(2011)\citenamefont {Drut},
  \citenamefont {L\"ahde},\ and\ \citenamefont {Ten}}]{PhysRevLett.106.205302}%
  \BibitemOpen
  \bibfield  {author} {\bibinfo {author} {\bibfnamefont {J.~E.}\ \bibnamefont
  {Drut}}, \bibinfo {author} {\bibfnamefont {T.~A.}\ \bibnamefont {L\"ahde}}, \
  and\ \bibinfo {author} {\bibfnamefont {T.}~\bibnamefont {Ten}},\ }\href
  {\doibase 10.1103/PhysRevLett.106.205302} {\bibfield  {journal} {\bibinfo
  {journal} {Phys. Rev. Lett.}\ }\textbf {\bibinfo {volume} {106}},\ \bibinfo
  {pages} {205302} (\bibinfo {year} {2011})}\BibitemShut {NoStop}%
\bibitem [{\citenamefont {Carlson}\ \emph {et~al.}(2011)\citenamefont
  {Carlson}, \citenamefont {Gandolfi}, \citenamefont {Schmidt},\ and\
  \citenamefont {Zhang}}]{PhysRevA.84.061602}%
  \BibitemOpen
  \bibfield  {author} {\bibinfo {author} {\bibfnamefont {J.}~\bibnamefont
  {Carlson}}, \bibinfo {author} {\bibfnamefont {S.}~\bibnamefont {Gandolfi}},
  \bibinfo {author} {\bibfnamefont {K.~E.}\ \bibnamefont {Schmidt}}, \ and\
  \bibinfo {author} {\bibfnamefont {S.}~\bibnamefont {Zhang}},\ }\href
  {\doibase 10.1103/PhysRevA.84.061602} {\bibfield  {journal} {\bibinfo
  {journal} {Phys. Rev. A}\ }\textbf {\bibinfo {volume} {84}},\ \bibinfo
  {pages} {061602} (\bibinfo {year} {2011})}\BibitemShut {NoStop}%
\bibitem [{\citenamefont {Haussmann}(1993)}]{HaussmannZPhys91}%
  \BibitemOpen
  \bibfield  {author} {\bibinfo {author} {\bibfnamefont {R.}~\bibnamefont
  {Haussmann}},\ }\href@noop {} {\bibfield  {journal} {\bibinfo  {journal} {Z.
  Phys. B}\ }\textbf {\bibinfo {volume} {91}},\ \bibinfo {pages} {291}
  (\bibinfo {year} {1993})}\BibitemShut {NoStop}%
\bibitem [{\citenamefont {Haussmann}(1994)}]{PhysRevB.49.12975}%
  \BibitemOpen
  \bibfield  {author} {\bibinfo {author} {\bibfnamefont {R.}~\bibnamefont
  {Haussmann}},\ }\href {\doibase 10.1103/PhysRevB.49.12975} {\bibfield
  {journal} {\bibinfo  {journal} {Phys. Rev. B}\ }\textbf {\bibinfo {volume}
  {49}},\ \bibinfo {pages} {12975} (\bibinfo {year} {1994})}\BibitemShut
  {NoStop}%
\bibitem [{\citenamefont {Haussmann}\ \emph {et~al.}(2007)\citenamefont
  {Haussmann}, \citenamefont {Rantner}, \citenamefont {Cerrito},\ and\
  \citenamefont {Zwerger}}]{PhysRevA.75.023610}%
  \BibitemOpen
  \bibfield  {author} {\bibinfo {author} {\bibfnamefont {R.}~\bibnamefont
  {Haussmann}}, \bibinfo {author} {\bibfnamefont {W.}~\bibnamefont {Rantner}},
  \bibinfo {author} {\bibfnamefont {S.}~\bibnamefont {Cerrito}}, \ and\
  \bibinfo {author} {\bibfnamefont {W.}~\bibnamefont {Zwerger}},\ }\href
  {\doibase 10.1103/PhysRevA.75.023610} {\bibfield  {journal} {\bibinfo
  {journal} {Phys. Rev. A}\ }\textbf {\bibinfo {volume} {75}},\ \bibinfo
  {pages} {023610} (\bibinfo {year} {2007})}\BibitemShut {NoStop}%
\bibitem [{\citenamefont {Haussmann}\ \emph {et~al.}(2009)\citenamefont
  {Haussmann}, \citenamefont {Punk},\ and\ \citenamefont
  {Zwerger}}]{PhysRevA.80.063612}%
  \BibitemOpen
  \bibfield  {author} {\bibinfo {author} {\bibfnamefont {R.}~\bibnamefont
  {Haussmann}}, \bibinfo {author} {\bibfnamefont {M.}~\bibnamefont {Punk}}, \
  and\ \bibinfo {author} {\bibfnamefont {W.}~\bibnamefont {Zwerger}},\ }\href
  {\doibase 10.1103/PhysRevA.80.063612} {\bibfield  {journal} {\bibinfo
  {journal} {Phys. Rev. A}\ }\textbf {\bibinfo {volume} {80}},\ \bibinfo
  {pages} {063612} (\bibinfo {year} {2009})}\BibitemShut {NoStop}%
\bibitem [{\citenamefont {Palestini}\ \emph {et~al.}(2010)\citenamefont
  {Palestini}, \citenamefont {Perali}, \citenamefont {Pieri},\ and\
  \citenamefont {Strinati}}]{PhysRevA.82.021605}%
  \BibitemOpen
  \bibfield  {author} {\bibinfo {author} {\bibfnamefont {F.}~\bibnamefont
  {Palestini}}, \bibinfo {author} {\bibfnamefont {A.}~\bibnamefont {Perali}},
  \bibinfo {author} {\bibfnamefont {P.}~\bibnamefont {Pieri}}, \ and\ \bibinfo
  {author} {\bibfnamefont {G.~C.}\ \bibnamefont {Strinati}},\ }\href {\doibase
  10.1103/PhysRevA.82.021605} {\bibfield  {journal} {\bibinfo  {journal} {Phys.
  Rev. A}\ }\textbf {\bibinfo {volume} {82}},\ \bibinfo {pages} {021605}
  (\bibinfo {year} {2010})}\BibitemShut {NoStop}%
\bibitem [{\citenamefont {Enss}\ \emph {et~al.}(2011)\citenamefont {Enss},
  \citenamefont {Haussmann},\ and\ \citenamefont {Zwerger}}]{Enss2011770}%
  \BibitemOpen
  \bibfield  {author} {\bibinfo {author} {\bibfnamefont {T.}~\bibnamefont
  {Enss}}, \bibinfo {author} {\bibfnamefont {R.}~\bibnamefont {Haussmann}}, \
  and\ \bibinfo {author} {\bibfnamefont {W.}~\bibnamefont {Zwerger}},\ }\href
  {\doibase 10.1016/j.aop.2010.10.002} {\bibfield  {journal} {\bibinfo
  {journal} {Annals of Physics}\ }\textbf {\bibinfo {volume} {326}},\ \bibinfo
  {pages} {770 } (\bibinfo {year} {2011})}\BibitemShut {NoStop}%
\bibitem [{\citenamefont {Hu}\ \emph {et~al.}(2011)\citenamefont {Hu},
  \citenamefont {Liu},\ and\ \citenamefont {Drummond}}]{HuDrummond}%
  \BibitemOpen
  \bibfield  {author} {\bibinfo {author} {\bibfnamefont {H.}~\bibnamefont
  {Hu}}, \bibinfo {author} {\bibfnamefont {X.-J.}\ \bibnamefont {Liu}}, \ and\
  \bibinfo {author} {\bibfnamefont {P.~D.}\ \bibnamefont {Drummond}},\ }\href
  {http://stacks.iop.org/1367-2630/13/i=3/a=035007} {\bibfield  {journal}
  {\bibinfo  {journal} {New Journal of Physics}\ }\textbf {\bibinfo {volume}
  {13}},\ \bibinfo {pages} {035007} (\bibinfo {year} {2011})}\BibitemShut
  {NoStop}%
\bibitem [{\citenamefont {Nishida}\ and\ \citenamefont
  {Son}(2006)}]{PhysRevLett.97.050403}%
  \BibitemOpen
  \bibfield  {author} {\bibinfo {author} {\bibfnamefont {Y.}~\bibnamefont
  {Nishida}}\ and\ \bibinfo {author} {\bibfnamefont {D.~T.}\ \bibnamefont
  {Son}},\ }\href {\doibase 10.1103/PhysRevLett.97.050403} {\bibfield
  {journal} {\bibinfo  {journal} {Phys. Rev. Lett.}\ }\textbf {\bibinfo
  {volume} {97}},\ \bibinfo {pages} {050403} (\bibinfo {year}
  {2006})}\BibitemShut {NoStop}%
\bibitem [{\citenamefont {Nishida}\ and\ \citenamefont
  {Son}(2007)}]{PhysRevA.75.063617}%
  \BibitemOpen
  \bibfield  {author} {\bibinfo {author} {\bibfnamefont {Y.}~\bibnamefont
  {Nishida}}\ and\ \bibinfo {author} {\bibfnamefont {D.~T.}\ \bibnamefont
  {Son}},\ }\href {\doibase 10.1103/PhysRevA.75.063617} {\bibfield  {journal}
  {\bibinfo  {journal} {Phys. Rev. A}\ }\textbf {\bibinfo {volume} {75}},\
  \bibinfo {pages} {063617} (\bibinfo {year} {2007})}\BibitemShut {NoStop}%
\bibitem [{\citenamefont {Nishida}(2007)}]{PhysRevA.75.063618}%
  \BibitemOpen
  \bibfield  {author} {\bibinfo {author} {\bibfnamefont {Y.}~\bibnamefont
  {Nishida}},\ }\href {\doibase 10.1103/PhysRevA.75.063618} {\bibfield
  {journal} {\bibinfo  {journal} {Phys. Rev. A}\ }\textbf {\bibinfo {volume}
  {75}},\ \bibinfo {pages} {063618} (\bibinfo {year} {2007})}\BibitemShut
  {NoStop}%
\bibitem [{\citenamefont {Enss}(2012)}]{PhysRevA.86.013616}%
  \BibitemOpen
  \bibfield  {author} {\bibinfo {author} {\bibfnamefont {T.}~\bibnamefont
  {Enss}},\ }\href {\doibase 10.1103/PhysRevA.86.013616} {\bibfield  {journal}
  {\bibinfo  {journal} {Phys. Rev. A}\ }\textbf {\bibinfo {volume} {86}},\
  \bibinfo {pages} {013616} (\bibinfo {year} {2012})}\BibitemShut {NoStop}%
\bibitem [{\citenamefont {Diehl}\ and\ \citenamefont
  {Wetterich}(2006)}]{PhysRevA.73.033615}%
  \BibitemOpen
  \bibfield  {author} {\bibinfo {author} {\bibfnamefont {S.}~\bibnamefont
  {Diehl}}\ and\ \bibinfo {author} {\bibfnamefont {C.}~\bibnamefont
  {Wetterich}},\ }\href {\doibase 10.1103/PhysRevA.73.033615} {\bibfield
  {journal} {\bibinfo  {journal} {Phys. Rev. A}\ }\textbf {\bibinfo {volume}
  {73}},\ \bibinfo {pages} {033615} (\bibinfo {year} {2006})}\BibitemShut
  {NoStop}%
\bibitem [{\citenamefont {Diehl}\ and\ \citenamefont
  {Wetterich}(2007)}]{Diehl:2005ae}%
  \BibitemOpen
  \bibfield  {author} {\bibinfo {author} {\bibfnamefont {S.}~\bibnamefont
  {Diehl}}\ and\ \bibinfo {author} {\bibfnamefont {C.}~\bibnamefont
  {Wetterich}},\ }\href {\doibase 10.1016/j.nuclphysb.2007.02.026} {\bibfield
  {journal} {\bibinfo  {journal} {Nucl.Phys.}\ }\textbf {\bibinfo {volume}
  {B770}},\ \bibinfo {pages} {206} (\bibinfo {year} {2007})}\BibitemShut
  {NoStop}%
\bibitem [{\citenamefont {Diener}\ \emph {et~al.}(2008)\citenamefont {Diener},
  \citenamefont {Sensarma},\ and\ \citenamefont
  {Randeria}}]{PhysRevA.77.023626}%
  \BibitemOpen
  \bibfield  {author} {\bibinfo {author} {\bibfnamefont {R.~B.}\ \bibnamefont
  {Diener}}, \bibinfo {author} {\bibfnamefont {R.}~\bibnamefont {Sensarma}}, \
  and\ \bibinfo {author} {\bibfnamefont {M.}~\bibnamefont {Randeria}},\ }\href
  {\doibase 10.1103/PhysRevA.77.023626} {\bibfield  {journal} {\bibinfo
  {journal} {Phys. Rev. A}\ }\textbf {\bibinfo {volume} {77}},\ \bibinfo
  {pages} {023626} (\bibinfo {year} {2008})}\BibitemShut {NoStop}%
\bibitem [{\citenamefont {Nikoli\ifmmode~\acute{c}\else \'{c}\fi{}}\ and\
  \citenamefont {Sachdev}(2007)}]{PhysRevA.75.033608}%
  \BibitemOpen
  \bibfield  {author} {\bibinfo {author} {\bibfnamefont {P.}~\bibnamefont
  {Nikoli\ifmmode~\acute{c}\else \'{c}\fi{}}}\ and\ \bibinfo {author}
  {\bibfnamefont {S.}~\bibnamefont {Sachdev}},\ }\href {\doibase
  10.1103/PhysRevA.75.033608} {\bibfield  {journal} {\bibinfo  {journal} {Phys.
  Rev. A}\ }\textbf {\bibinfo {volume} {75}},\ \bibinfo {pages} {033608}
  (\bibinfo {year} {2007})}\BibitemShut {NoStop}%
\bibitem [{\citenamefont {Gubbels}\ and\ \citenamefont
  {Stoof}(2008)}]{PhysRevLett.100.140407}%
  \BibitemOpen
  \bibfield  {author} {\bibinfo {author} {\bibfnamefont {K.~B.}\ \bibnamefont
  {Gubbels}}\ and\ \bibinfo {author} {\bibfnamefont {H.~T.~C.}\ \bibnamefont
  {Stoof}},\ }\href {\doibase 10.1103/PhysRevLett.100.140407} {\bibfield
  {journal} {\bibinfo  {journal} {Phys. Rev. Lett.}\ }\textbf {\bibinfo
  {volume} {100}},\ \bibinfo {pages} {140407} (\bibinfo {year}
  {2008})}\BibitemShut {NoStop}%
\bibitem [{\citenamefont {Gubbels}\ and\ \citenamefont
  {Stoof}(2011)}]{PhysRevA.84.013610}%
  \BibitemOpen
  \bibfield  {author} {\bibinfo {author} {\bibfnamefont {K.~B.}\ \bibnamefont
  {Gubbels}}\ and\ \bibinfo {author} {\bibfnamefont {H.~T.~C.}\ \bibnamefont
  {Stoof}},\ }\href {\doibase 10.1103/PhysRevA.84.013610} {\bibfield  {journal}
  {\bibinfo  {journal} {Phys. Rev. A}\ }\textbf {\bibinfo {volume} {84}},\
  \bibinfo {pages} {013610} (\bibinfo {year} {2011})}\BibitemShut {NoStop}%
\bibitem [{\citenamefont {Gubbels}\ and\ \citenamefont
  {Stoof}(2013)}]{Gubbels:2013mda}%
  \BibitemOpen
  \bibfield  {author} {\bibinfo {author} {\bibfnamefont {K.}~\bibnamefont
  {Gubbels}}\ and\ \bibinfo {author} {\bibfnamefont {H.}~\bibnamefont
  {Stoof}},\ }\href {\doibase 10.1016/j.physrep.2012.11.004} {\bibfield
  {journal} {\bibinfo  {journal} {Phys.Rept.}\ }\textbf {\bibinfo {volume}
  {525}},\ \bibinfo {pages} {255} (\bibinfo {year} {2013})}\BibitemShut
  {NoStop}%
\bibitem [{\citenamefont {Birse}\ \emph {et~al.}(2005)\citenamefont {Birse},
  \citenamefont {Krippa}, \citenamefont {McGovern},\ and\ \citenamefont
  {Walet}}]{Birse:2004ha}%
  \BibitemOpen
  \bibfield  {author} {\bibinfo {author} {\bibfnamefont {M.~C.}\ \bibnamefont
  {Birse}}, \bibinfo {author} {\bibfnamefont {B.}~\bibnamefont {Krippa}},
  \bibinfo {author} {\bibfnamefont {J.~A.}\ \bibnamefont {McGovern}}, \ and\
  \bibinfo {author} {\bibfnamefont {N.~R.}\ \bibnamefont {Walet}},\ }\href
  {\doibase 10.1016/j.physletb.2004.11.044} {\bibfield  {journal} {\bibinfo
  {journal} {Phys.Lett.}\ }\textbf {\bibinfo {volume} {B605}},\ \bibinfo
  {pages} {287} (\bibinfo {year} {2005})}\BibitemShut {NoStop}%
\bibitem [{\citenamefont {Diehl}\ \emph
  {et~al.}(2007{\natexlab{a}})\citenamefont {Diehl}, \citenamefont {Gies},
  \citenamefont {Pawlowski},\ and\ \citenamefont
  {Wetterich}}]{PhysRevA.76.021602}%
  \BibitemOpen
  \bibfield  {author} {\bibinfo {author} {\bibfnamefont {S.}~\bibnamefont
  {Diehl}}, \bibinfo {author} {\bibfnamefont {H.}~\bibnamefont {Gies}},
  \bibinfo {author} {\bibfnamefont {J.~M.}\ \bibnamefont {Pawlowski}}, \ and\
  \bibinfo {author} {\bibfnamefont {C.}~\bibnamefont {Wetterich}},\ }\href
  {\doibase 10.1103/PhysRevA.76.021602} {\bibfield  {journal} {\bibinfo
  {journal} {Phys. Rev. A}\ }\textbf {\bibinfo {volume} {76}},\ \bibinfo
  {pages} {021602} (\bibinfo {year} {2007}{\natexlab{a}})}\BibitemShut
  {NoStop}%
\bibitem [{\citenamefont {Diehl}\ \emph
  {et~al.}(2007{\natexlab{b}})\citenamefont {Diehl}, \citenamefont {Gies},
  \citenamefont {Pawlowski},\ and\ \citenamefont
  {Wetterich}}]{PhysRevA.76.053627}%
  \BibitemOpen
  \bibfield  {author} {\bibinfo {author} {\bibfnamefont {S.}~\bibnamefont
  {Diehl}}, \bibinfo {author} {\bibfnamefont {H.}~\bibnamefont {Gies}},
  \bibinfo {author} {\bibfnamefont {J.~M.}\ \bibnamefont {Pawlowski}}, \ and\
  \bibinfo {author} {\bibfnamefont {C.}~\bibnamefont {Wetterich}},\ }\href
  {\doibase 10.1103/PhysRevA.76.053627} {\bibfield  {journal} {\bibinfo
  {journal} {Phys. Rev. A}\ }\textbf {\bibinfo {volume} {76}},\ \bibinfo
  {pages} {053627} (\bibinfo {year} {2007}{\natexlab{b}})}\BibitemShut
  {NoStop}%
\bibitem [{\citenamefont {Floerchinger}\ \emph {et~al.}(2008)\citenamefont
  {Floerchinger}, \citenamefont {Scherer}, \citenamefont {Diehl},\ and\
  \citenamefont {Wetterich}}]{PhysRevB.78.174528}%
  \BibitemOpen
  \bibfield  {author} {\bibinfo {author} {\bibfnamefont {S.}~\bibnamefont
  {Floerchinger}}, \bibinfo {author} {\bibfnamefont {M.}~\bibnamefont
  {Scherer}}, \bibinfo {author} {\bibfnamefont {S.}~\bibnamefont {Diehl}}, \
  and\ \bibinfo {author} {\bibfnamefont {C.}~\bibnamefont {Wetterich}},\ }\href
  {\doibase 10.1103/PhysRevB.78.174528} {\bibfield  {journal} {\bibinfo
  {journal} {Phys. Rev. B}\ }\textbf {\bibinfo {volume} {78}},\ \bibinfo
  {pages} {174528} (\bibinfo {year} {2008})}\BibitemShut {NoStop}%
\bibitem [{\citenamefont {Diehl}\ \emph {et~al.}(2010)\citenamefont {Diehl},
  \citenamefont {Floerchinger}, \citenamefont {Gies}, \citenamefont
  {Pawlowkski},\ and\ \citenamefont {Wetterich}}]{ANDP:ANDP201010458}%
  \BibitemOpen
  \bibfield  {author} {\bibinfo {author} {\bibfnamefont {S.}~\bibnamefont
  {Diehl}}, \bibinfo {author} {\bibfnamefont {S.}~\bibnamefont {Floerchinger}},
  \bibinfo {author} {\bibfnamefont {H.}~\bibnamefont {Gies}}, \bibinfo {author}
  {\bibfnamefont {J.}~\bibnamefont {Pawlowkski}}, \ and\ \bibinfo {author}
  {\bibfnamefont {C.}~\bibnamefont {Wetterich}},\ }\href {\doibase
  10.1002/andp.201010458} {\bibfield  {journal} {\bibinfo  {journal} {Annalen
  der Physik}\ }\textbf {\bibinfo {volume} {522}},\ \bibinfo {pages} {615}
  (\bibinfo {year} {2010})}\BibitemShut {NoStop}%
\bibitem [{\citenamefont {Tanizaki}\ \emph
  {et~al.}(2013{\natexlab{a}})\citenamefont {Tanizaki}, \citenamefont
  {Fejős},\ and\ \citenamefont {Hatsuda}}]{Tanizaki:2013doa}%
  \BibitemOpen
  \bibfield  {author} {\bibinfo {author} {\bibfnamefont {Y.}~\bibnamefont
  {Tanizaki}}, \bibinfo {author} {\bibfnamefont {G.}~\bibnamefont {Fejős}}, \
  and\ \bibinfo {author} {\bibfnamefont {T.}~\bibnamefont {Hatsuda}},\
  }\href@noop {} {\  (\bibinfo {year} {2013}{\natexlab{a}})},\ \Eprint
  {http://arxiv.org/abs/1310.5800} {arXiv:1310.5800 [cond-mat.quant-gas]}
  \BibitemShut {NoStop}%
\bibitem [{\citenamefont {Tanizaki}\ \emph
  {et~al.}(2013{\natexlab{b}})\citenamefont {Tanizaki}, \citenamefont
  {Fejős},\ and\ \citenamefont {Hatsuda}}]{Tanizaki:2013fba}%
  \BibitemOpen
  \bibfield  {author} {\bibinfo {author} {\bibfnamefont {Y.}~\bibnamefont
  {Tanizaki}}, \bibinfo {author} {\bibfnamefont {G.}~\bibnamefont {Fejős}}, \
  and\ \bibinfo {author} {\bibfnamefont {T.}~\bibnamefont {Hatsuda}},\
  }\href@noop {} {\  (\bibinfo {year} {2013}{\natexlab{b}})},\ \Eprint
  {http://arxiv.org/abs/1311.4229} {arXiv:1311.4229 [cond-mat.quant-gas]}
  \BibitemShut {NoStop}%
\bibitem [{\citenamefont {Tanizaki}(2013)}]{Tanizaki:2013yba}%
  \BibitemOpen
  \bibfield  {author} {\bibinfo {author} {\bibfnamefont {Y.}~\bibnamefont
  {Tanizaki}},\ }\href@noop {} {\  (\bibinfo {year} {2013})},\ \Eprint
  {http://arxiv.org/abs/1311.4157} {arXiv:1311.4157 [cond-mat.quant-gas]}
  \BibitemShut {NoStop}%
\bibitem [{\citenamefont {Gies}\ and\ \citenamefont
  {Wetterich}(2002)}]{PhysRevD.65.065001}%
  \BibitemOpen
  \bibfield  {author} {\bibinfo {author} {\bibfnamefont {H.}~\bibnamefont
  {Gies}}\ and\ \bibinfo {author} {\bibfnamefont {C.}~\bibnamefont
  {Wetterich}},\ }\href {\doibase 10.1103/PhysRevD.65.065001} {\bibfield
  {journal} {\bibinfo  {journal} {Phys. Rev. D}\ }\textbf {\bibinfo {volume}
  {65}},\ \bibinfo {pages} {065001} (\bibinfo {year} {2002})}\BibitemShut
  {NoStop}%
\bibitem [{\citenamefont {Floerchinger}(2010)}]{Floerchinger:2010da}%
  \BibitemOpen
  \bibfield  {author} {\bibinfo {author} {\bibfnamefont {S.}~\bibnamefont
  {Floerchinger}},\ }\href {\doibase 10.1140/epjc/s10052-010-1361-z} {\bibfield
   {journal} {\bibinfo  {journal} {Eur.Phys.J.}\ }\textbf {\bibinfo {volume}
  {C69}},\ \bibinfo {pages} {119} (\bibinfo {year} {2010})}\BibitemShut
  {NoStop}%
\bibitem [{\citenamefont {Pawlowski}(2007)}]{Pawlowski20072831}%
  \BibitemOpen
  \bibfield  {author} {\bibinfo {author} {\bibfnamefont {J.~M.}\ \bibnamefont
  {Pawlowski}},\ }\href {\doibase 10.1016/j.aop.2007.01.007} {\bibfield
  {journal} {\bibinfo  {journal} {Annals of Physics}\ }\textbf {\bibinfo
  {volume} {322}},\ \bibinfo {pages} {2831 } (\bibinfo {year}
  {2007})}\BibitemShut {NoStop}%
\bibitem [{\citenamefont {Wetterich}(1991)}]{Wetterich:1989xg}%
  \BibitemOpen
  \bibfield  {author} {\bibinfo {author} {\bibfnamefont {C.}~\bibnamefont
  {Wetterich}},\ }\href {\doibase 10.1016/0550-3213(91)90099-J} {\bibfield
  {journal} {\bibinfo  {journal} {Nucl.Phys.}\ }\textbf {\bibinfo {volume}
  {B352}},\ \bibinfo {pages} {529} (\bibinfo {year} {1991})}\BibitemShut
  {NoStop}%
\bibitem [{\citenamefont {Wetterich}(1993{\natexlab{a}})}]{Wetterich:1993ne}%
  \BibitemOpen
  \bibfield  {author} {\bibinfo {author} {\bibfnamefont {C.}~\bibnamefont
  {Wetterich}},\ }\href {\doibase 10.1007/BF01560044} {\bibfield  {journal}
  {\bibinfo  {journal} {Z.Phys.}\ }\textbf {\bibinfo {volume} {C60}},\ \bibinfo
  {pages} {461} (\bibinfo {year} {1993}{\natexlab{a}})}\BibitemShut {NoStop}%
\bibitem [{\citenamefont {Polchinski}(1984)}]{Polchinski:1983gv}%
  \BibitemOpen
  \bibfield  {author} {\bibinfo {author} {\bibfnamefont {J.}~\bibnamefont
  {Polchinski}},\ }\href {\doibase 10.1016/0550-3213(84)90287-6} {\bibfield
  {journal} {\bibinfo  {journal} {Nucl.Phys.}\ }\textbf {\bibinfo {volume}
  {B231}},\ \bibinfo {pages} {269} (\bibinfo {year} {1984})}\BibitemShut
  {NoStop}%
\bibitem [{\citenamefont {Wetterich}(1993{\natexlab{b}})}]{Wetterich1993}%
  \BibitemOpen
  \bibfield  {author} {\bibinfo {author} {\bibfnamefont {C.}~\bibnamefont
  {Wetterich}},\ }\href {\doibase 10.1016/0370-2693(93)90726-X} {\bibfield
  {journal} {\bibinfo  {journal} {Physics Letters B}\ }\textbf {\bibinfo
  {volume} {301}},\ \bibinfo {pages} {90} (\bibinfo {year}
  {1993}{\natexlab{b}})}\BibitemShut {NoStop}%
\bibitem [{\citenamefont {Berges}\ \emph {et~al.}(2002)\citenamefont {Berges},
  \citenamefont {Tetradis},\ and\ \citenamefont {Wetterich}}]{Berges:2000ew}%
  \BibitemOpen
  \bibfield  {author} {\bibinfo {author} {\bibfnamefont {J.}~\bibnamefont
  {Berges}}, \bibinfo {author} {\bibfnamefont {N.}~\bibnamefont {Tetradis}}, \
  and\ \bibinfo {author} {\bibfnamefont {C.}~\bibnamefont {Wetterich}},\ }\href
  {\doibase 10.1016/S0370-1573(01)00098-9} {\bibfield  {journal} {\bibinfo
  {journal} {Phys.Rept.}\ }\textbf {\bibinfo {volume} {363}},\ \bibinfo {pages}
  {223} (\bibinfo {year} {2002})}\BibitemShut {NoStop}%
\bibitem [{\citenamefont {Wetterich}(2001)}]{Wetterich:2001kra}%
  \BibitemOpen
  \bibfield  {author} {\bibinfo {author} {\bibfnamefont {C.}~\bibnamefont
  {Wetterich}},\ }\href {\doibase 10.1142/S0217751X01004591} {\bibfield
  {journal} {\bibinfo  {journal} {Int.J.Mod.Phys.}\ }\textbf {\bibinfo {volume}
  {A16}},\ \bibinfo {pages} {1951} (\bibinfo {year} {2001})}\BibitemShut
  {NoStop}%
\bibitem [{\citenamefont {Gies}(2012)}]{Gies:2006wv}%
  \BibitemOpen
  \bibfield  {author} {\bibinfo {author} {\bibfnamefont {H.}~\bibnamefont
  {Gies}},\ }\href {\doibase 10.1007/978-3-642-27320-9_6} {\bibfield  {journal}
  {\bibinfo  {journal} {Lect.Notes Phys.}\ }\textbf {\bibinfo {volume} {852}},\
  \bibinfo {pages} {287} (\bibinfo {year} {2012})}\BibitemShut {NoStop}%
\bibitem [{\citenamefont {Schaefer}\ and\ \citenamefont
  {Wambach}(2008)}]{Schaefer:2006sr}%
  \BibitemOpen
  \bibfield  {author} {\bibinfo {author} {\bibfnamefont {B.-J.}\ \bibnamefont
  {Schaefer}}\ and\ \bibinfo {author} {\bibfnamefont {J.}~\bibnamefont
  {Wambach}},\ }\href {\doibase 10.1134/S1063779608070083} {\bibfield
  {journal} {\bibinfo  {journal} {Phys.Part.Nucl.}\ }\textbf {\bibinfo {volume}
  {39}},\ \bibinfo {pages} {1025} (\bibinfo {year} {2008})}\BibitemShut
  {NoStop}%
\bibitem [{\citenamefont {Delamotte}(2012)}]{Delamotte:2007pf}%
  \BibitemOpen
  \bibfield  {author} {\bibinfo {author} {\bibfnamefont {B.}~\bibnamefont
  {Delamotte}},\ }\href {\doibase 10.1007/978-3-642-27320-9_2} {\bibfield
  {journal} {\bibinfo  {journal} {Lect.Notes Phys.}\ }\textbf {\bibinfo
  {volume} {852}},\ \bibinfo {pages} {49} (\bibinfo {year} {2012})}\BibitemShut
  {NoStop}%
\bibitem [{\citenamefont {Kopietz}\ \emph {et~al.}(2010)\citenamefont
  {Kopietz}, \citenamefont {Bartosch},\ and\ \citenamefont
  {Sch\"{u}tz}}]{Kopietz2010}%
  \BibitemOpen
  \bibfield  {author} {\bibinfo {author} {\bibfnamefont {P.}~\bibnamefont
  {Kopietz}}, \bibinfo {author} {\bibfnamefont {L.}~\bibnamefont {Bartosch}}, \
  and\ \bibinfo {author} {\bibfnamefont {F.}~\bibnamefont {Sch\"{u}tz}},\
  }\href@noop {} {\emph {\bibinfo {title} {{Introduction to the Functional
  Renormalization Group}}}}\ (\bibinfo  {publisher} {Springer, Berlin},\
  \bibinfo {year} {2010})\BibitemShut {NoStop}%
\bibitem [{\citenamefont {Metzner}\ \emph {et~al.}(2012)\citenamefont
  {Metzner}, \citenamefont {Salmhofer}, \citenamefont {Honerkamp},
  \citenamefont {Meden},\ and\ \citenamefont {Schonhammer}}]{Metzner:2011cw}%
  \BibitemOpen
  \bibfield  {author} {\bibinfo {author} {\bibfnamefont {W.}~\bibnamefont
  {Metzner}}, \bibinfo {author} {\bibfnamefont {M.}~\bibnamefont {Salmhofer}},
  \bibinfo {author} {\bibfnamefont {C.}~\bibnamefont {Honerkamp}}, \bibinfo
  {author} {\bibfnamefont {V.}~\bibnamefont {Meden}}, \ and\ \bibinfo {author}
  {\bibfnamefont {K.}~\bibnamefont {Schonhammer}},\ }\href@noop {} {\bibfield
  {journal} {\bibinfo  {journal} {Rev.Mod.Phys.}\ }\textbf {\bibinfo {volume}
  {84}},\ \bibinfo {pages} {299} (\bibinfo {year} {2012})}\BibitemShut
  {NoStop}%
\bibitem [{\citenamefont {Braun}(2012)}]{Braun:2011pp}%
  \BibitemOpen
  \bibfield  {author} {\bibinfo {author} {\bibfnamefont {J.}~\bibnamefont
  {Braun}},\ }\href {\doibase 10.1088/0954-3899/39/3/033001} {\bibfield
  {journal} {\bibinfo  {journal} {J.Phys.}\ }\textbf {\bibinfo {volume}
  {G39}},\ \bibinfo {pages} {033001} (\bibinfo {year} {2012})}\BibitemShut
  {NoStop}%
\bibitem [{\citenamefont {Scherer}\ \emph {et~al.}(2011)\citenamefont
  {Scherer}, \citenamefont {Floerchinger},\ and\ \citenamefont
  {Gies}}]{Scherer:2010sv}%
  \BibitemOpen
  \bibfield  {author} {\bibinfo {author} {\bibfnamefont {M.~M.}\ \bibnamefont
  {Scherer}}, \bibinfo {author} {\bibfnamefont {S.}~\bibnamefont
  {Floerchinger}}, \ and\ \bibinfo {author} {\bibfnamefont {H.}~\bibnamefont
  {Gies}},\ }\href@noop {} {\bibfield  {journal} {\bibinfo  {journal} {Phil.
  Trans. R. Soc. A}\ }\textbf {\bibinfo {volume} {368}},\ \bibinfo {pages}
  {2779} (\bibinfo {year} {2011})}\BibitemShut {NoStop}%
\bibitem [{\citenamefont {Boettcher}\ \emph {et~al.}(2012)\citenamefont
  {Boettcher}, \citenamefont {Pawlowski},\ and\ \citenamefont
  {Diehl}}]{Boettcher:2012cm}%
  \BibitemOpen
  \bibfield  {author} {\bibinfo {author} {\bibfnamefont {I.}~\bibnamefont
  {Boettcher}}, \bibinfo {author} {\bibfnamefont {J.~M.}\ \bibnamefont
  {Pawlowski}}, \ and\ \bibinfo {author} {\bibfnamefont {S.}~\bibnamefont
  {Diehl}},\ }\href {\doibase 10.1016/j.nuclphysbps.2012.06.004} {\bibfield
  {journal} {\bibinfo  {journal} {Nucl.Phys.Proc.Suppl.}\ }\textbf {\bibinfo
  {volume} {228}},\ \bibinfo {pages} {63} (\bibinfo {year} {2012})}\BibitemShut
  {NoStop}%
\bibitem [{\citenamefont {Canet}\ \emph {et~al.}(2010)\citenamefont {Canet},
  \citenamefont {Chat\'e}, \citenamefont {Delamotte},\ and\ \citenamefont
  {Wschebor}}]{PhysRevLett.104.150601}%
  \BibitemOpen
  \bibfield  {author} {\bibinfo {author} {\bibfnamefont {L.}~\bibnamefont
  {Canet}}, \bibinfo {author} {\bibfnamefont {H.}~\bibnamefont {Chat\'e}},
  \bibinfo {author} {\bibfnamefont {B.}~\bibnamefont {Delamotte}}, \ and\
  \bibinfo {author} {\bibfnamefont {N.}~\bibnamefont {Wschebor}},\ }\href
  {\doibase 10.1103/PhysRevLett.104.150601} {\bibfield  {journal} {\bibinfo
  {journal} {Phys. Rev. Lett.}\ }\textbf {\bibinfo {volume} {104}},\ \bibinfo
  {pages} {150601} (\bibinfo {year} {2010})}\BibitemShut {NoStop}%
\bibitem [{\citenamefont {Canet}\ \emph {et~al.}(2011)\citenamefont {Canet},
  \citenamefont {Chat\'e}, \citenamefont {Delamotte},\ and\ \citenamefont
  {Wschebor}}]{PhysRevE.84.061128}%
  \BibitemOpen
  \bibfield  {author} {\bibinfo {author} {\bibfnamefont {L.}~\bibnamefont
  {Canet}}, \bibinfo {author} {\bibfnamefont {H.}~\bibnamefont {Chat\'e}},
  \bibinfo {author} {\bibfnamefont {B.}~\bibnamefont {Delamotte}}, \ and\
  \bibinfo {author} {\bibfnamefont {N.}~\bibnamefont {Wschebor}},\ }\href
  {\doibase 10.1103/PhysRevE.84.061128} {\bibfield  {journal} {\bibinfo
  {journal} {Phys. Rev. E}\ }\textbf {\bibinfo {volume} {84}},\ \bibinfo
  {pages} {061128} (\bibinfo {year} {2011})}\BibitemShut {NoStop}%
\bibitem [{\citenamefont {Kloss}\ \emph {et~al.}(2012)\citenamefont {Kloss},
  \citenamefont {Canet},\ and\ \citenamefont {Wschebor}}]{PhysRevE.86.051124}%
  \BibitemOpen
  \bibfield  {author} {\bibinfo {author} {\bibfnamefont {T.}~\bibnamefont
  {Kloss}}, \bibinfo {author} {\bibfnamefont {L.}~\bibnamefont {Canet}}, \ and\
  \bibinfo {author} {\bibfnamefont {N.}~\bibnamefont {Wschebor}},\ }\href
  {\doibase 10.1103/PhysRevE.86.051124} {\bibfield  {journal} {\bibinfo
  {journal} {Phys. Rev. E}\ }\textbf {\bibinfo {volume} {86}},\ \bibinfo
  {pages} {051124} (\bibinfo {year} {2012})}\BibitemShut {NoStop}%
\bibitem [{\citenamefont {Floerchinger}(2012)}]{Floerchinger:2011sc}%
  \BibitemOpen
  \bibfield  {author} {\bibinfo {author} {\bibfnamefont {S.}~\bibnamefont
  {Floerchinger}},\ }\href {\doibase 10.1007/JHEP05(2012)021} {\bibfield
  {journal} {\bibinfo  {journal} {JHEP}\ }\textbf {\bibinfo {volume} {1205}},\
  \bibinfo {pages} {021} (\bibinfo {year} {2012})}\BibitemShut {NoStop}%
\bibitem [{\citenamefont {Floerchinger}(2013)}]{Floerchinger:2013tp}%
  \BibitemOpen
  \bibfield  {author} {\bibinfo {author} {\bibfnamefont {S.}~\bibnamefont
  {Floerchinger}},\ }\href@noop {} {\  (\bibinfo {year} {2013})},\ \Eprint
  {http://arxiv.org/abs/1301.6542} {arXiv:1301.6542 [nucl-th]} \BibitemShut
  {NoStop}%
\bibitem [{\citenamefont {Diehl}\ \emph {et~al.}(2008)\citenamefont {Diehl},
  \citenamefont {Krahl},\ and\ \citenamefont {Scherer}}]{PhysRevC.78.034001}%
  \BibitemOpen
  \bibfield  {author} {\bibinfo {author} {\bibfnamefont {S.}~\bibnamefont
  {Diehl}}, \bibinfo {author} {\bibfnamefont {H.~C.}\ \bibnamefont {Krahl}}, \
  and\ \bibinfo {author} {\bibfnamefont {M.}~\bibnamefont {Scherer}},\ }\href
  {\doibase 10.1103/PhysRevC.78.034001} {\bibfield  {journal} {\bibinfo
  {journal} {Phys. Rev. C}\ }\textbf {\bibinfo {volume} {78}},\ \bibinfo
  {pages} {034001} (\bibinfo {year} {2008})}\BibitemShut {NoStop}%
\bibitem [{\citenamefont {Krippa}\ \emph {et~al.}(2010)\citenamefont {Krippa},
  \citenamefont {Walet},\ and\ \citenamefont {Birse}}]{PhysRevA.81.043628}%
  \BibitemOpen
  \bibfield  {author} {\bibinfo {author} {\bibfnamefont {B.}~\bibnamefont
  {Krippa}}, \bibinfo {author} {\bibfnamefont {N.~R.}\ \bibnamefont {Walet}}, \
  and\ \bibinfo {author} {\bibfnamefont {M.~C.}\ \bibnamefont {Birse}},\ }\href
  {\doibase 10.1103/PhysRevA.81.043628} {\bibfield  {journal} {\bibinfo
  {journal} {Phys. Rev. A}\ }\textbf {\bibinfo {volume} {81}},\ \bibinfo
  {pages} {043628} (\bibinfo {year} {2010})}\BibitemShut {NoStop}%
\bibitem [{\citenamefont {Birse}\ \emph {et~al.}(2011)\citenamefont {Birse},
  \citenamefont {Krippa},\ and\ \citenamefont {Walet}}]{PhysRevA.83.023621}%
  \BibitemOpen
  \bibfield  {author} {\bibinfo {author} {\bibfnamefont {M.~C.}\ \bibnamefont
  {Birse}}, \bibinfo {author} {\bibfnamefont {B.}~\bibnamefont {Krippa}}, \
  and\ \bibinfo {author} {\bibfnamefont {N.~R.}\ \bibnamefont {Walet}},\ }\href
  {\doibase 10.1103/PhysRevA.83.023621} {\bibfield  {journal} {\bibinfo
  {journal} {Phys. Rev. A}\ }\textbf {\bibinfo {volume} {83}},\ \bibinfo
  {pages} {023621} (\bibinfo {year} {2011})}\BibitemShut {NoStop}%
\bibitem [{\citenamefont {Schnoerr}\ \emph {et~al.}(2013)\citenamefont
  {Schnoerr}, \citenamefont {Boettcher}, \citenamefont {Pawlowski},\ and\
  \citenamefont {Wetterich}}]{Schnoerr:2013bk}%
  \BibitemOpen
  \bibfield  {author} {\bibinfo {author} {\bibfnamefont {D.}~\bibnamefont
  {Schnoerr}}, \bibinfo {author} {\bibfnamefont {I.}~\bibnamefont {Boettcher}},
  \bibinfo {author} {\bibfnamefont {J.~M.}\ \bibnamefont {Pawlowski}}, \ and\
  \bibinfo {author} {\bibfnamefont {C.}~\bibnamefont {Wetterich}},\ }\href
  {\doibase 10.1016/j.aop.2013.03.013} {\bibfield  {journal} {\bibinfo
  {journal} {Annals Phys.}\ }\textbf {\bibinfo {volume} {334}},\ \bibinfo
  {pages} {83} (\bibinfo {year} {2013})}\BibitemShut {NoStop}%
\bibitem [{\citenamefont {Wetterich}(2008)}]{PhysRevB.77.064504}%
  \BibitemOpen
  \bibfield  {author} {\bibinfo {author} {\bibfnamefont {C.}~\bibnamefont
  {Wetterich}},\ }\href {\doibase 10.1103/PhysRevB.77.064504} {\bibfield
  {journal} {\bibinfo  {journal} {Phys. Rev. B}\ }\textbf {\bibinfo {volume}
  {77}},\ \bibinfo {pages} {064504} (\bibinfo {year} {2008})}\BibitemShut
  {NoStop}%
\bibitem [{\citenamefont {Dupuis}(2009)}]{PhysRevLett.102.190401}%
  \BibitemOpen
  \bibfield  {author} {\bibinfo {author} {\bibfnamefont {N.}~\bibnamefont
  {Dupuis}},\ }\href {\doibase 10.1103/PhysRevLett.102.190401} {\bibfield
  {journal} {\bibinfo  {journal} {Phys. Rev. Lett.}\ }\textbf {\bibinfo
  {volume} {102}},\ \bibinfo {pages} {190401} (\bibinfo {year}
  {2009})}\BibitemShut {NoStop}%
\bibitem [{\citenamefont {Boettcher}\ \emph {et~al.}(2013)\citenamefont
  {Boettcher}, \citenamefont {Diehl}, \citenamefont {Pawlowski},\ and\
  \citenamefont {Wetterich}}]{PhysRevA.87.023606}%
  \BibitemOpen
  \bibfield  {author} {\bibinfo {author} {\bibfnamefont {I.}~\bibnamefont
  {Boettcher}}, \bibinfo {author} {\bibfnamefont {S.}~\bibnamefont {Diehl}},
  \bibinfo {author} {\bibfnamefont {J.~M.}\ \bibnamefont {Pawlowski}}, \ and\
  \bibinfo {author} {\bibfnamefont {C.}~\bibnamefont {Wetterich}},\ }\href
  {\doibase 10.1103/PhysRevA.87.023606} {\bibfield  {journal} {\bibinfo
  {journal} {Phys. Rev. A}\ }\textbf {\bibinfo {volume} {87}},\ \bibinfo
  {pages} {023606} (\bibinfo {year} {2013})}\BibitemShut {NoStop}%
\bibitem [{\citenamefont {Litim}\ \emph {et~al.}(2006)\citenamefont {Litim},
  \citenamefont {Pawlowski},\ and\ \citenamefont {Vergara}}]{Litim:2006nn}%
  \BibitemOpen
  \bibfield  {author} {\bibinfo {author} {\bibfnamefont {D.~F.}\ \bibnamefont
  {Litim}}, \bibinfo {author} {\bibfnamefont {J.~M.}\ \bibnamefont
  {Pawlowski}}, \ and\ \bibinfo {author} {\bibfnamefont {L.}~\bibnamefont
  {Vergara}},\ }\href@noop {} {\  (\bibinfo {year} {2006})},\ \Eprint
  {http://arxiv.org/abs/hep-th/0602140} {arXiv:hep-th/0602140 [hep-th]}
  \BibitemShut {NoStop}%
\bibitem [{\citenamefont {Floerchinger}\ \emph {et~al.}(2010)\citenamefont
  {Floerchinger}, \citenamefont {Scherer},\ and\ \citenamefont
  {Wetterich}}]{PhysRevA.81.063619}%
  \BibitemOpen
  \bibfield  {author} {\bibinfo {author} {\bibfnamefont {S.}~\bibnamefont
  {Floerchinger}}, \bibinfo {author} {\bibfnamefont {M.~M.}\ \bibnamefont
  {Scherer}}, \ and\ \bibinfo {author} {\bibfnamefont {C.}~\bibnamefont
  {Wetterich}},\ }\href {\doibase 10.1103/PhysRevA.81.063619} {\bibfield
  {journal} {\bibinfo  {journal} {Phys. Rev. A}\ }\textbf {\bibinfo {volume}
  {81}},\ \bibinfo {pages} {063619} (\bibinfo {year} {2010})}\BibitemShut
  {NoStop}%
\bibitem [{\citenamefont {Schirotzek}\ \emph {et~al.}(2008)\citenamefont
  {Schirotzek}, \citenamefont {Shin}, \citenamefont {Schunck},\ and\
  \citenamefont {Ketterle}}]{PhysRevLett.101.140403}%
  \BibitemOpen
  \bibfield  {author} {\bibinfo {author} {\bibfnamefont {A.}~\bibnamefont
  {Schirotzek}}, \bibinfo {author} {\bibfnamefont {Y.-i.}\ \bibnamefont
  {Shin}}, \bibinfo {author} {\bibfnamefont {C.~H.}\ \bibnamefont {Schunck}}, \
  and\ \bibinfo {author} {\bibfnamefont {W.}~\bibnamefont {Ketterle}},\ }\href
  {\doibase 10.1103/PhysRevLett.101.140403} {\bibfield  {journal} {\bibinfo
  {journal} {Phys. Rev. Lett.}\ }\textbf {\bibinfo {volume} {101}},\ \bibinfo
  {pages} {140403} (\bibinfo {year} {2008})}\BibitemShut {NoStop}%
\bibitem [{\citenamefont {Bartosch}\ \emph {et~al.}(2009)\citenamefont
  {Bartosch}, \citenamefont {Kopietz},\ and\ \citenamefont
  {Ferraz}}]{PhysRevB.80.104514}%
  \BibitemOpen
  \bibfield  {author} {\bibinfo {author} {\bibfnamefont {L.}~\bibnamefont
  {Bartosch}}, \bibinfo {author} {\bibfnamefont {P.}~\bibnamefont {Kopietz}}, \
  and\ \bibinfo {author} {\bibfnamefont {A.}~\bibnamefont {Ferraz}},\ }\href
  {\doibase 10.1103/PhysRevB.80.104514} {\bibfield  {journal} {\bibinfo
  {journal} {Phys. Rev. B}\ }\textbf {\bibinfo {volume} {80}},\ \bibinfo
  {pages} {104514} (\bibinfo {year} {2009})}\BibitemShut {NoStop}%
\bibitem [{\citenamefont {Eichhorn}\ \emph {et~al.}(2013)\citenamefont
  {Eichhorn}, \citenamefont {Mesterházy},\ and\ \citenamefont
  {Scherer}}]{Eichhorn:2013zza}%
  \BibitemOpen
  \bibfield  {author} {\bibinfo {author} {\bibfnamefont {A.}~\bibnamefont
  {Eichhorn}}, \bibinfo {author} {\bibfnamefont {D.}~\bibnamefont
  {Mesterházy}}, \ and\ \bibinfo {author} {\bibfnamefont {M.~M.}\ \bibnamefont
  {Scherer}},\ }\href {\doibase 10.1103/PhysRevE.88.042141} {\bibfield
  {journal} {\bibinfo  {journal} {Phys.Rev.}\ }\textbf {\bibinfo {volume}
  {E88}},\ \bibinfo {pages} {042141} (\bibinfo {year} {2013})}\BibitemShut
  {NoStop}%
\bibitem [{\citenamefont {Schaefer}\ and\ \citenamefont
  {Wambach}(2005)}]{Schaefer:2004en}%
  \BibitemOpen
  \bibfield  {author} {\bibinfo {author} {\bibfnamefont {B.-J.}\ \bibnamefont
  {Schaefer}}\ and\ \bibinfo {author} {\bibfnamefont {J.}~\bibnamefont
  {Wambach}},\ }\href {\doibase 10.1016/j.nuclphysa.2005.04.012} {\bibfield
  {journal} {\bibinfo  {journal} {Nucl.Phys.}\ }\textbf {\bibinfo {volume}
  {A757}},\ \bibinfo {pages} {479} (\bibinfo {year} {2005})}\BibitemShut
  {NoStop}%
\bibitem [{\citenamefont {Herbst}\ \emph {et~al.}(2011)\citenamefont {Herbst},
  \citenamefont {Pawlowski},\ and\ \citenamefont {Schaefer}}]{Herbst:2010rf}%
  \BibitemOpen
  \bibfield  {author} {\bibinfo {author} {\bibfnamefont {T.~K.}\ \bibnamefont
  {Herbst}}, \bibinfo {author} {\bibfnamefont {J.~M.}\ \bibnamefont
  {Pawlowski}}, \ and\ \bibinfo {author} {\bibfnamefont {B.-J.}\ \bibnamefont
  {Schaefer}},\ }\href {\doibase 10.1016/j.physletb.2010.12.003} {\bibfield
  {journal} {\bibinfo  {journal} {Phys.Lett.}\ }\textbf {\bibinfo {volume}
  {B696}},\ \bibinfo {pages} {58} (\bibinfo {year} {2011})}\BibitemShut
  {NoStop}%
\bibitem [{\citenamefont {Herbst}\ \emph {et~al.}(2013)\citenamefont {Herbst},
  \citenamefont {Pawlowski},\ and\ \citenamefont
  {Schaefer}}]{PhysRevD.88.014007}%
  \BibitemOpen
  \bibfield  {author} {\bibinfo {author} {\bibfnamefont {T.~K.}\ \bibnamefont
  {Herbst}}, \bibinfo {author} {\bibfnamefont {J.~M.}\ \bibnamefont
  {Pawlowski}}, \ and\ \bibinfo {author} {\bibfnamefont {B.-J.}\ \bibnamefont
  {Schaefer}},\ }\href {\doibase 10.1103/PhysRevD.88.014007} {\bibfield
  {journal} {\bibinfo  {journal} {Phys. Rev. D}\ }\textbf {\bibinfo {volume}
  {88}},\ \bibinfo {pages} {014007} (\bibinfo {year} {2013})}\BibitemShut
  {NoStop}%
\bibitem [{\citenamefont {Rennecke}\ and\ \citenamefont
  {Pawlowski}(2013)}]{Fabian}%
  \BibitemOpen
  \bibfield  {author} {\bibinfo {author} {\bibfnamefont {F.}~\bibnamefont
  {Rennecke}}\ and\ \bibinfo {author} {\bibfnamefont {J.~M.}\ \bibnamefont
  {Pawlowski}},\ }\href@noop {} {\bibfield  {journal} {\bibinfo  {journal} {in
  preparation}\ } (\bibinfo {year} {2013})}\BibitemShut {NoStop}%
\bibitem [{\citenamefont {Litim}(2000)}]{Litim:2000ci}%
  \BibitemOpen
  \bibfield  {author} {\bibinfo {author} {\bibfnamefont {D.~F.}\ \bibnamefont
  {Litim}},\ }\href {\doibase 10.1016/S0370-2693(00)00748-6} {\bibfield
  {journal} {\bibinfo  {journal} {Phys.Lett.}\ }\textbf {\bibinfo {volume}
  {B486}},\ \bibinfo {pages} {92} (\bibinfo {year} {2000})}\BibitemShut
  {NoStop}%
\bibitem [{\citenamefont {Litim}(2001{\natexlab{a}})}]{Litim:2001up}%
  \BibitemOpen
  \bibfield  {author} {\bibinfo {author} {\bibfnamefont {D.~F.}\ \bibnamefont
  {Litim}},\ }\href {\doibase 10.1103/PhysRevD.64.105007} {\bibfield  {journal}
  {\bibinfo  {journal} {Phys.Rev.}\ }\textbf {\bibinfo {volume} {D64}},\
  \bibinfo {pages} {105007} (\bibinfo {year} {2001}{\natexlab{a}})}\BibitemShut
  {NoStop}%
\bibitem [{\citenamefont {Litim}(2001{\natexlab{b}})}]{Litim:2001fd}%
  \BibitemOpen
  \bibfield  {author} {\bibinfo {author} {\bibfnamefont {D.~F.}\ \bibnamefont
  {Litim}},\ }\href {\doibase 10.1142/S0217751X01004748} {\bibfield  {journal}
  {\bibinfo  {journal} {Int.J.Mod.Phys.}\ }\textbf {\bibinfo {volume} {A16}},\
  \bibinfo {pages} {2081} (\bibinfo {year} {2001}{\natexlab{b}})}\BibitemShut
  {NoStop}%
\bibitem [{\citenamefont {Ran\ifmmode~\mbox{\c{c}}\else \c{c}\fi{}on}\ and\
  \citenamefont {Dupuis}(2012{\natexlab{a}})}]{PhysRevA.85.063607}%
  \BibitemOpen
  \bibfield  {author} {\bibinfo {author} {\bibfnamefont {A.}~\bibnamefont
  {Ran\ifmmode~\mbox{\c{c}}\else \c{c}\fi{}on}}\ and\ \bibinfo {author}
  {\bibfnamefont {N.}~\bibnamefont {Dupuis}},\ }\href {\doibase
  10.1103/PhysRevA.85.063607} {\bibfield  {journal} {\bibinfo  {journal} {Phys.
  Rev. A}\ }\textbf {\bibinfo {volume} {85}},\ \bibinfo {pages} {063607}
  (\bibinfo {year} {2012}{\natexlab{a}})}\BibitemShut {NoStop}%
\bibitem [{\citenamefont {Ran\ifmmode~\mbox{\c{c}}\else \c{c}\fi{}on}\ and\
  \citenamefont {Dupuis}(2011)}]{PhysRevB.84.174513}%
  \BibitemOpen
  \bibfield  {author} {\bibinfo {author} {\bibfnamefont {A.}~\bibnamefont
  {Ran\ifmmode~\mbox{\c{c}}\else \c{c}\fi{}on}}\ and\ \bibinfo {author}
  {\bibfnamefont {N.}~\bibnamefont {Dupuis}},\ }\href {\doibase
  10.1103/PhysRevB.84.174513} {\bibfield  {journal} {\bibinfo  {journal} {Phys.
  Rev. B}\ }\textbf {\bibinfo {volume} {84}},\ \bibinfo {pages} {174513}
  (\bibinfo {year} {2011})}\BibitemShut {NoStop}%
\bibitem [{\citenamefont {Ran\ifmmode~\mbox{\c{c}}\else \c{c}\fi{}on}\ and\
  \citenamefont {Dupuis}(2012{\natexlab{b}})}]{PhysRevA.86.043624}%
  \BibitemOpen
  \bibfield  {author} {\bibinfo {author} {\bibfnamefont {A.}~\bibnamefont
  {Ran\ifmmode~\mbox{\c{c}}\else \c{c}\fi{}on}}\ and\ \bibinfo {author}
  {\bibfnamefont {N.}~\bibnamefont {Dupuis}},\ }\href {\doibase
  10.1103/PhysRevA.86.043624} {\bibfield  {journal} {\bibinfo  {journal} {Phys.
  Rev. A}\ }\textbf {\bibinfo {volume} {86}},\ \bibinfo {pages} {043624}
  (\bibinfo {year} {2012}{\natexlab{b}})}\BibitemShut {NoStop}%
\end{thebibliography}%

\end{document}